\documentclass[journal=mamobx]{achemso}

\SectionNumbersOn
\mciteErrorOnUnknownfalse

\usepackage{amsmath, mathtools, environ}
\usepackage{amstext}
\usepackage{amssymb}
\usepackage{graphicx}
\usepackage{import}
\usepackage{appendix}
\usepackage{lscape}
\usepackage{subcaption}
\usepackage[boxed]{algorithm}
\usepackage{algorithm}
\usepackage{algpseudocode}
\usepackage{multirow}

\usepackage{longtable}

\usepackage{mleftright}

\newcommand{\etal}{\textit{et al.\ }}
\newcommand{\pd}[2]{\frac{\partial #1}{\partial #2}}

\newcommand{\vect}[1]{\boldsymbol{#1}}

\newcommand{\hf}{{\frac12}}

\newcommand{\diff}[1]{\, d#1}

\newcommand{\dims}[0]{\mathcal{D}}

\usepackage[dvipsnames]{xcolor}

\newcommand{\mylinelabel}[1]{}
\newcommand{\abs}[1]{\left| #1 \right|}
\newcommand{\ddn}[2][]{\partial_{\vect{n}_{#1}} #2}

\newcommand{\lap}{\nabla^2}

\newcommand{\range}[2]{\left[ #1, #2 \right]}
\newcommand{\rin}[1][\vect{r}]{\vect{r} \in}

\newcommand{\vn}{v_{\vect{n}}}
\newcommand{\vecr}{\vect{r}}

\newcommand{\of}[1]{\mleft( #1 \mright)}

\newcommand{\fd}[2]{\frac{\delta #1}{\delta #2}}

\newcommand{\Diff}[1]{\mathcal{D}#1}

\newcommand{\qf}{q}
\newcommand{\qb}{\hat{q}}

\newcommand{\deriv}[2]{\frac{d #1}{d #2}}
\newcommand{\ham}{H}
\newcommand{\lag}{\mathcal{L}}
\newcommand{\Q}{Q}

\newcommand{\lb}{\hat{\lambda}}

\newcommand{\dds}[1]{\partial_s #1}
\newcommand{\myint}{\int\limits}
\newcommand{\tand}{\textrm{and}}
\newcommand{\heav}{H}
\newcommand{\dirac}{\delta}

\newcommand{\XN}[1]{\chi_{#1} N}

\newcommand{\detail}[1]{}

\newcommand{\patchspaces}[1]{
\NewEnviron{my#1}{%
\par\begingroup\setstretch{1.0}%
\begin{#1}
\BODY
\end{#1}%
\endgroup}}

\patchspaces{align}
\patchspaces{align*}
\patchspaces{multline}
\patchspaces{multline*}

\NewEnviron{myalignat}[1]{%
\par\begingroup\setstretch{1.0}%
\begin{alignat}{#1}
\BODY
\end{alignat}%
\endgroup}

\NewEnviron{myalignat*}[1]{%
\par\begingroup\setstretch{1.0}%
\begin{alignat*}{#1}
\BODY
\end{alignat*}%
\endgroup}

\newcommand{\maxwidth}[1]{%
\ifdim\width>#1%
#1%
\else%
\width%
\fi%
}

\title{Equilibrium of Free Surfaces and Nanoparticles in Self-Consistent Field Theory of Block Copolymers}
\author{Daniil Bochkov}
\email{bochkov.ds@gmail.com}
\affiliation{Department of Mechanical Engineering, University of California, Santa Barbara, Santa Barbara, CA, 93106}
\author{Ivana Bagaric}
\affiliation{Institute of Fluid Dynamics, Swiss Federal Institute of Technology Zurich, Zurich, Switzerland}
\author{Gaddiel Ouaknin}
\affiliation{Department of Mechanical Engineering, University of California, Santa Barbara, Santa Barbara, CA, 93106}
\author{Fr\'ed\'eric Gibou}
\affiliation{Department of Mechanical Engineering, University of California, Santa Barbara, Santa Barbara, CA, 93106}
\alsoaffiliation{Department of Computer Science, University of California, Santa Barbara, Santa Barbara, CA, 93106}
\begin{document}

\begin{abstract}
This work presents a general and unified theory describing block copolymer self-assembly in the presence of free surfaces and nanoparticles in the context of Self-Consistent Filed Theory. Specifically, the derived theory applies to free and tethered polymer chains, nanoparticles of any shape, arbitrary non-uniform surface energies and grafting densities, and takes into account a possible formation of triple-junction points (e.g., polymer-air-substrate). One of the main ingredients of the proposed theory is a simple procedure for a consistent imposition of boundary conditions on surfaces with non-zero surface energies and/or non-zero grafting densities that results in singularity-free pressure-like fields, which is crucial for the calculation of forces. The generality of the theory is demonstrated using several representative examples such as the meniscus formation in the graphoepitaxy and the co-assembly of ``polarized'' nano-rods and diblock copolymer melts.

\end{abstract}

\maketitle

\section{Introduction}
Self-Consistent Field Theory\cite{matsen2001standard,Fredrickson:06:The-equilibrium-theo} (SCFT) is a powerful framework that has become a standard approach for modeling block copolymer (BCP) self-assembly. In its standard form, SCFT is suitable for investigating periodic (bulk) morphologies of self-assembling BCPs or in static confinements. However, in certain situations, such as in the presence of polymer-air interface and/or nanoparticles, the confining boundaries are free to move and in fact must be considered as part of the solution. 

The importance of simulating free surface block copolymer melts comes from the fact that it is not uncommon in practical applications for the polymer material to be exposed to a gaseous or liquid phase. Free surfaces provide a channel for releasing internal stresses of polymer melts. This can lead not only to alterations in spatial dimensions of self-assembled features, but to the potential relaxation into a different, more energetically stable morphology.
For example, besides the meniscus formation near confining walls, this can also lead to the formation of unique free surface features of block copolymers such as holes, islands, and terraces \cite{Croll2006,maher2016structure}. The simulation of block copolymer free surfaces is challenging due to the two-way coupling between the free surface and the internal polymer morphology. Within the SCFT framework, free polymer surfaces are typically modeled by introducing into the system an additional chemical species that play the role of the air \cite{Kim;Matsen:09:Droplets-of-structur,Stasiak;McGraw;Dalnoki-Veress;Etal:12:Step-edges-in-thin-f,Carpenter2015}. While being simple and straightforward to implement, this approach is not free of caveats. Among them is the fact that only relatively low interaction strengths between dissimilar species can be used in the SCFT for numerical calculations to remain stable. This leads to the inability to impose physically high incompatibility between the polymer material and the surrounding air as well as to a smeared-out polymer-air interface. In Ouaknin \etal\cite{Ouaknin;Laachi;Bochkov;Etal:17:Functional-level-set}, a conceptually different framework has been proposed, where the polymer-air interface is treated in a sharp fashion and its governing equations are derived based on the PDE-constrained shape sensitivity analysis. However, results in Ouaknin \etal\cite{Ouaknin;Laachi;Bochkov;Etal:17:Functional-level-set} are limited to neutral polymer-air interfaces due to the well-known inconsistency within the SCFT to model selectively attractive/repulsive boundaries\cite{Fredrickson:06:The-equilibrium-theo}. 

Mixing of BCPs and nanoparticles constitutes a promising avenue for the creation of functional materials and advanced polymer self-assembly applications \cite{bockstaller2005block}. Similar to the case of free polymer surfaces, the theoretical description of these systems is complicated by  two-way couplings: on the one hand, the self-assembling polymer structures guide the placement of nanoparticles but, on the other hand, the presence of nanoparticles influence the resulting polymer morphologies. The problem of co-assembly of BCPs and nanoparticles in the SCFT framework has been approached from explicit and implicit perspectives. In the former case, each nanoparticle in the system is described explicitly while in the latter case, the distribution of nanoparticles is described in terms of density fields. Each framework has its own advantages: the first methodology allows for a straightforward incorporation of such features as excluded volume effects, arbitrary shapes of particles, complex particle-particle interactions; the second methodology is more efficient at obtaining statistical information about particles configurations. The explicit approach was employed in the hybrid particle-field method proposed in Sides \etal\cite{sides2006hybrid}, in which nanoparticles were modeled by the so-called cavity function that represents a gradual transition between polymer and particle's material and the relaxation of nanoparticles towards the equilibrium state is accomplished using an approximate derivatives of the system's energy with respect to the particles' positions. Among implicit approaches are the hybrid SCFT-DFT method proposed in Thompson \etal\cite{thompson2001predicting} and the field-theoretic method described in Koski \etal \cite{koski2013field}.

In this work we derive a generalized and unified theory in the context of SCFT describing equilibrium of polymer free surfaces and submerged nanoparticles extending the ideas introduced in Ouaknin \etal\cite{Ouaknin;Laachi;Bochkov;Etal:17:Functional-level-set}. The interfaces between polymer melt and surrounding material are considered to be sharp. Surface energies and surface grafting densities are modeled in a unified consistent approach that results in pressure-like fields free of singularities near domain boundaries. The theory is derived for an arbitrary mixture of linear BCP chains containing both free and tethered polymer chains in the case of arbitrary non-uniform surface energies and grafting densities. The shapes of substrate and nanoparticles are considered to be arbitrary as well and three-phase (e.g., polyer-air-substrate) triple junctions are explicitly taken into account. The theory is implemented numerically in the Level-Set Method framework along the lines of Ouaknin \etal\cite{Ouaknin;Laachi;Delaney;etal:16:Self-consistent-fiel}, validated using benchmark tests, and applied to model the meniscus formation in graphoepitaxy and the co-assembly of nano-rods and diblock copolymer melts.

The rest of this manuscript is organized as follows. Section \ref{sec:model} presents a consistent approach for imposing arbitrary surface energies and grafting densities for incompressible BCP melts in the SCFT framework. In sections \ref{sec:free-surface} we derive the expressions for forces acting on polymer free surface and submerged nanoparticles. In section \ref{sec:numerics}, we briefly discuss the numerical implementation used in this work. In section \ref{sec:results} we present a number of examples demonstrating the capabilities of the proposed methods. Finally, section \ref{sec:conclusion} draws conclusions and discusses future research directions.

\section{A consistent approach for imposing arbitrary surface energies and grafting densities in SCFT}\label{sec:model}

\begin{figure}[!h]
\centering
\includegraphics[width=.3\textwidth]{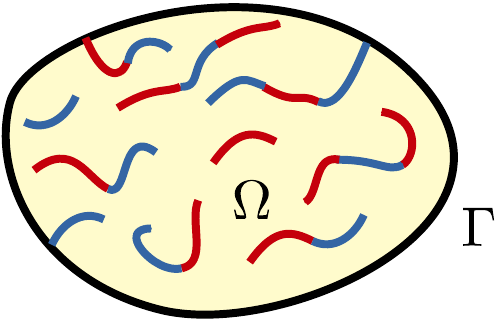}
\caption{Problem geometry and notation used in section \ref{sec:model}.}
\label{fig:scft:problem-statement}
\end{figure}

Let us consider an incompressible mixture of $n$ linear block copolymer chains composed of two chemical species denoted as A and B in a region $\Omega$ with boundary $\Gamma$ as pictured in Figure~\ref{fig:scft:problem-statement}. Specifically, we assume that there exist $M$ types (e.g., AB, ABA, BABA, etc) of polymer chains and that the fraction of chains belonging to a particular type is given by $\phi_\alpha$, $\alpha = 1,\ldots, M$. The length of chains for a given type $\alpha$ is given by $l_\alpha N$, where $N$ is a representative number of statistical segments in polymer chains. The individual shapes of polymer chains are parametrized by the contour variable $s$ and are denoted as $\vect{r}_{\alpha, j}=\vect{r}_{\alpha, j}\of{s}$, such that $s\in\range{0}{l_\alpha}$ for $\alpha \in \range{1}{M}$, $j\in\range{1}{\phi_\alpha n}$. The precise composition of chains for a given type $\alpha$ is given by function $\zeta_\alpha\of{s}$ taking values $-1$ and $1$, such that $\zeta_\alpha\of{s}=-1$ corresponds to an A segment at location $s$ and $\zeta_\alpha\of{s}=1$ corresponds to a B segment at location $s$. We assume that fraction $G_\alpha$ of polymer chains of type $\alpha$ are grafted with density $g_\alpha\of{\vect{r}}$ onto domain boundary $\Gamma$. We also denote the surface energy (potentially, nonuniform) of species $A$ and $B$ on boundary $\Gamma$ as $\gamma_A =\gamma_A\of{\vect{r}}$ and $\gamma_B = \gamma_B\of{\vect{r}}$, correspondingly. Furthermore, as a common practice it is assumed that $A$ and $B$ statistical segments have equal volumes $v_0$. In the following we, first, briefly outline the derivation of a mean-field approximation for this system (see, e.g., Fredrickson\cite{Fredrickson:06:The-equilibrium-theo} for a more detailed discussion) and then describe a general procedure to ensure non-singular behavior of the pressure-like chemical field.

The total stretching energy of polymer chains can be written as:
\detail{
\begin{myalign*}
U_0 = \beta^{-1}\sum_{j=1}^{n} \frac{3}{2Nb^2}\myint_{0}^{1}\left\vert \frac{d \vect{r}_j\of{s}}{ds} \right\vert^2 \diff{s} 
\end{myalign*}}
\begin{myalign*}
U_0 = 
\beta^{-1} \sum_{\alpha=1}^{M} \sum_{j=1}^{n_\alpha} \myint_{0}^{l_\alpha} \frac{1}{4} \left\vert \frac{d \vect{r}_{\alpha,j}\of{s}}{ds} \right\vert^2 \diff{s} 
\end{myalign*}
where $\beta = \frac{1}{k_B T}$. Note that spatial quantities are measured in units of a representative gyration radius $R_g = b \sqrt{\frac{N}{6}}$, where $b$ is the statistical length of a single polymer bead (assumed equal for all beads in a polymer chain).

The interaction energy between dissimilar chemical blocks is taken into account through a Flory-Higgins effective interaction parameter $\chi_{AB}$:  
\begin{myalign*}
U_1 = \beta^{-1}\frac{1}{v_0} \chi_{AB} \myint_{\Omega} \hat{\rho}_A \hat{\rho}_B \diff{\vect{r}},
\end{myalign*}
where the microscopic number concentrations, that is, fraction of segments at any point, are given by:
\begin{myalign*}
\hat{\rho}_A &= 
v_0 N \sum_{\alpha=1}^{M} \sum_{j=1}^{n_\alpha} 
\myint_0^{l_\alpha} 
\heav\of{-\zeta_\alpha\of{s}} \delta ( \vect{r} - \vect{r}_{j}\of{s} ) 
\diff{s} 
\\ 
\hat{\rho}_B &=
v_0 N \sum_{\alpha=1}^{M} \sum_{j=1}^{n_\alpha} 
\myint_0^{l_\alpha} 
\heav\of{\zeta_\alpha\of{s}} \delta ( \vect{r} - \vect{r}_{j}\of{s} ) 
\diff{s}, 
\end{myalign*}
where $\heav\of{x}$ is the Heaviside step function.

Finally, the total surface energy of the polymer melt can be expressed as:
\begin{myalign*}
U_2 = \myint_{\Gamma} \left( \gamma_A \hat{\rho}_A + \gamma_B \hat{\rho}_B \right)\diff{\vect{r}}
= \myint_{\Omega} \left( \gamma_A \hat{\rho}_A + \gamma_B \hat{\rho}_B \right) \delta_\Gamma \diff{\vect{r}},
\end{myalign*}
where it is assumed that the local surface energy of the polymer material is equal to a weighted averaged of surface energies of $A$ and $B$ species, and where $\delta_\Gamma$ represents the surface delta function associated with domain boundary $\Gamma$.

Taking into account stretching, interaction, and surface energies the total energy of the system $U$ can be expressed as:
\begin{myalign*}
U = U_0 + U_1 + U_2
\end{myalign*}

Using the Hubbard-Stratonovich transformation, the canonical partition function can be expressed as:
\begin{myalign*}
  \mathcal{Z}_C = \mathcal{Z}_0 \iint \exp{\left( -\ham[\mu_+, \mu_-] \right)} \Diff{\mu_+} \Diff{\mu_-},
\end{myalign*}
where $\mathcal{Z}_0$ denotes the ideal gas partition function, $\mu_+$ and $\mu_-$ are fluctuating pressure-like and exchange-like potential fields, and the effective Hamiltonian $\ham$ is given by:
\begin{multline*}
\frac{\ham}{n} = 
\frac{1}{V} \myint_\Omega \left( \frac{\mu_-^2}{\chi_{AB} N} - \mu_+ \right) \diff{\vect{r}}
- \sum_{\alpha=1}^{M} (1-G_\alpha) \varphi_\alpha \log\of{\Q^f_\alpha [\mu_-, \mu_+]}
\\
- 
\sum_{\alpha=1}^{M} G_\alpha\varphi_\alpha \myint_{\Gamma} g_\alpha\of{\vect{r}} \log\of{\Q^t_\alpha [\vect{r}; \mu_-, \mu_+]} \diff{\vect{r}}.
\end{multline*} 
The single-chain partition functions of a free chain $\Q^f_\alpha$ and a tethered one $\Q^t_\alpha$ can be computed from the respective chain propagators $q^f_\alpha = q^f_\alpha\of{s, \vect{r}}$ and $q^t_\alpha = q^t_\alpha \of{s, \vect{r}}$ as\cite{muller2002phase}:
\begin{myalign}\label{eq:Q}
\Q^f_\alpha
= \frac{1}{V} \myint_{\Omega} q^f_\alpha\of{l_\alpha, \vect{r}} \diff{\vect{r}}
\quad
\textrm{and}
\quad
\Q^t_\alpha\of{\vect{r}}
= q^t_\alpha\of{l_\alpha, \vect{r}}
\end{myalign}
Chain propagators $q^f_\alpha$ and $q^t_\alpha$ satisfy the so-called modified diffusion equations which, in this particular case of tethered and untethered polymer chains having the same composition, are identical. Thus, we denote both of them by a single symbol $q_\alpha=q^f_\alpha=q^t_\alpha$ where: 
\begin{myalign}\label{eq:qf:delta}
\left\{
\begin{aligned}
\dds{\qf_\alpha} + \left(\mu_\alpha\of{s} + \eta \gamma_\alpha\of{s}\delta_\Gamma \right) \qf_\alpha &= \lap \qf_\alpha, 
&\vect{r} &\in \Omega, & 0 &< s \leq l_\alpha, \\
\ddn{\qf_\alpha} &= 0, 
&\vect{r} &\in\Gamma, & 0 &< s \leq l_\alpha, \\
\qf_\alpha &= 1, 
&\vect{r} &\in \Omega, & s &= 0
\end{aligned}
\right.
\end{myalign}
where $\eta = \frac{N v_0}{R_g} \frac{1}{k_B T}$ and 
\begin{myalign*} 
\mu_\alpha\of{s}
&=
\mu_+ + \zeta_\alpha\of{s} \mu_-
\\ 
\gamma_\alpha\of{s}
&=
\frac{\gamma_B+\gamma_A}{2} + \zeta_\alpha\of{s} \frac{\gamma_B-\gamma_A}{2}
\end{myalign*}
The total fractions of block A and B at any point in space can be computed as:
\begin{myalign}\label{eq:scft:densities}
\rho_A\of{\vect{r}}
&= 
\sum_{\alpha=1}^{M}
\varphi_\alpha
\myint_{0}^{l_\alpha}
\heav\of{-\zeta_\alpha\of{s}} \qf_\alpha\of{s,\vect{r}} \qb_\alpha \of{s,\vect{r}}
\diff{s},
\\
\rho_B\of{\vect{r}}
&= 
\sum_{\alpha=1}^{M}
\varphi_\alpha
\myint_{0}^{l_\alpha}
\heav\of{\zeta_\alpha\of{s}} \qf_\alpha\of{s,\vect{r}} \qb_\alpha \of{s,\vect{r}}
\diff{s},
\end{myalign}
where $\qb_\alpha = (1-G_\alpha)\frac{\qb^f_\alpha}{\Q^f}  + G_\alpha V \qb^t_\alpha$ is a weighted sum of the so-called \textit{complimentary chain propagators} for free and tethered chains, $\qb^f_\alpha$ and $\qb^t_\alpha$, of type $\alpha$ and satisfies:
\begin{myalign}\label{eq:qb:delta}
\left\{
\begin{aligned}
-\dds{\qb_\alpha} + \left(\mu_\alpha\of{s} + \eta \gamma_\alpha\of{s}\delta_\Gamma \right) \qb_\alpha &= \lap \qb_\alpha, 
&\vect{r} &\in \Omega, & 0 &\leq s < l_\alpha, \\
\ddn{\qb_\alpha} &= 0, 
&\vect{r} &\in\Gamma, & 0 &\leq s < l_\alpha, \\
\qb_\alpha &= \qb^0_\alpha, 
&\vect{r} &\in \Omega, & s &= l_\alpha.
\end{aligned}
\right.
\end{myalign}
The expression for initial conditions in the above PDE $\qb^0_\alpha = \frac{1-G_\alpha}{\Q^f} + \frac{G_\alpha g_\alpha\of{\vecr}}{\Q^t} V \delta_\Gamma$ contains a uniform part $\frac{1-G_\alpha}{\Q^f}$, corresponding to free chains originating at any point with equal probability, and a singular one $\frac{G_\alpha g_\alpha\of{\vecr}}{\Q^t} V \delta_\Gamma$, corresponding to tethered chains originating only at grafting surfaces and according to grafting density $g_\alpha\of{\vecr}$.

A mean-field approximation for this system is obtained by accounting only for instances of fields $\mu_+ = \mu_+^*$ and $\mu_- = \mu_-^*$ that have the largest contribution in $\mathcal{Z}_C$, that is:
\begin{myalign*}
  \mathcal{Z}_C \approx \exp \left( -\ham [\mu_+^*, \mu_-^*] \right),
\end{myalign*}
where
\begin{myalign}\label{eq:scft:saddle}
  \fd{\ham}{\mu_+} [\mu_+^*, \mu_-^*] = 0
  \quad \text{and} \quad
  \fd{\ham}{\mu_-} [\mu_+^*, \mu_-^*] = 0.
\end{myalign}
This results in the following so-called SCFT equations:
\begin{myalign}
\label{eq:scft:pressure}
\rho_A + \rho_B &= 1, \\ 
\label{eq:scft:exchange}
\rho_A - \rho_B &= \frac{2 \mu_-^\ast}{\chi N},
\end{myalign} 

Note that equations \eqref{eq:qf:delta} and \eqref{eq:qb:delta} involve delta-like potentials which may be difficult to deal with from the numerical point of view. Some of the previous works avoided this issue by approximating the delta-like potential by a smooth function\cite{matsen1997thin}. Such an approach, however, introduces an artificial scale over which the delta function is smoothed. Alternatively, it can be shown (see Appendix \ref{app:weak}) that the diffusion equations \eqref{eq:qf:delta} and \eqref{eq:qb:delta} are equivalent to:
\begin{myalign}\label{eq:qf:robin}
\left\{
\begin{aligned}
\dds{\qf_\alpha} + \mu_\alpha\of{s} \qf_\alpha &= \lap \qf_\alpha, 
&\vect{r} &\in \Omega, & 0 &< s \leq l_\alpha, \\
\ddn{\qf_\alpha} + \eta \gamma_\alpha\of{s} \qf_\alpha&= 0, 
&\vect{r} &\in\Gamma, & 0 &< s \leq l_\alpha, \\
\qf_\alpha &= 1, 
&\vect{r} &\in \Omega, & s &= 0
\end{aligned}
\right.
\end{myalign}
 and
\begin{myalign}\label{eq:qb:robin}
\left\{
\begin{aligned}
-\dds{\qb_\alpha} + \mu_\alpha\of{s} \qb_\alpha &= \lap \qb_\alpha, 
&\vect{r} &\in \Omega, & 0 &\leq s < l_\alpha, \\
\ddn{\qb_\alpha} + \eta \gamma_\alpha\of{s} \qb_\alpha &= 0, 
&\vect{r} &\in\Gamma, & 0 &\leq s < l_\alpha, \\
\qb_\alpha &= \qb^0_\alpha, 
&\vect{r} &\in \Omega, & s &= l_\alpha
\end{aligned}
\right.
\end{myalign}
where the surface energy terms now enter as a Robin (mixed) boundary conditions (this form can also be obtained directly from fully-fluctuating theory). This formulation is significantly more amenable to standard numerical methods without additional approximations, the only remaining non-smooth feature of the above equations is the delta-like distribution in the initial conditions for $\qb_\alpha$. However, given the diffusive nature of the PDE, such a singular feature does not persist, in fact the solution is infinitely smooth for $s>0$, thus only the initial condition needs a special treatment. Specifically, in this work, we employ a finite volume method in which volumertic integration of the delta-like distribution leads to a well-defined surface integral as discussed in later sections.

While the solution to the forward and backward diffusion equations \eqref{eq:qf:robin}-\eqref{eq:qb:robin} is expected to produce smooth quantities, it turns out that solving the entire system of the SCFT equation might still produce a singular behavior \cite{Fredrickson:06:The-equilibrium-theo,matsen2009melt,chantawansri2011spectral}. Mathematically, this is due to the fact that Robin boundary conditions representing surface energies and initial conditions for $\qb$ representing tethered ends are not consistent with equation \eqref{eq:scft:pressure} representing the incompressibility of the system. In order to see this more clearly, we, first, note that equation \eqref{eq:qb:robin} can be equivalently written as
\begin{myalign}\label{eq:qb:robin2}
\left\{
\begin{aligned}
-\dds{\qb_\alpha} + \left(\mu_\alpha\of{s} + \eta \gamma_\alpha\of{s}\delta_\Gamma \right) \qb_\alpha &= \lap \qb_\alpha, 
&\vect{r} &\in \Omega, & 0 &\leq s < l_\alpha, \\
\ddn{\qb_\alpha} + \eta \gamma_\alpha\of{s} \qb_\alpha &= \frac{G_\alpha g_\alpha\of{\vecr}}{\Q^t} V \delta\of{s-l_\alpha}, 
&\vect{r} &\in\Gamma, & 0 &\leq s < l_\alpha, \\
\qb_\alpha &= \frac{1-G_\alpha}{\Q^f}, 
&\vect{r} &\in \Omega, & s &= l_\alpha
\end{aligned}
\right.
\end{myalign}
where now the inhomogeneity of Robin boundary conditions model the effect of tethered chain ends.
Now, taking the derivative of \eqref{eq:scft:densities} on the domain boundary in the normal direction and using the boundary conditions from \eqref{eq:qf:robin} and \eqref{eq:qb:robin} one obtains:
\begin{myalign*}
\ddn{\left( \rho_A + \rho_B \right)} 
= 
- 2\eta\left( \gamma_A \rho_A + \gamma_B \rho_B \right) 
+ V \sum_{\alpha=1}^{M} G_\alpha \varphi_\alpha g_\alpha\of{\vect{r}},
\end{myalign*} 
while differentiating the incompressibility equation \eqref{eq:scft:pressure} leads to:
\begin{myalign*}
\ddn{\left( \rho_A + \rho_B \right)} = 0.
\end{myalign*} 
It is clear that the two results are inconsistent whenever the surface potentials are nonzero ($\gamma_A \neq 0$ and/or $\gamma_B \neq 0$) or tethered polymer chains are present ($g_\alpha\of{\vect{r}} \neq 0$).
Attempting to solve the entire system of SCFT equations with chain propagators satisfying the above diffusion equations produces pressure fields with a singular behavior near domain boundaries. Physically, one can interpret this behavior in the following way: crowding of tethered polymer chain ends or non-zero surface potentials results in emerging of a compressing force which is acting on an incompressible substance. 

While for certain applications such a singular behavior does not have an adverse effect, it is crucial to have well-defined values of the pressure field for modeling polymer-air and polymer-particle interactions. We now show that the described inconsistency is only apparent and can be completely removed.

Using, again, the results of \ref{app:weak}, we can write the diffusion equations \eqref{eq:qf:robin} and \eqref{eq:qb:robin} in yet another equivalent form: 
\begin{myalign}\label{eq:qf:almost}
\left\{
\begin{aligned}
\dds{\qf_\alpha} + \left(\mu_\alpha\of{s} + \eta \gamma_{\text{c}} \right) \qf_\alpha &= \lap \qf_\alpha, 
&\vect{r} &\in \Omega, & 0 &< s \leq l_\alpha, \\
\ddn{\qf_\alpha} + \eta \left( \gamma_\alpha\of{s} - \gamma_{\text{c}} \right) \qf_\alpha&= 0, 
&\vect{r} &\in\Gamma, & 0 &< s \leq l_\alpha, \\
\qf_\alpha &= 1, 
&\vect{r} &\in \Omega, & s &= 0
\end{aligned}
\right.
\end{myalign}
and
\begin{myalign}\label{eq:qb:almost}
\left\{
\begin{aligned}
-\dds{\qb_\alpha} + \left(\mu_\alpha\of{s} + \eta \gamma_{\text{c}} \right) \qb_\alpha &= \lap \qb_\alpha, 
&\vect{r} &\in \Omega, & 0 &\leq s < l_\alpha, \\
\ddn{\qb_\alpha} + \eta \left( \gamma_\alpha\of{s} - \gamma_{\text{c}} \right)\qb_\alpha &= \frac{G_\alpha g_\alpha\of{\vecr}}{\Q^t} V \delta\of{s-l_\alpha}, 
&\vect{r} &\in\Gamma, & 0 &\leq s < l_\alpha, \\
\qb_\alpha &= \frac{1-G_\alpha}{\Q^f}, 
&\vect{r} &\in \Omega, & s &= l_\alpha,
\end{aligned}
\right.
\end{myalign}
where we added and subtracted an arbitrary function $\gamma_{\text{c}}=\gamma_{\text{c}}\of{\vecr}$ defined on the domain's boundary $\Gamma$ from the surface energy terms of the boundary conditions. Having this freedom in choosing the exact form of $\gamma_{\text{c}}$ allows one to ensure that the boundary conditions in equations \eqref{eq:qf:almost} and \eqref{eq:qb:almost} are consistent with the incompressibility equation \eqref{eq:scft:pressure}. Indeed, taking the normal derivative of the total density on the domain boundary and equating it to zero leads to the following equation for $\gamma_{\text{c}}$:
\begin{myalign*}
- 2\eta\left( \left(\gamma_A - \gamma_{\text{c}}\right) \rho_A + \left(\gamma_B - \gamma_{\text{c}}\right) \rho_B \right) 
+ V \sum_{\alpha=1}^{M} G_\alpha \varphi_\alpha g_\alpha\of{\vect{r}}
= 0,
\end{myalign*}
from which it follows that if $\gamma_{\text{c}}$ is chosen as:
\begin{myalign*}
\gamma_{\text{c}} = 
\frac{\left( \gamma_A \rho_A  + \gamma_B  \rho_B \right)
-\hf \eta^{-1} V \sum_{\alpha=1}^{M} G_\alpha \varphi_\alpha g_\alpha\of{\vect{r}}}{\rho_A +  \rho_B},
\end{myalign*}
then there is no inconsistency between the incompressibility equation \eqref{eq:scft:pressure} and the boundary conditions in \eqref{eq:qf:almost} and \eqref{eq:qb:almost}.

Finally, redefining the pressure field as $\mu_+ \rightarrow \mu_+ - \eta \gamma_{\text{c}} \delta_\Gamma$, or in other words, explicitly subtracting from it the singular part, one arrives to the following equations for chain propagators:
\begin{myalign}\label{eq:qf:final}
\left\{
\begin{aligned}
\dds{\qf_\alpha} + \mu_\alpha\of{s} \qf_\alpha &= \lap \qf_\alpha, 
&\vect{r} &\in \Omega, & 0 &< s \leq l_\alpha, \\
\ddn{\qf_\alpha} + \eta \left( \gamma_\alpha\of{s} - \gamma_{\text{c}} \right) \qf_\alpha&= 0, 
&\vect{r} &\in\Gamma, & 0 &< s \leq l_\alpha, \\
\qf_\alpha &= 1, 
&\vect{r} &\in \Omega, & s &= 0
\end{aligned}
\right.
\end{myalign}
and
\begin{myalign}\label{eq:qb:final}
\left\{
\begin{aligned}
-\dds{\qb_\alpha} + \mu_\alpha\of{s} \qb_\alpha &= \lap \qb_\alpha, 
&\vect{r} &\in \Omega, & 0 &\leq s < l_\alpha, \\
\ddn{\qb_\alpha} + \eta \left( \gamma_\alpha\of{s} - \gamma_{\text{c}} \right)\qb_\alpha &= \frac{G_\alpha g_\alpha\of{\vecr}}{\Q^t} V \delta\of{s-l_\alpha}, 
&\vect{r} &\in\Gamma, & 0 &\leq s < l_\alpha, \\
\qb_\alpha &= \frac{1-G_\alpha}{\Q^f}, 
&\vect{r} &\in \Omega, & s &= l_\alpha,
\end{aligned}
\right.
\end{myalign}
and the system's energy:
\begin{multline}\label{eq:scft:energy}
\frac{\ham}{n} = 
\frac{1}{V} \myint_\Omega \left( \frac{\mu_-^2}{\chi_{AB} N} - \mu_+ \right) \diff{\vect{r}}
+ \eta \frac{1}{V} \myint_{\Gamma} \gamma_{\text{c}} \diff{\Gamma}
\\
- \sum_{\alpha=1}^{M} \varphi_\alpha 
\left[
(1-G_\alpha) \log\of{\Q_\alpha [\mu_-, \mu_+]}
+
G_\alpha \myint_{\Gamma} g_\alpha\of{\vect{r}} \log\of{\Q_\alpha [\vect{r}; \mu_-, \mu_+]} \diff{\vect{r}}
\right],
\end{multline}
which are completely free of inconsistencies. 
From this result, $\gamma_{\text{c}}$ can be interpreted as the effective surface energy of the polymer material. Note that it contains two contributions: one from the nominal surface energies of A and B blocks ($\frac{\gamma_A \rho_A  + \gamma_B  \rho_B }{\rho_A +  \rho_B}$), and the second one due to tethered polymer ends ($-\frac{\hf \eta^{-1} V \sum_{\alpha=1}^{M} \varphi_\alpha G_\alpha g_\alpha\of{\vect{r}}}{\rho_A +  \rho_B}$).

\textbf{Remark.} We note that in Matsen \etal\cite{matsen2009melt} a similar approach was proposed to subtract off a singularity in the pressure field. However, singularities only due to tethered chain ends on planar geometries were considered. The magnitude of the singularity was deduced from a in-depth analysis of the Strong Segregation Theory. The described above derivation offers a simpler and unified approach for computing the magnitude of pressure singularities which is straightforward to apply in a general multidimensional setting for domains of arbitrary shapes.

\textbf{Remark.} In the numerical experiments, we characterize the strength of surface energy using the expression for interfacial tension of a symmetric polymer-polymer interface estimated in the case of strong segregation \cite{Fredrickson:06:The-equilibrium-theo}:
\begin{myalign*}
  \gamma = \frac{1}{\eta} \sqrt{\chi N}.
\end{myalign*}
Imposing surface energies using the above formula in terms of $\chi N$ allows straightforward comparison with $A$-$B$ interaction strength $\XN{AB}$.

\subsection{Solving the SCFT equations for $\mu_+^*$ and $\mu_-^*$}
The common approach for obtaining a saddle point $\left(\mu_+, \mu_-\right) = \left(\mu_+^*, \mu_-^*\right)$ of the Hamiltonian $\ham$ is to start from some initial guess (seed) $\left(\mu_+, \mu_-\right) = \left( \mu_+^{(0)}, \mu_-^{(0)} \right)$ and to iteratively evolve the fields $\left(\mu_+, \mu_-\right)$ in the steepest descent/ascent directions until a saddle point is reached within a specified tolerance $\varepsilon_\text{tol}$, that is, given values at the $k$th iteration fields at the $(k+1)$th iteration are updated as:
\begin{myalign}\label{eq:scft:solving}
\begin{aligned}
  \mu_+^{(k+1)} &= \mu_+^{(k)} + \lambda_+ \fd{\ham}{\mu_+} [\mu_+^{(k)}, \mu_-^{(k)}] = \mu_+^{(k)} + \lambda_+ \left( \rho_A^{(k)} + \rho_B^{(k)} - 1 \right),
  \\
  \mu_-^{(k+1)} &= \mu_-^{(k)} - \lambda_- \fd{\ham}{\mu_-} [\mu_+^{(k)}, \mu_-^{(k)}] = \mu_-^{(k)} - \lambda_- \left( \frac{2\mu_-^{(k)}}{\chi_{AB} N} - \rho_A^{(k)} + \rho_B^{(k)} \right),
\end{aligned}
\end{myalign}
where $\lambda_+$ and $\lambda_-$ are step sizes in moving along steepest descent/ascent directions (typically, taken as $\lambda_+ = \lambda_- = 1$), and iterations are terminated once
\begin{myalign}\label{eq:scft:convergence}
  \frac{1}{V}
  \sqrt{
  \myint_{\Omega}
  \left( \fd{\ham}{\mu_+} \right)^2 \diff{\vect{r}}
  } < \varepsilon_\text{tol}
  \quad \text{and} \quad
  \frac{1}{V}
  \sqrt{
    \myint_{\Omega}
    \left( \fd{\ham}{\mu_-} \right)^2 \diff{\vect{r}}
  } < \varepsilon_\text{tol}.
\end{myalign} 

The solution of the proposed modified system of the SCFT equations accounting for surface energies is further complicated by the presence of terms depending on $\gamma_{\text{c}}$, which in turn is a function of the local densities $\rho_A$ and $\rho_B$. One possible strategy could be simply to recalculate the values for $\gamma_{\text{c}}$ every few fields updates $\mu_-$ and $\mu_+$. However, we observed that this approach still results in numerical artifacts in the pressure field $\mu_+$. This seems to be linked to the fact that in such a case, the density fields $\rho_A$ and $\rho_B$ generally do not satisfy the condition $\ddn{\left(\rho_A + \rho_B\right)} = 0$ during such iterations. Once the pressure field is polluted with features of small enough scale they persist for extremely long times. Our experience suggested that a better strategy is to iteratively find $\gamma_{\text{c}}$ that satisfies $\ddn{\left(\rho_A + \rho_B\right)} = 0$ (we observed that a small number of iterations $n_\gamma$ = 2-4 is sufficient) before every update of fields $\mu_-$ and $\mu_+$. The overall procedure for solving modified system of SCFT equations can be summarized as follows:
\begin{enumerate}
\item Set a desired seed for $\mu_-^{(0)}$ and $\mu_+^{(0)} = 0$.
\item Iterate in $k$ until conditions \eqref{eq:scft:convergence} are satisfied:
\begin{enumerate}
\item Iterate $n_\gamma$ times:
\begin{enumerate}
\item Set $\gamma_{\text{c}} = 
\dfrac{\left( \gamma_A \rho_A  + \gamma_B  \rho_B \right)
-\hf \eta^{-1} V \sum_{\alpha=1}^{M} \varphi_\alpha G_\alpha g_\alpha\of{\vect{r}}}{\rho_A +  \rho_B}$
\item Solve diffusion equations \eqref{eq:qf:final} and \eqref{eq:qb:final}
\item Compute density fields according to \eqref{eq:scft:densities}
\end{enumerate}
\item Compute $\mu_+^{(k+1)}$ and $\mu_-^{(k+1)}$ using \eqref{eq:scft:solving}
\end{enumerate}
\end{enumerate}

\section{Forces acting on free surfaces and nanoparticles}\label{sec:free-surface}

\begin{figure}[!h]
\centering
\includegraphics[width=.4\textwidth]{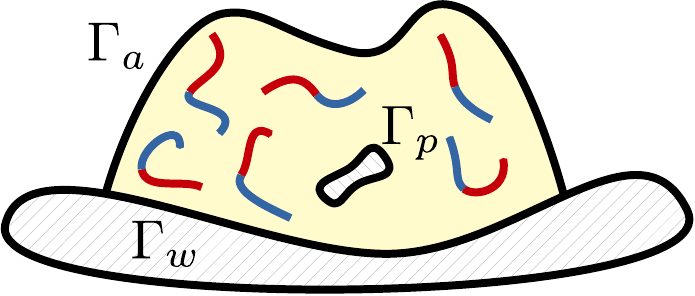}
\caption{Problem geometry and notation used in section \ref{sec:free-surface}.}
\label{fig:free-surface:problem-statement}
\end{figure}

Without loss of generality, let us now consider a polymer droplet containing a nanoparticle of arbitrary shape on a (possibly curved) substrate as illustrated in figure~\ref{fig:free-surface:problem-statement}. 
The generalizations to the case of multiple particles is straightforward. For convenience, we refer to the surrounding substance as ``air'', however it represents any gaseous or liquid non-solvent material. Similarly, the term ``substrate'' stand for a body partially submerged in the polymer material, which could be a stationary substrate or a movable partially sunk nanoparticle.  
As in previous section, we denote the region occupied by the polymer material as $\Omega$. 
The boundary of this region, denoted as $\Gamma$, consists of the polymer-wall interface $\Gamma_w$, polymer-air interface $\Gamma_a$, and polymer-particle interface $\Gamma_p$, that is, $\Gamma = \Gamma_w \cup \Gamma_a \cup \Gamma_p$. 
We denote the surface energy of the $A$ and $B$ blocks on these interfaces as $\gamma_{wA},\gamma_{wB}$; 
$\gamma_{aA},\gamma_{aB}$; and 
$\gamma_{pA},\gamma_{pB}$; correspondingly. 
The surface energy of the air-wall interface is denoted as $\gamma_{aw}$.
The air related surface energies, $\gamma_{aA}$, $\gamma_{aB}$, and $\gamma_{aw}$, are assumed to be constants, while the polymer-wall and polymer-particle surface energies can be spatially variable quantities 
$\gamma_{wA} = \gamma_{wA}\of{\vect{r}}$, 
$\gamma_{wB}= \gamma_{wB}\of{\vect{r}}$, 
$\gamma_{pA}= \gamma_{pA}\of{\vect{r}}$, 
$\gamma_{pB}= \gamma_{pB}\of{\vect{r}}$. 
We assume that polymer chain can only be grafted onto the nanoparticle, that is, $g_\alpha\of{\vect{x}} = 0$ for $\vect{x} \notin \Gamma_p$.

Given the results of the previous section, the energy of such a system per polymer chain can be expressed as:
\begin{multline*}
\frac{\ham V}{n} = 
\myint_\Omega \left( \frac{\mu_-^2}{\chi_{AB} N} - \mu_+ \right) \diff{\vect{r}}
- V\sum_{\alpha=1}^{M} \varphi_\alpha  \log\of{\Q_\alpha [\mu_-, \mu_+]}
\\
- V\sum_{\alpha=1}^{M} \varphi_\alpha \myint_{\Gamma_p} g_\alpha\of{\vect{r}} \log\of{\Q_\alpha [\vect{r}; \mu_-, \mu_+]} \diff{\vect{r}}
+ 
\eta \sum_{\nu=a,p,w} \myint_{\Gamma_\nu} \gamma_{\nu, c} \diff{\Gamma}
- 
\eta \myint_{\Gamma_a} \gamma_{aw} \diff{\Gamma},
\end{multline*}
where
\begin{myalign*}
\gamma_{w, \text{c}} &= 
\frac{\left( \gamma_{wA} \rho_A  + \gamma_{wB} \rho_B \right)}{\rho_A +  \rho_B},
\\
\gamma_{p, \text{c}} &= 
\frac{\left( \gamma_{pA} \rho_A  + \gamma_{pB} \rho_B \right)
-\hf \eta^{-1} V \sum_{\alpha=1}^{M} \varphi_\alpha G_\alpha g_\alpha\of{\vect{r}}}{\rho_A +  \rho_B},
\\
\gamma_{a, \text{c}} &= 
\frac{\left( \gamma_{aA} \rho_A  + \gamma_{aB} \rho_B \right)}{\rho_A +  \rho_B},
\end{myalign*}
single chain partition functions $\Q_f$ and $\Q_t$ are given by \eqref{eq:Q} and where chain propagators $\qf_f$ ans $\qf_t$ satisfy the modified diffusion equations \eqref{eq:qf:final} provided ${\Gamma = \Gamma_a \cup \Gamma_w \cup \Gamma_a}$ and 
\begin{myalign*}
\gamma_A =
\begin{cases}
\gamma_{aA}, &\vecr\in\Gamma_a, \\
\gamma_{wA}, &\vecr\in\Gamma_w, \\
\gamma_{pA}, &\vecr\in\Gamma_p,
\end{cases}
\quad
\gamma_B =
\begin{cases}
\gamma_{aB}, &\vecr\in\Gamma_a, \\
\gamma_{wB}, &\vecr\in\Gamma_w, \\
\gamma_{pB}, &\vecr\in\Gamma_p.
\end{cases}
\end{myalign*}

\begin{figure}[!h]
\centering
\includegraphics[width=.4\textwidth]{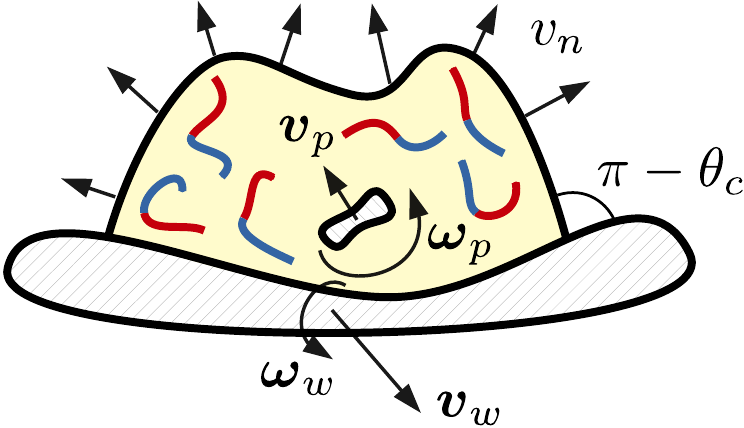}
\caption{Deformation of free surface in normal velocity field $\vn$.}
\label{fig:free-surface:deformation}
\end{figure}

To derive expressions for the forces acting on the free surface and the nanoparticle, we consider a situation in which the free surface is being deformed with an arbitrary normal velocity $v_{\vect{n}}$, the nanoparticle is moving with translational and rotational velocities $\vect{v}_p$ and $\vect{\omega}_p$, and the substrate\footnote{While the motion of the substrate does not correspond to a practically relevant situation, it allows to apply the derived below theory in the case when the term ``substrate'' represents a partially submerged nanoparticle.} is moving with translational and rotational velocities $\vect{v}_w$ and $\vect{\omega}_w$ in fictitious time $\tau$ as depicted in Figure \ref{fig:free-surface:deformation}, and compute analytically the derivative of the system's energy with respect to $\tau$. 
In doing so, one has to take into account three effects that an evolving problem geometry incurs: (1) changes in domains of integration (assuming all integrands are independent of $\tau$); (2) changes in chain propagators $\qf_f = \qf_f\of{\tau}$ and $\qf_t = \qf_t\of{\tau}$ (assuming $\mu_-$ and $\mu_+$ are independent of $\tau$); (3) changes in the overall SCFT solutions represented by fields $\mu_- = \mu_-\of{\tau}$ and $\mu_+ = \mu_+\of{\tau}$. First, we note that due to the nature of the problem, specifically, that $\mu_-$ and $\mu_+$ satisfy saddle-point conditions, the contribution of the third effect in the change of system's energy is zero. Indeed, by applying the chain rule one obtains:
\begin{myalign*}
\frac{d\ham}{d \tau}  &= 
\myint_\Omega \underbrace{\fd{\ham}{\mu_+}}_{=0} \frac{d\mu_+}{d \tau} \diff{\vect{r}}
+
\myint_\Omega \underbrace{\fd{\ham}{\mu_-}}_{=0} \frac{d\mu_-}{d \tau} \diff{\vect{r}}
+
\left. \frac{d \ham}{d \tau} \right|_{\mu = \text{const}}
=
\left. \frac{d \ham}{d \tau} \right|_{\mu = \text{const}}
\end{myalign*}
This means that only the first two effects need to be taken into account while deriving the derivative of system's energy with respect to $\tau$, which simplifies the task significantly. To take into account the second effect we use a Lagrangian multipliers approach where we use partial differential equations for chains propagators $\qf_\alpha$ in their weak formulations (see appendix \ref{app:weak}) as constraints. That is, we introduce a Lagrangian $\lag$ as 
\begin{mymultline*}
\lag = 
\frac{\ham V}{n}
+
\sum_{\alpha=1}^{M}
\Bigg[
\myint_{\Omega}
\myint_{0}^{l_\alpha}
\left(
-\qf_\alpha \partial_s \lb_\alpha + \mu_\alpha\of{s} \qf_\alpha \lb_\alpha + \nabla \qf_\alpha \cdot \nabla \lb_\alpha
\right)
\diff{s}
\diff{\vect{r}}
\\
+
\myint_{\Omega}
\left(
\qf_\alpha\of{l_\alpha} \lb_\alpha\of{l_\alpha} - \lb_\alpha\of{0}
\right)
\diff{\vect{r}} 
+  
\sum_{\nu=a,p,w}
\myint_{\Gamma_\nu}
\myint_{0}^{l_\alpha}
\eta (\gamma_{\nu, \alpha} \of{s} - \gamma_{\nu, c}) \qf_\alpha \lb_\alpha
\diff{s}
\diff{\vect{r}}
\Bigg],
\end{mymultline*}
where $\lb_\alpha$ are Lagrangian multipliers associated with the PDE-constraints for $\qf_\alpha$. The full derivative of $\ham$ with respect to $\tau$ is then equal to the partial derivative of $\lag$, that is, $\deriv{\ham}{\tau} = \pd{\lag}{\tau}$, provided that the following optimality conditions are satisfied:
\begin{myalign*}
\delta_{\lb_\alpha} \lag 
=
\int 
\frac{\delta \lag}{\delta \lb_\alpha}
\delta \lb_\alpha
&= 0, \quad \forall \lb_\alpha,  
\\
\delta_{\qf_\alpha} \lag 
=
\int 
\frac{\delta \lag}{\delta \qf_\alpha}
\delta \qf_\alpha
&= 0, \quad \forall \qf_\alpha.
\end{myalign*} 
It is trivial to show that the first condition recovers the weak form of \eqref{eq:qf:final}. The second condition provides equations for Lagrangian multipliers $\lb_\alpha$. After explicitly taking the variation of $\lag$ with respect to $\qf_\alpha$, one obtains the following expression:
\begin{mymultline*}
\delta_{\qf_\alpha} \lag = 
\myint_{\Omega}
\myint_{0}^{l_\alpha}
\left(
-\delta \qf_\alpha \partial_s \lb_\alpha + \mu_\alpha\of{s} \delta \qf_\alpha \lb_\alpha + \nabla \delta \qf_\alpha \cdot \nabla \lb_\alpha
\right)
\diff{s}
\diff{\vect{r}}
\\
+
\myint_{\Omega}
\left(
\lb_\alpha\of{l_\alpha} - \frac{\varphi_\alpha(1-G_\alpha)}{\Q_\alpha}
\right) \delta \qf_\alpha\of{l_\alpha}
\diff{\vect{r}} 
- 
V \varphi_\alpha G_\alpha
\myint_{\Gamma_p} g_\alpha\of{\vect{r}} \frac{\delta \qf_\alpha\of{l_\alpha}}{\qf_\alpha\of{l_\alpha}} \diff{\vect{r}}
\\
+  
\sum_{\nu=a,p,w}
\myint_{\Gamma_\nu}
\myint_{0}^{l_\alpha}
\eta (\gamma_{\nu, \alpha}\of{s} - \gamma_{\nu, c}) \delta \qf_f \lb_f
\diff{s}
\diff{\vect{r}}
\end{mymultline*}
Using the integration by parts formula \eqref{eq:weak:byparts} leads to 
which are weak formulations of the following diffusion equations:
\begin{mymultline*}
\delta_{\qf_\alpha} \lag 
= 
\myint_{\Omega}
\myint_{0}^{l_\alpha}
\left(
-\delta \qf_\alpha \partial_s \lb_\alpha + \mu_\alpha\of{s} \delta \qf_\alpha \lb_\alpha - \lap \lb_\alpha
\right) \delta \qf_\alpha
\diff{s}
\diff{\vect{r}}
\\
+
\myint_{\Omega}
\left(
\lb_\alpha\of{l_\alpha} - \frac{\varphi_\alpha(1-G_\alpha)}{\Q_\alpha}
\right) \delta \qf_\alpha\of{l_\alpha}
\diff{\vect{r}} 
+  
\sum_{\nu=a,w}
\myint_{\Gamma_\nu}
\myint_{0}^{l_\alpha}
\left(
\ddn{\lb_\alpha} 
+
\eta (\gamma_{\nu, \alpha}\of{s} - \gamma_{\nu, c}) \lb_\alpha
\right) \delta \qf_\alpha
\diff{s}
\diff{\vect{r}}.
\\
+  
\myint_{\Gamma_p}
\myint_{0}^{l_\alpha}
\left(
\ddn{\lb_\alpha} 
+
\eta (\gamma_{p, \alpha}\of{s} - \gamma_{p, c}) \lb_\alpha
-
V G_\alpha \varphi_\alpha g_\alpha\of{\vect{r}} \frac{\delta\of{s-l_\alpha}}{\qf_\alpha\of{l_\alpha}}
\right) \delta \qf_\alpha
\diff{s}
\diff{\vect{r}}
\end{mymultline*}
As one can see $\delta_{\qf_\alpha} \lag = 0$, $\forall \qf_\alpha$, leads to a weak formulation of the diffusion equation for the Lagrange multiplier $\lb_\alpha$ that is completely analogous to the equation for the complimentary chain propagator $\qb_\alpha$ \eqref{eq:qb:final} with the exception that the initial and boundary conditions are scaled by $\varphi_\alpha$. Thus, we have $\lb_\alpha = \varphi_\alpha \qb_\alpha$.

We now turn our attention to computing the partial derivative of Lagrangian $\lag$ with respect to $\tau$. The only terms in $\lag$ that are explicit functions of the problem geometry are domains of integration $\Omega=\Omega\of{\tau}$, $\Gamma_a=\Gamma_a\of{\tau}$, $\Gamma_p=\Gamma_p\of{\tau}$, $\Gamma_w=\Gamma_w\of{\tau}$,  surface energies $\gamma_{pA}=\gamma_{pA}\of{\vecr, \tau}$, $\gamma_{pB}=\gamma_{pB}\of{\vecr, \tau}$, $\gamma_{wA}=\gamma_{wA}\of{\vecr, \tau}$, $\gamma_{wB}=\gamma_{wB}\of{\vecr, \tau}$ and grafting density $g= g_\alpha\of{\vecr,\tau}$.  It can be shown (see appendix \ref{app:shape-deriv}) that for integrals of the form:
\begin{myalign}\label{eq:free:integrals}
\myint_{\Omega\of{\tau}} a\of{\vecr} \diff{\vecr}, \quad 
\myint_{\Gamma_a\of{\tau}} a\of{\vecr} \diff{\vecr}, 
\myint_{\Gamma_p\of{\tau}} a\of{\vecr} b\of{\vecr, \tau} \diff{\vecr},\quad \textrm{and} \quad
\myint_{\Gamma_w\of{\tau}} a\of{\vecr} c\of{\vecr, \tau} \diff{\vecr},
\end{myalign}
where $a\of{\vecr}$ is a field independent of $\tau$, while $b\of{\vecr,\tau}$ and $c\of{\vecr,\tau}$ are fields that follow the nanoparticle and the substrate, correspondingly, the following equations hold true:
\begin{myalign}\label{eq:free:integrals:derivs}
\begin{aligned}
\pd{}{\tau} \myint_{\Omega} a \diff{\vecr}   
&= \myint_{\Gamma_a} a \vn \diff{\vecr}
+ 
\myint_{\Gamma_p} 
a 
\left( \vect{v}_p - \vect{\omega}_p \times \vecr \right) \cdot \vect{n} 
\diff{\Gamma}
+ 
\myint_{\Gamma_w} 
a 
\left( \vect{v}_w - \vect{\omega}_w \times \vecr \right) \cdot \vect{n} 
\diff{\Gamma} , 
\\
\pd{}{\tau} \myint_{\Gamma_a} a \diff{\vecr} 
&= 
\myint_{\Gamma_a} \vn \left( \kappa + \ddn{} \right) a \diff{\vecr} 
+ 
\myint_{\Gamma_a \cap \Gamma_w} a \frac{ \vn \cos\of{\theta_c} + \left( \vect{v}_w - \vect{\omega}_w \times \vecr \right) \cdot \vect{n}_w}{\sin\of{\theta_c}} \diff{\vecr},
\\
\pd{}{\tau} \myint_{\Gamma_p} a b \diff{\vecr} 
&= 
\myint_{\Gamma_p} 
\left( \vect{v}_p - \vect{\omega}_p \times \vecr \right) \cdot
\left( \vect{n} \kappa a b + \vect{n} b \ddn{a}
- a \nabla_\perp b
\right)
\diff{\vecr},
\\
\pd{}{\tau} \myint_{\Gamma_w} a c \diff{\vecr} 
&= 
\myint_{\Gamma_w} 
\left( \vect{v}_w - \vect{\omega}_w \times \vecr \right) \cdot
\left( \vect{n} \kappa a c + \vect{n} c \ddn{a}
- a \nabla_\perp c
\right)
\diff{\vecr}
\\
&+
\myint_{\Gamma_a \cap \Gamma_w} a c \frac{ \vn + \left( \vect{v}_w - \vect{\omega}_w \times \vecr \right) \cdot \vect{n}_w \cos\of{\theta_c}}{\sin\of{\theta_c}} \diff{\vecr}, 
\end{aligned}
\end{myalign}
where $\kappa$ is the curvature of the polymer-air interface $\Gamma_a$, $\theta_c$ is the angle between $\Gamma_a$ and $\Gamma_w$ at their intersection, and tangential gradient $\nabla_\perp$ is defined as $\nabla_\perp = \nabla - \vect{n} \ddn{}$. Applying these formulas while differentiating $\lag$ with respect to $\tau$, one arrives at (assuming the volume of the system $V$ is constant):
\begin{myalign}\label{eq:dhdt}
\frac{d}{d\tau} \left( \frac{\ham V}{n} \right) 
&= 
\pd{}{\tau}\lag 
\\
&= 
\myint_{\Gamma_a} \sigma \vn \diff{\vecr}
+ 
\myint_{\Gamma_a \cap \Gamma_w} f^c_a \vn \diff{\vecr}
+ 
\vect{v}_p \cdot
\myint_{\Gamma_p} \vect{f}_p \diff{\vecr}
+
\vect{\omega}_p \cdot
\myint_{\Gamma_p}  \vect{f}_p \times \vecr \diff{\vecr}
\\
&+ 
\vect{v}_w \cdot
\left[
\myint_{\Gamma_w} \vect{f}_w \diff{\vecr}
+ 
\myint_{\Gamma_a \cap \Gamma_w} 
f^c_w \vect{n}_w \diff{\vecr}
\right]
+
\vect{\omega}_w \cdot
\left[
\myint_{\Gamma_w}  \vect{f}_w \times \vecr \diff{\vecr}
+ 
\myint_{\Gamma_a \cap \Gamma_w} 
f^c_w \vect{n}_w \times \vecr \diff{\vecr}
\right],
\end{myalign}
where the surface force densities acting on polymer-air, polymer-particle, and polymer-wall are given by
\begin{align*}
\sigma &= \Pi + \eta \left( \kappa + \ddn{} \right) \gamma_{a,\text{eff}},
\\
\vect{f}_p &=
\Pi \vect{n}_p
+
\eta \vect{n}_p
\left(
\kappa \gamma_{p,\text{eff}}
+
\gamma_{pA} \ddn{\rho_A} + \gamma_{pB} \ddn{\rho_B}
- 
\sum_{\alpha=1}^{M}
G_\alpha \varphi_\alpha V \eta^{-1} g_\alpha\of{\vect{r}} \frac{\ddn{\Q^t_\alpha}}{\Q^t_\alpha}
\right)
\\
&- 
\eta
\left( 
\rho_A \nabla_\perp\gamma_{pA} + \rho_B \nabla_\perp\gamma_{pB}
- 
\sum_{\alpha=1}^{M}
G_\alpha \varphi_\alpha V \eta^{-1} \nabla_\perp g_\alpha\of{\vect{r}} \log\of{\Q^t_\alpha}
\right),
\\
\vect{f}_w &= 
\Pi \vect{n}_w
+
\eta\vect{n}_w \left( \kappa \left( \gamma_{w,\text{eff}} - \gamma_{aw} \right)
+
\gamma_{wA} \ddn{\rho_A} + \gamma_{wB} \ddn{\rho_B} \right)
- 
\left( \rho_A \nabla_\perp \gamma_{wA} + \rho_B \nabla_\perp \gamma_{wB} \right),
\end{align*}
the line force densities acting on polymer-air and polymer-wall surface at the triple junction are given by
\begin{align*}
f^c_a &= \eta \frac{ \left(\gamma_{w,\text{eff}} - \gamma_{aw}\right) + \gamma_{a,\text{eff}} \cos{\theta_c} }{\sin{\theta_c}},
\\
f^c_w &= 
\eta 
\frac{ \gamma_{a,\text{eff}} + \left( \gamma_{w,\text{eff}} - \gamma_{aw} \right) \cos{\theta_c} }{\sin{\theta_c}},
\end{align*}
and where
\begin{align*}
\Pi &= \frac{\mu_-^2}{\chi_{AB} N} - \mu_+
- 
\sum_{\alpha=1}^{M} \varphi_\alpha 
\hf
\left[
\frac{1-G_\alpha}{\Q^f_\alpha} \qf_\alpha \of{l_\alpha}
+
\qb_\alpha\of{0}
\right],
\\
\gamma_{a,\text{eff}} &= \gamma_{aA} \rho_A + \gamma_{aB} \rho_B,
\\
\gamma_{w,\text{eff}} &= \gamma_{wA} \rho_A + \gamma_{wB} \rho_B,
\\
\gamma_{p,\text{eff}} &= \gamma_{pA} \rho_A + \gamma_{pB} \rho_B
- 
\sum_{\alpha=1}^{M}
G_\alpha \varphi_\alpha V \eta^{-1} g_\alpha\of{\vect{r}} \log\of{\Q^t_\alpha}.
\end{align*}

The system is in an equilibrium state when forces and moments acting on the nanoparticle and the substrate are zero:
\begin{myalign*}
\vect{F}_p &= 
\myint_{\Gamma_p} \vect{f}_p \diff{\vecr} 
= 0,
&
\vect{M}_p &= 
\myint_{\Gamma_p}  \vect{f}_p \times \vecr \diff{\vecr}
= 0,
\\
\vect{F}_w &= 
\left[
\myint_{\Gamma_w} \vect{f}_w \diff{\vecr}
+ 
\myint_{\Gamma_a \cap \Gamma_w} 
f^c_w \vect{n}_w \diff{\vecr}
\right]
= 0,
&
\vect{M}_w &= 
\left[
\myint_{\Gamma_w}  \vect{f}_w \times \vecr \diff{\vecr}
+ 
\myint_{\Gamma_a \cap \Gamma_w} 
f^c_w \vect{n}_w \times \vecr \diff{\vecr}
\right]
= 0,
\end{myalign*}
the normal force density acting on the polymer-air interface is uniform:
\begin{myalign*}
\sigma - \frac{1}{V} \myint_{\Gamma} \sigma \diff{\Gamma} = 0
\end{myalign*}
and the contact angle at the triple junction satisfies the Young's equation
\begin{myalign*}
\gamma^{\text{eff}}_w - \gamma_{aw} + \gamma^{\text{eff}}_a \cos(\theta_c) &= 0.
\end{myalign*}

Numerically, such an equilibrium state can be achieved iteratively by starting from a non-equilibrium configuration and evolving the system's geometry in the direction of energy descent according to:
\begin{myalign*}
\vn &= - \left( \sigma - \frac{1}{V} \myint_{\Gamma} \sigma \diff{\Gamma}
\right),\,
\vect{v}_p = - \vect{F}_p,\,
\vect{\omega}_p = - \vect{M}_p,\,
\vect{v}_w = - \vect{F}_w,\,
\vect{\omega}_w = - \vect{M}_w,
\end{myalign*}
while imposing the contact angle:
\begin{myalign*}
\theta_c = \cos^{-1}\of{\frac{\gamma_{aw} - \gamma^{\text{eff}}_w}{\gamma^{\text{eff}}_w}}.
\end{myalign*}
and an appropriate care for preserving the volume of polymer material.
In such a way, the energy of the system is forced to decrease until it reaches a local minimum.

\section{Numerical aspects}\label{sec:numerics}
The application of the presented framework for investigating non-trivial situation requires advanced numerical capabilities. Specifically, one needs tools for the description and evolution of irregular interfaces as well as capabilities to solve modified diffusion equations \eqref{eq:qf:final} and \eqref{eq:qb:final} subject to Robin boundary conditions on such irregular interfaces. In this work, we use a combination of the Level-Set Method, adaptive Cartesian grids and sharp-interface finite-volume methods for solving PDE along the lines of Ouaknin \etal \cite{Ouaknin;Laachi;Delaney;etal:16:Self-consistent-fiel}, with the exception that the diffusion equations are solved with a more accurate schemes presented in Bochkov and Gibou\cite{Bochkov;Gibou:19:Solving-the-Poisson-}. The evolution of the free surface in the normal direction is performed by the semi-implicit advection scheme\cite{smereka2003semi} as described in Min and Gibou\cite{Min;Gibou:07:A-second-order-accur}. We note, however, that the proposed methodology can also be implemented in other numerical frameworks, for example, an interface-fitted finite element method.

\section{Results}\label{sec:results}
In this section, we present a number of numerical examples to validate the proposed framework and demonstrate its capabilities. In all examples we consider AB-type diblock copolymer chains of length $N$ and fraction of A segments $f$.

\subsection{Imposing surface energies}\label{sec:scft:test}
We begin with demonstrating the importance of the consistent approach for imposing surface energies presented in section \ref{sec:model} on the following three examples. 

In the first example, we consider a flower-shaped domain defined by the zero-isocontour of the following function:
\begin{myalign*}
\phi_1\of{x,y} = \sqrt{ x^2 + y^2 } - 5 \left( 1 + 0.3 \cos\of{5\theta} \right),
\end{myalign*}
where $\theta = \arctan\of{\frac{x}{y}}$ is the polar angle. In this example the surface energies are uniform and equal to $\gamma_{A} = \frac{1}{2\eta}$ and $\gamma_{B} = - \frac{1}{2\eta}$, that is, the domain's boundary preferentially attracts the polymer component $B$.

In the second example, we consider a domain formed by the intersection of a disk of radius $r_2 = 6.5 R_g$ and a horizontal stripe of width $w_2 = 8 R_g$. The straight segments of the domain's boundary preferentially attract the polymer component $B$, such that the surface energies there are equal to $\gamma_{As} = \frac{1}{2\eta}$ and $\gamma_{Bs} = -\frac{1}{2\eta}$. The curved segments of the domain's boundary preferentially attract the polymer component $A$, such that the surface energies there are equal to $\gamma_{Ac} = - \frac{1}{2\eta}$ and $\gamma_{Bc} = \frac{1}{2\eta}$.

Finally, in the third example we consider a circular disk of radius $r_3 = 6.5 R_g$ located at $\left( x_3, y_3 \right) = \left( 0.1 R_g, 2.1 R_g \right)$ that is cut by a corrugated horizontal line defined as the zero-isocontour of function:
\begin{myalign*}
\phi_3\of{x,y} = -y+0.01 - 0.25\cos\of{2 \pi \frac{x}{\lambda}},
\end{myalign*} 
where $\lambda =  4\sqrt[6]{\frac{8 \XN{AB}}{ 3 \pi^4}}$. The circular section of the domain's boundary has uniform surface energies $\gamma_{Ac} = \frac{1}{2\eta}$ and $ \gamma_{Bc} = - \frac{1}{2\eta}$ (attraction of the component $B$), while the bottom segment has nonuniform surface energies given by expressions:
\begin{myalign*}
\gamma_{Ab} &= - 2.5\cos\of{2\pi\frac{x}{\lambda}}, \\
\gamma_{Bb} &=   2.5\cos\of{2\pi\frac{x}{\lambda}}.
\end{myalign*}
As a result, the grooves attract the component $B$ while the hills attract the component $A$.

In all the examples, the parameters of the polymer material are set to $\XN{AB} = 30$, $f=0.33$. Polymer chains are discretized using $n_s=60$ beads and the grid resolution is $20/256 R_g$. The simulations begin with the field $\mu_-$ initialized with random values.

\begin{figure}[!h]
\centering
\begin{subfigure}{.3\textwidth}
\centering
\includegraphics[scale=0.15]{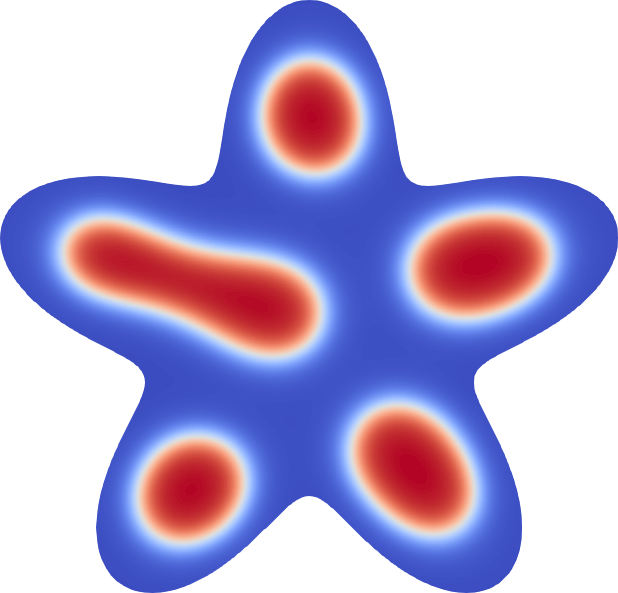}
\end{subfigure}
\begin{subfigure}{.3\textwidth}
\centering
\includegraphics[scale=0.25]{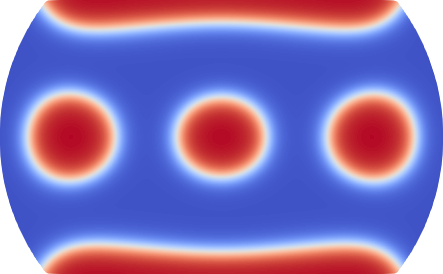}
\end{subfigure}
\begin{subfigure}{.3\textwidth}
\centering
\includegraphics[scale=0.25]{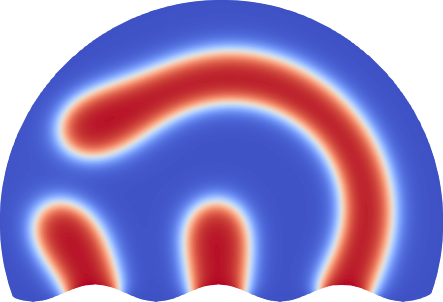}
\end{subfigure}
\caption{Resulting polymer morphologies obtained using the standard SCFT approach for imposing surface energies.}
\label{sec:scft:test:density:simple}
\end{figure}

\begin{figure}[!h]
\centering
\begin{subfigure}{.3\textwidth}
\centering
\includegraphics[scale=0.15]{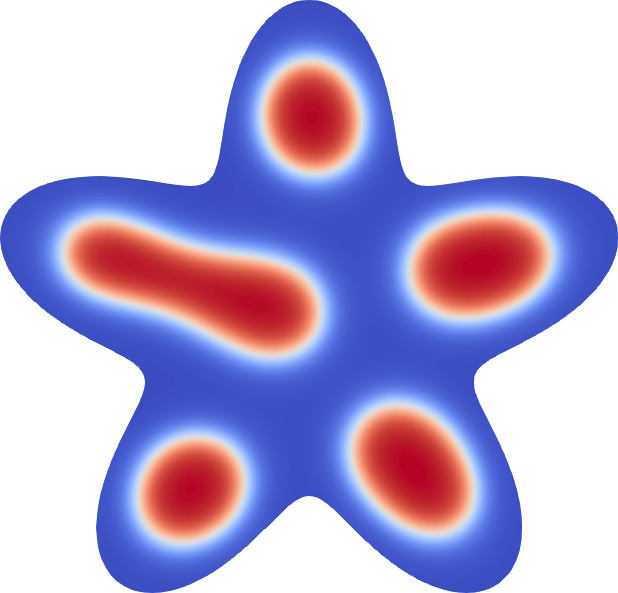}
\end{subfigure}
\begin{subfigure}{.3\textwidth}
\centering
\includegraphics[scale=0.25]{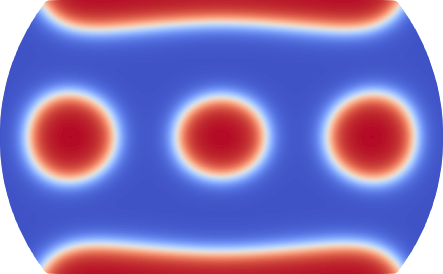}
\end{subfigure}
\begin{subfigure}{.3\textwidth}
\centering
\includegraphics[scale=0.25]{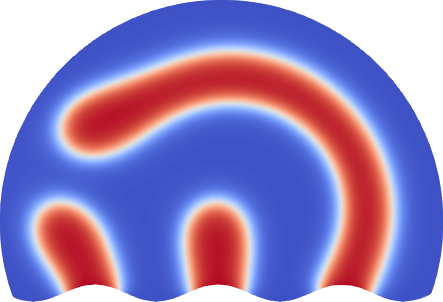}
\end{subfigure}
\caption{Resulting polymer morphologies obtained using the proposed approach for imposing surface energies.}
\label{sec:scft:test:density:smart}
\end{figure}

\begin{figure}[!h]
\centering
\begin{subfigure}{.3\textwidth}
\centering
\includegraphics[scale=0.2]{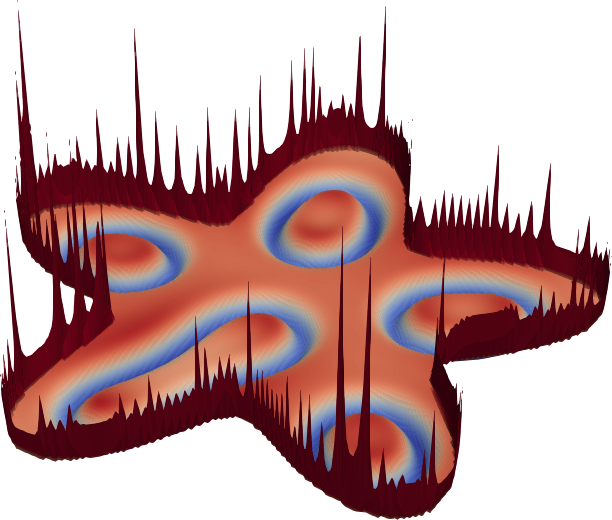}
\end{subfigure}
\begin{subfigure}{.3\textwidth}
\centering
\includegraphics[scale=0.15]{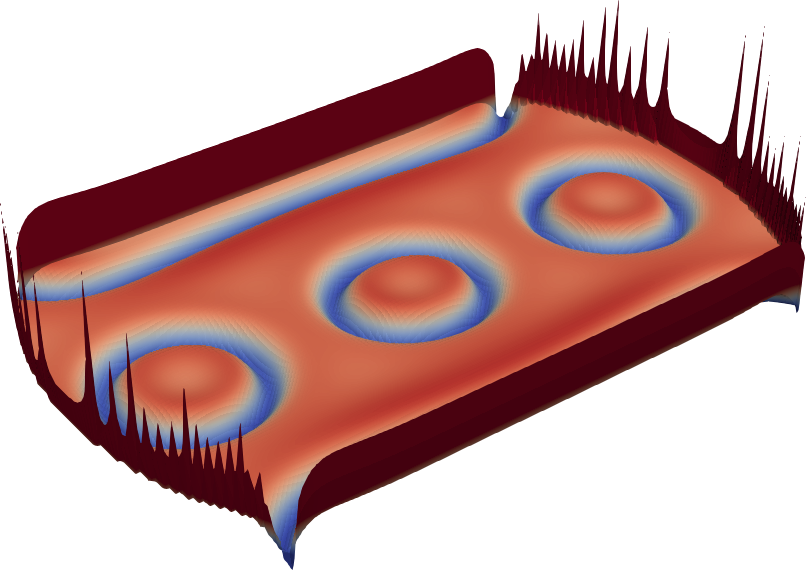}
\end{subfigure}
\begin{subfigure}{.3\textwidth}
\centering
\includegraphics[scale=0.15]{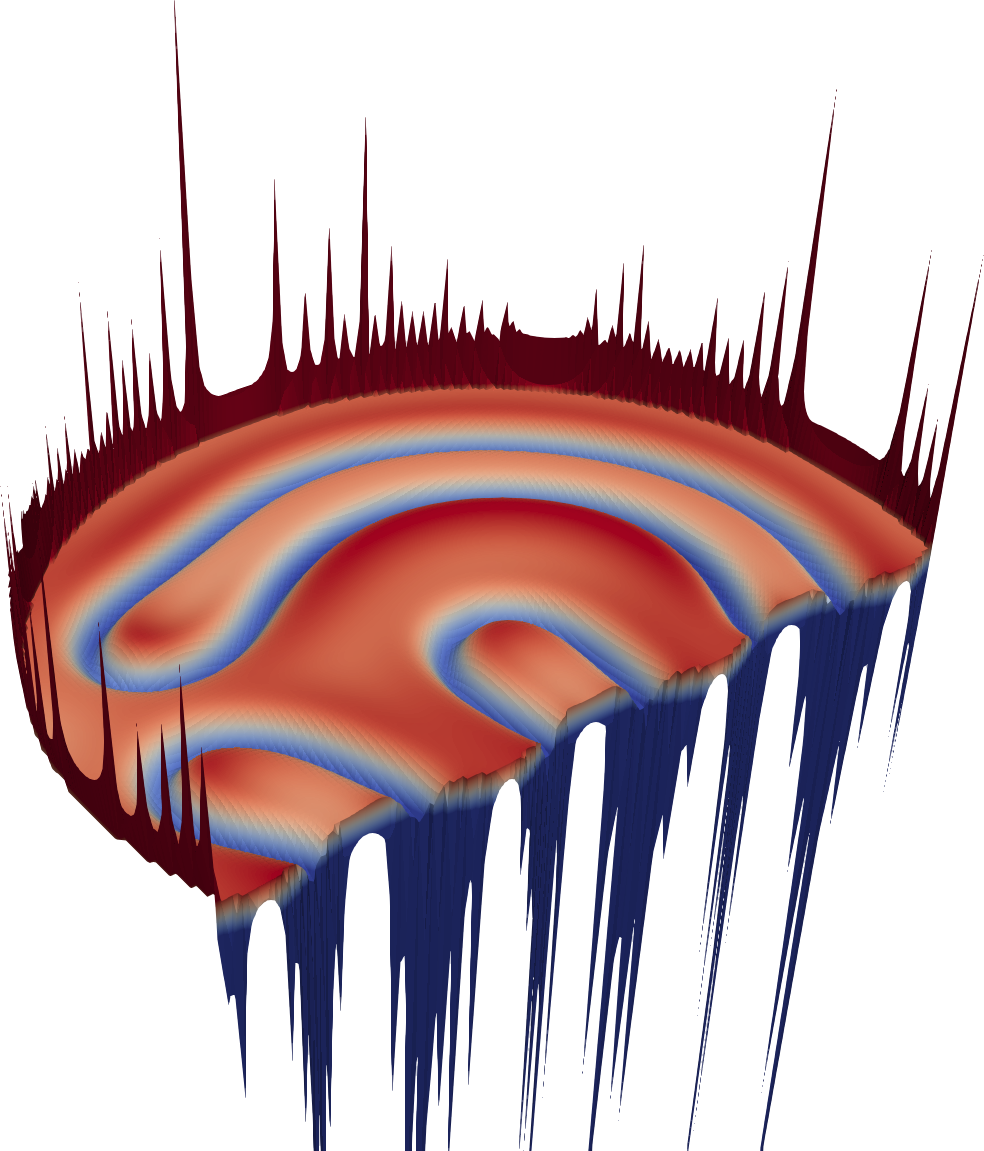}
\end{subfigure}
\caption{Resulting pressure field $\mu_+$ obtained using the standard SCFT approach for imposing surface energies.}
\label{sec:scft:test:pressure:simple}
\end{figure}

\begin{figure}[!h]
\centering
\begin{subfigure}{.3\textwidth}
\centering
\includegraphics[scale=0.2]{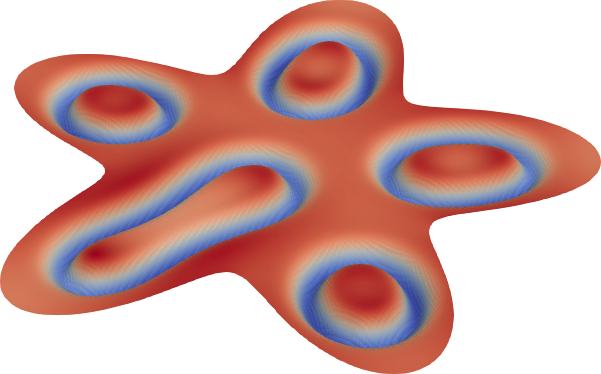}
\end{subfigure}
\begin{subfigure}{.3\textwidth}
\centering
\includegraphics[scale=0.15]{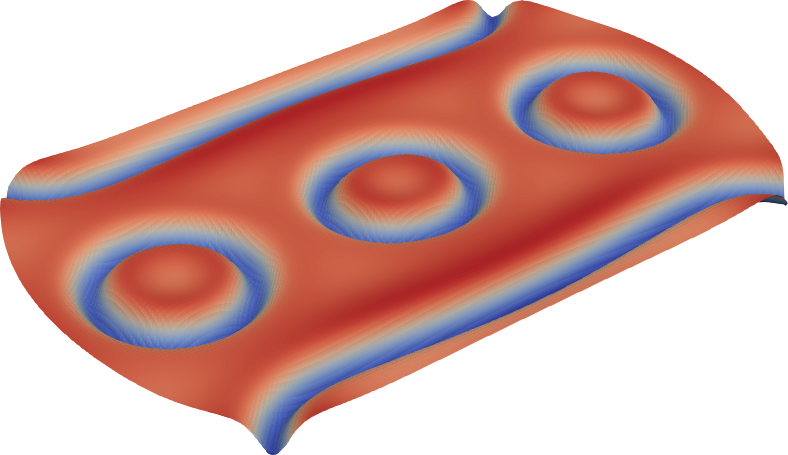}
\end{subfigure}
\begin{subfigure}{.3\textwidth}
\centering
\includegraphics[scale=0.15]{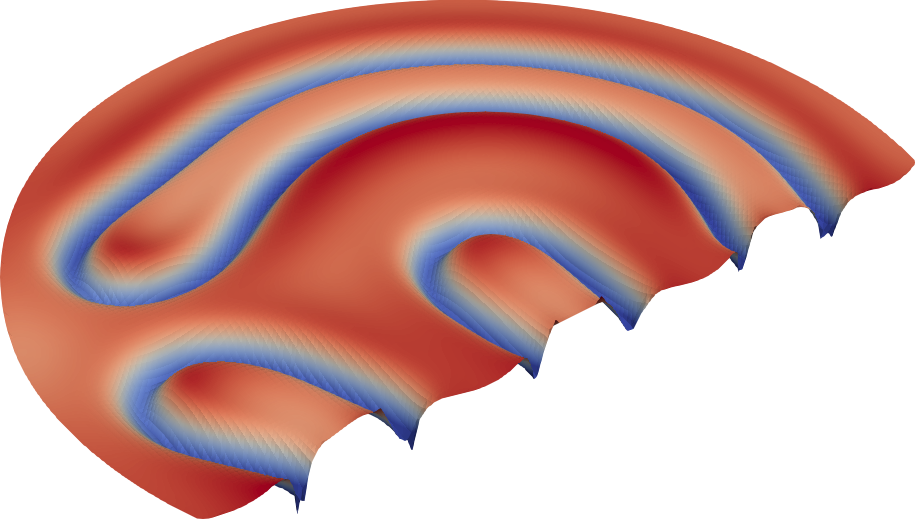}
\end{subfigure}
\caption{Resulting pressure field $\mu_+$ obtained using the proposed approach for imposing surface energies.}
\label{sec:scft:test:pressure:smart}
\end{figure}

Figures \ref{sec:scft:test:density:simple} and \ref{sec:scft:test:density:smart} depict the confining geometries considered and the resulting density fields obtained using the original system of SCFT equations and the modified system in \eqref{eq:qf:final}-\eqref{eq:qb:final}, while figures \ref{sec:scft:test:pressure:simple} and \ref{sec:scft:test:pressure:smart} demonstrate the resulting pressure fields obtained using the two approaches. As one can see, the solution of the two systems of equations leads to practically indistinguishable density fields, however, the standard approach leads to a singular behavior of $\mu_+$ near domains boundaries, while the proposed approach results in pressure fields with well-defined values everywhere.

\subsection{Validation tests}\label{sec:results:free:test}
\begin{figure}[!h]
\centering
\includegraphics[width=.99\textwidth]{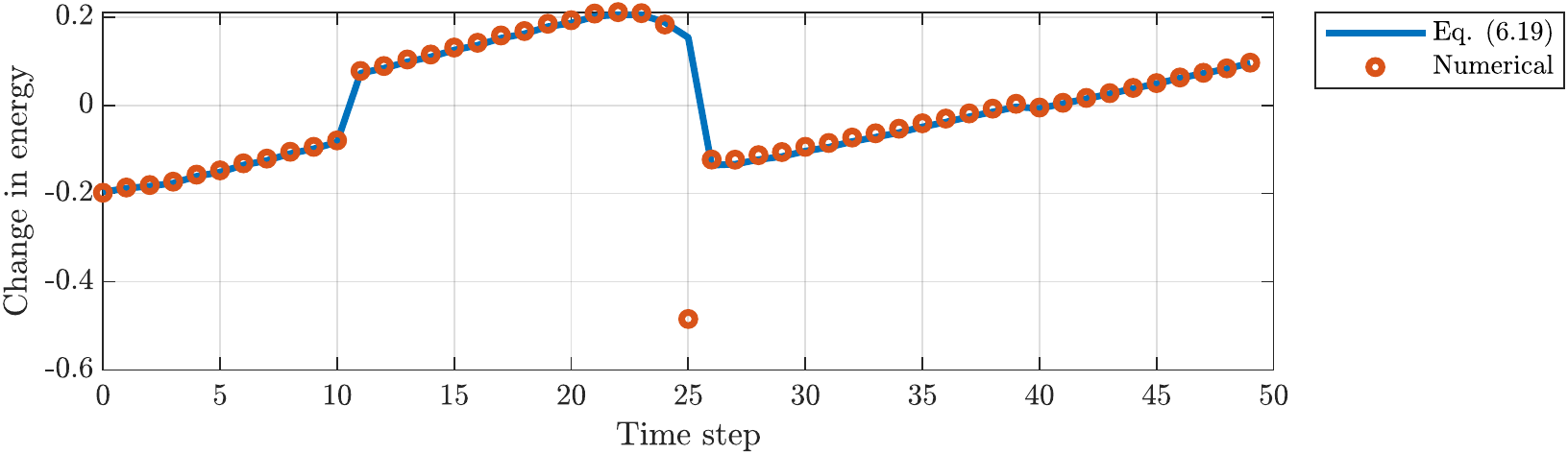}
\caption{Comparison between changes in energy computed using \eqref{eq:dhdt} and numerically for first example in section \ref{sec:results:free:test}.}
\label{fig:results:free:test:compare}
\end{figure}

To validate the obtained expressions for the system energy sensitivity to the free surface shape, we consider a synthetic test in which we externally impose a velocity field for the free surface, we track the change in energy at each iteration and we compare the numerical results to the prediction given in \eqref{eq:dhdt}. Specifically, we consider a polymer droplet of radius $r = 0.611R_g$ and deform it in the velocity field $(v_x, v_y)$:
\begin{myalign}\label{eq:results:free:velo}
\begin{aligned}
v_x &=  x \cos\of{5\tau}, \\
v_y &= -y \cos\of{5\tau}.
\end{aligned}
\end{myalign}
\begin{figure}[!h]
\centering
\includegraphics[width=.15\textwidth]{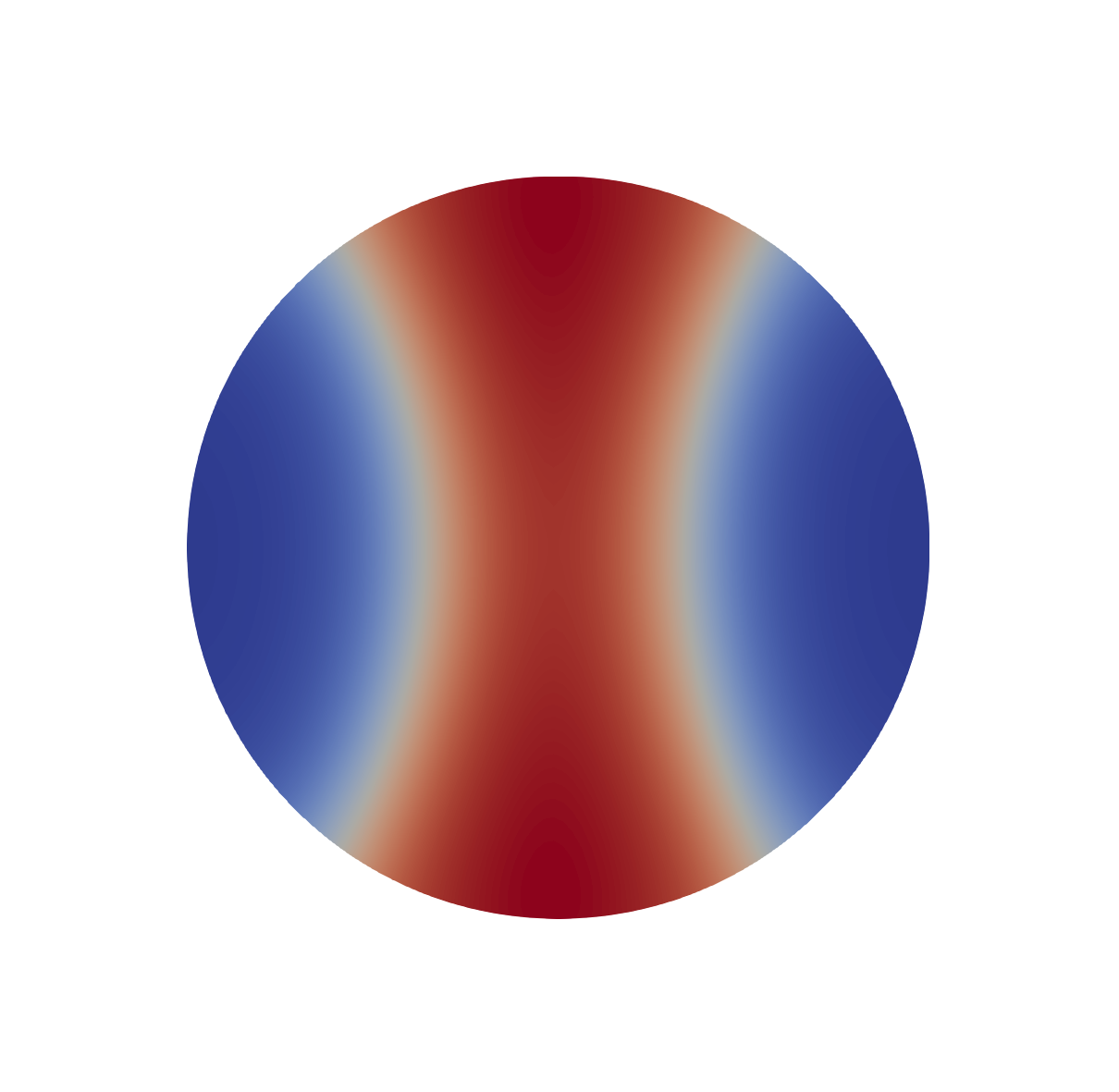}
\includegraphics[width=.15\textwidth]{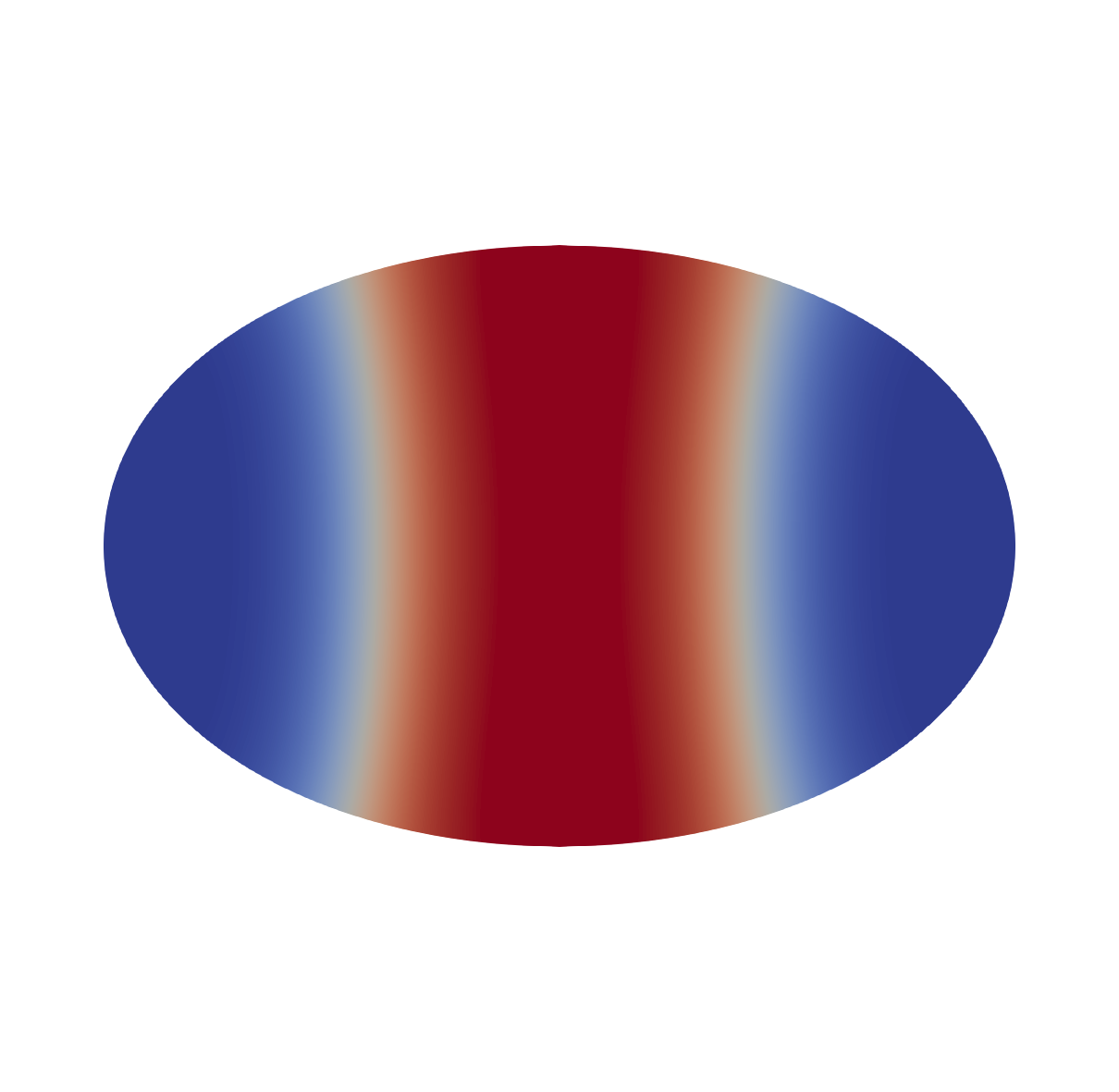}
\includegraphics[width=.15\textwidth]{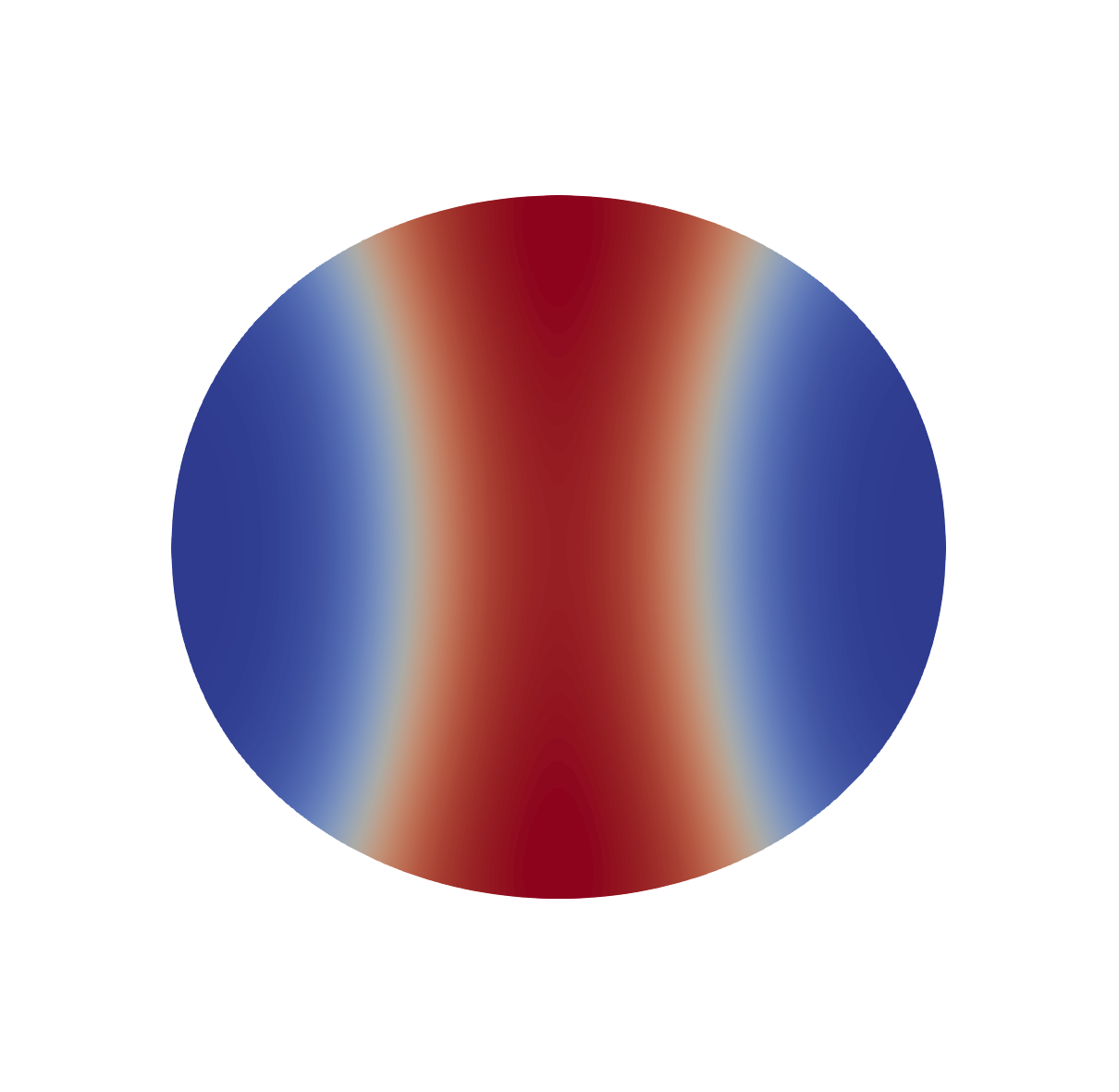}
\includegraphics[width=.15\textwidth]{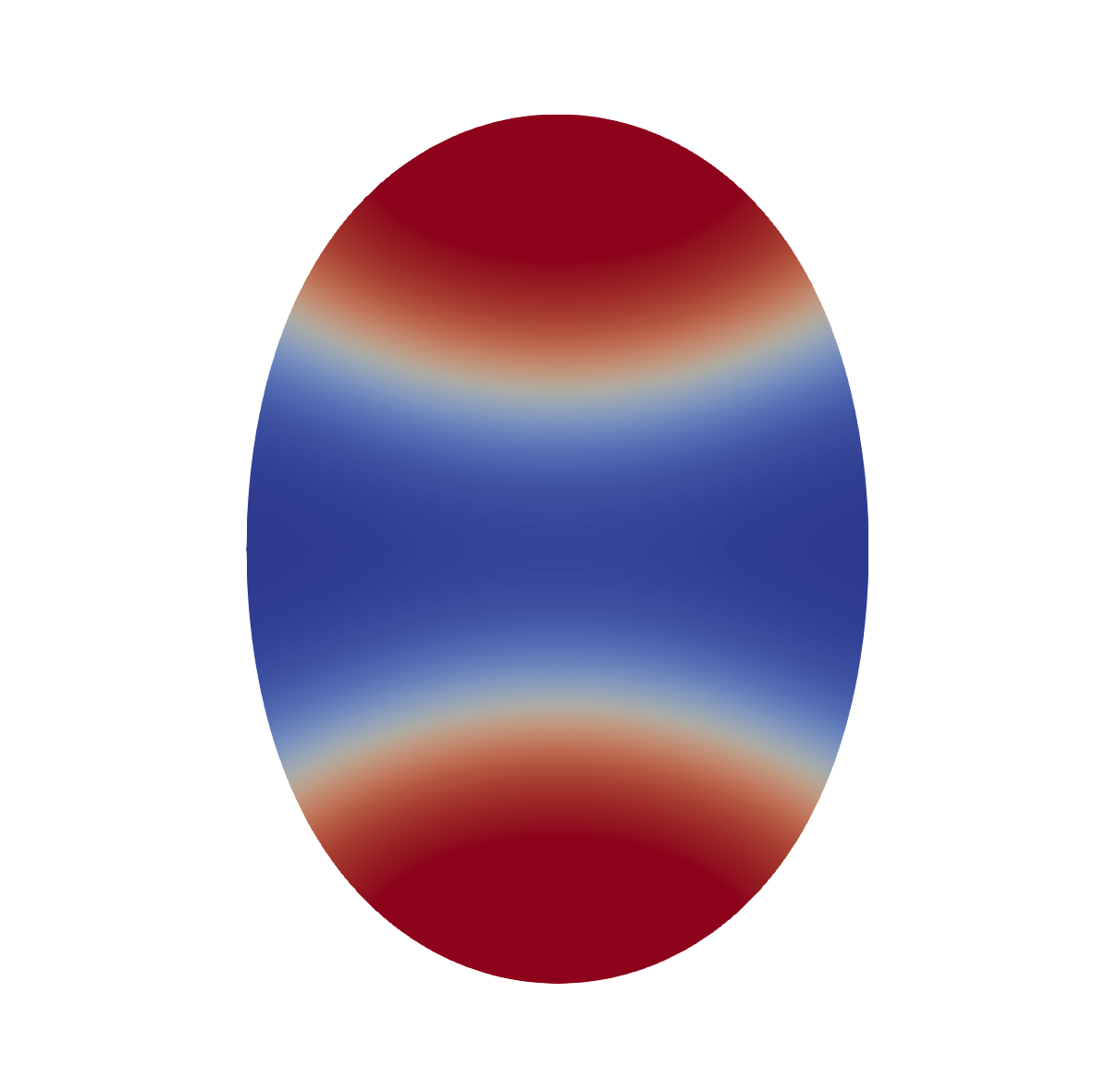}
\includegraphics[width=.15\textwidth]{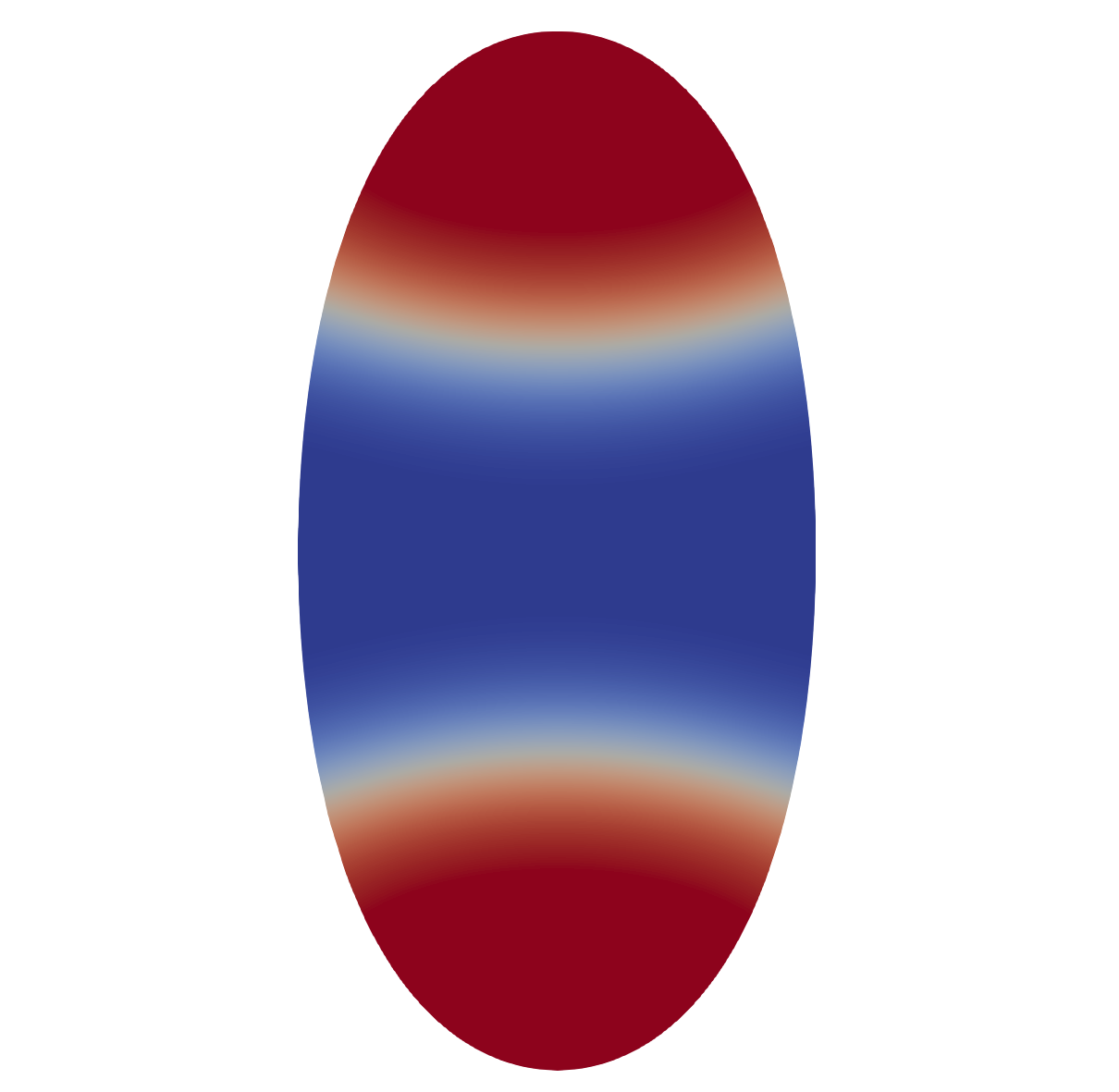}
\includegraphics[width=.15\textwidth]{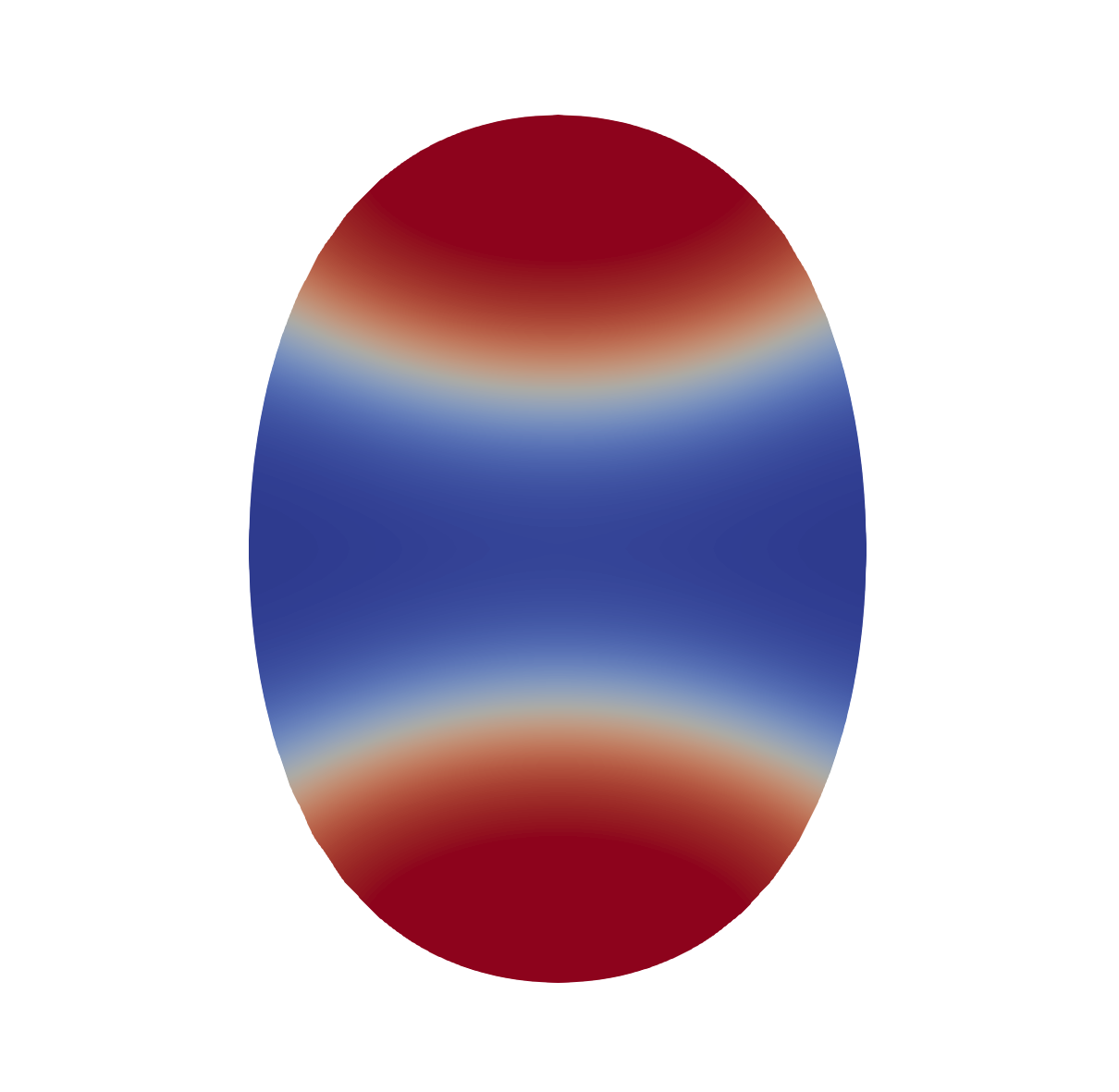}
\caption{Visualization of imposed deformations in first example from section \ref{sec:results:free:test}.}
\label{fig:results:free:test:visual}
\end{figure}
Figure \ref{fig:results:free:test:visual} visualize deformations that the polymer droplet undergoes and figure \ref{fig:results:free:test:compare} compares the predicted and actual changes in energy for each iteration. The predicted values match the actual values very well almost everywhere, expect for two spikes that correspond to topology changes in the polymer morphologies where the derived theory does not apply.

\begin{figure}[!h]
\centering
\includegraphics[width=.15\textwidth]{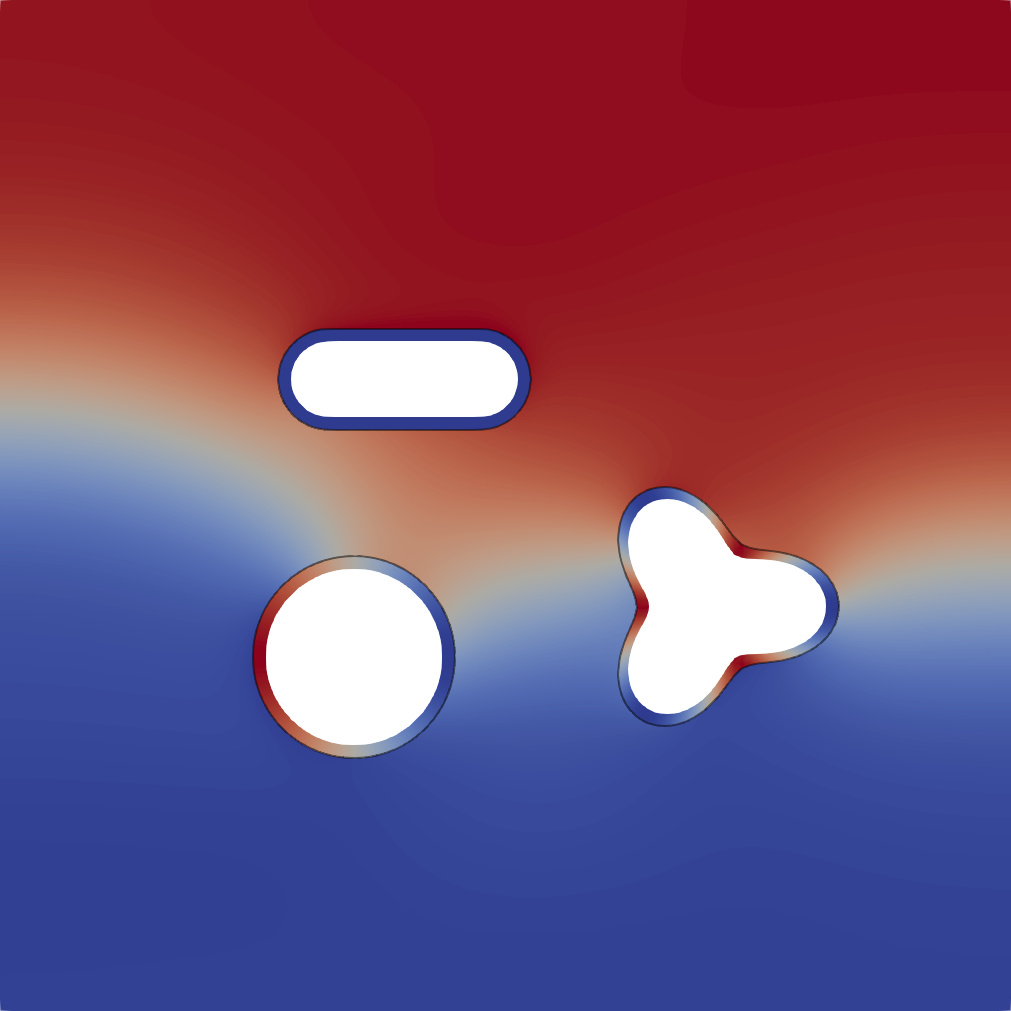}
\includegraphics[width=.15\textwidth]{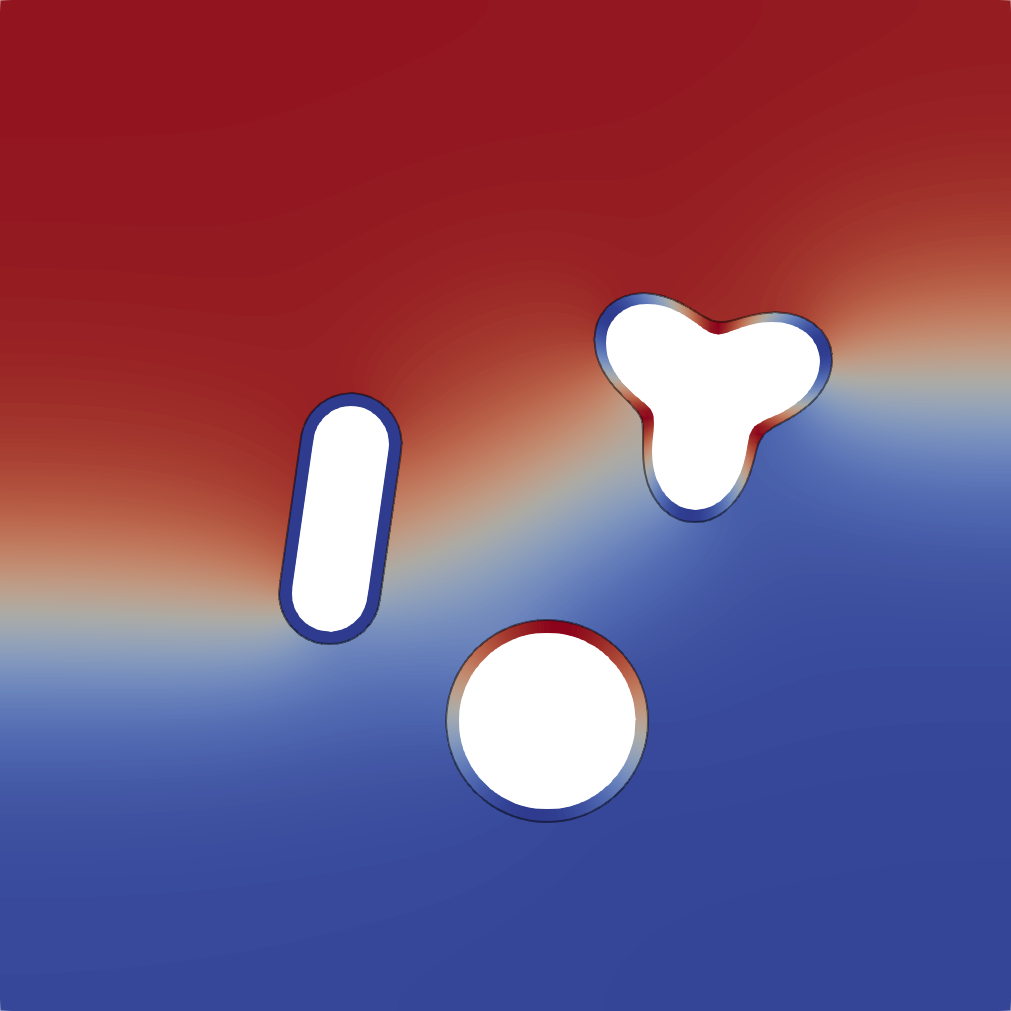}
\includegraphics[width=.15\textwidth]{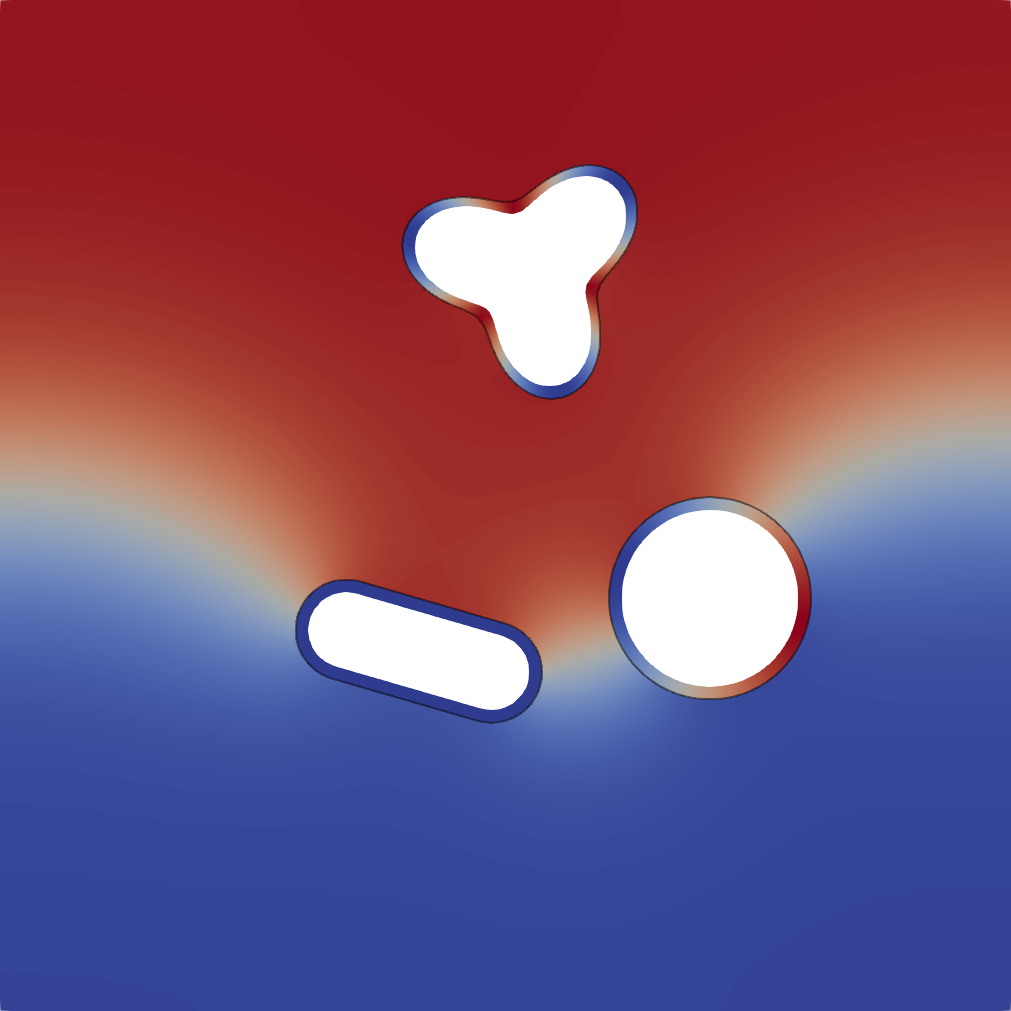}
\includegraphics[width=.15\textwidth]{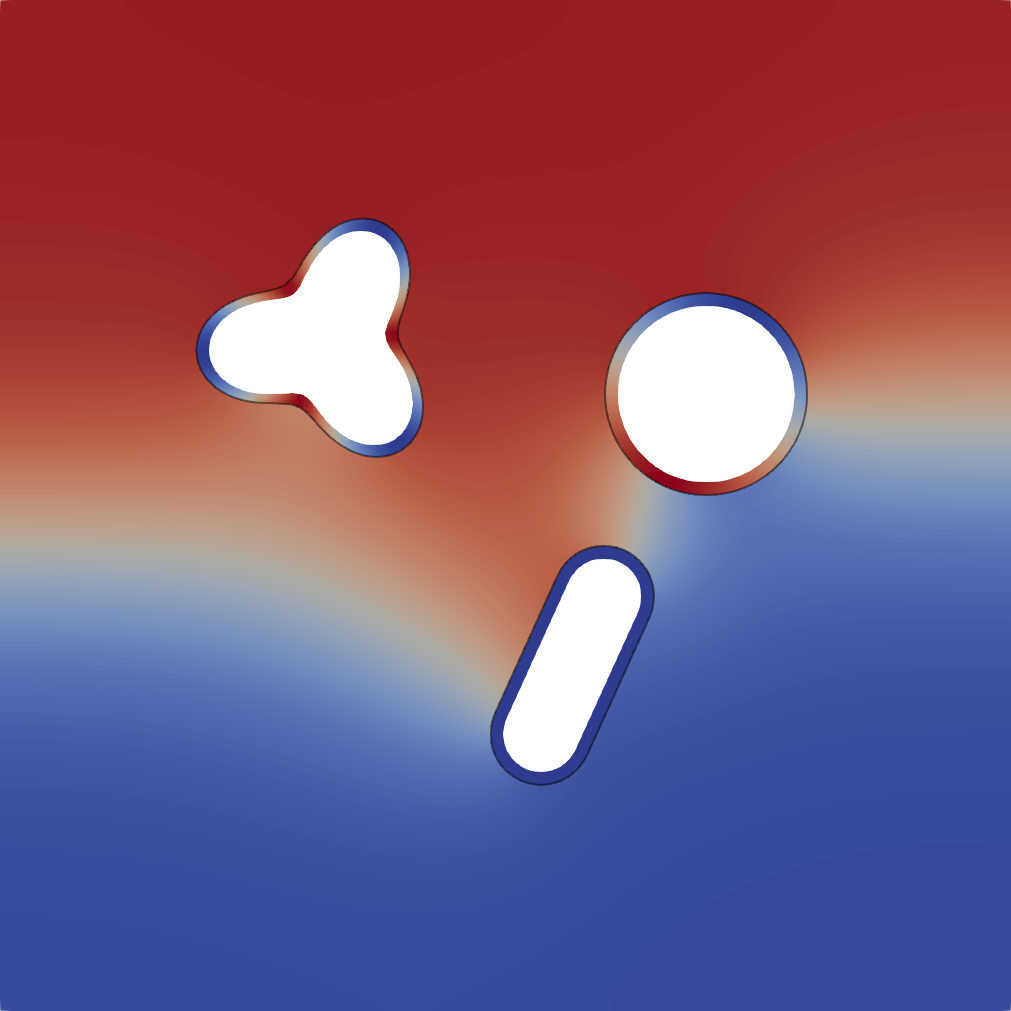}
\includegraphics[width=.15\textwidth]{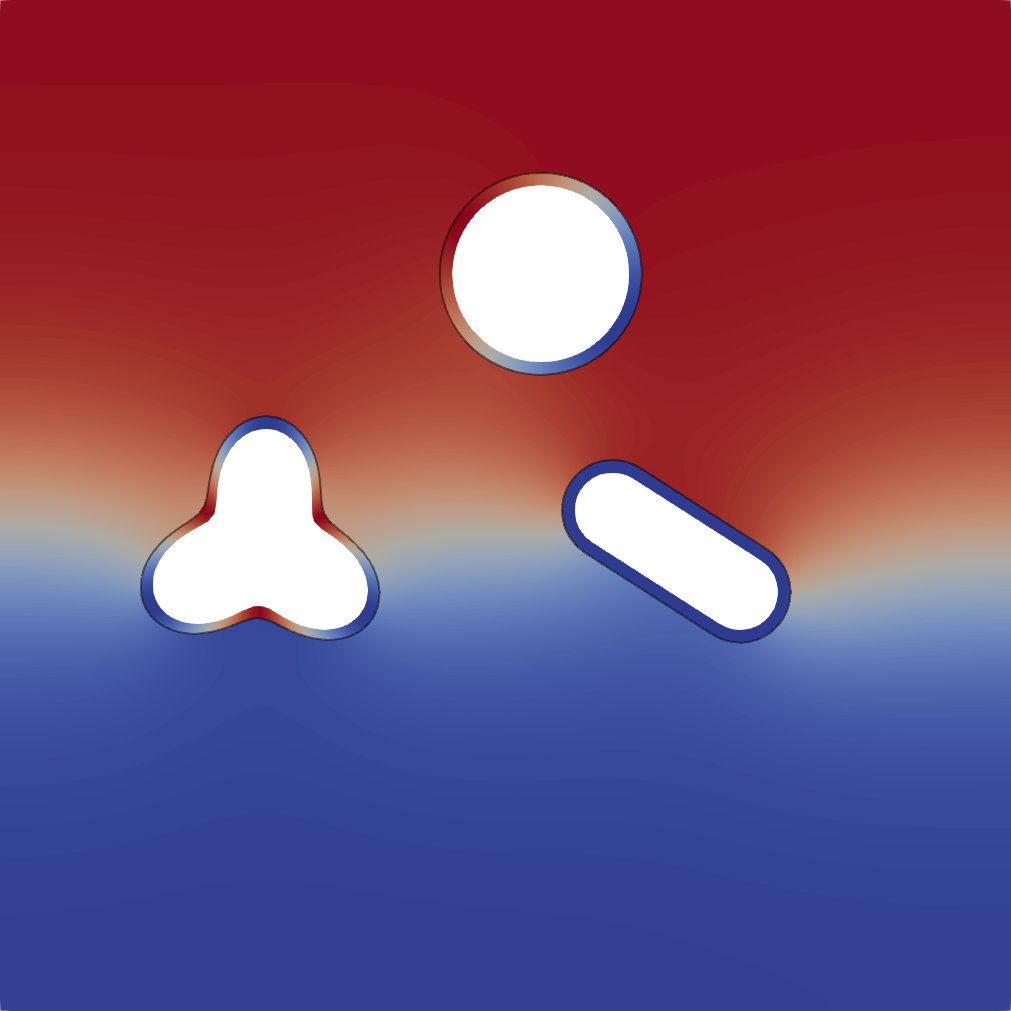}
\includegraphics[width=.15\textwidth]{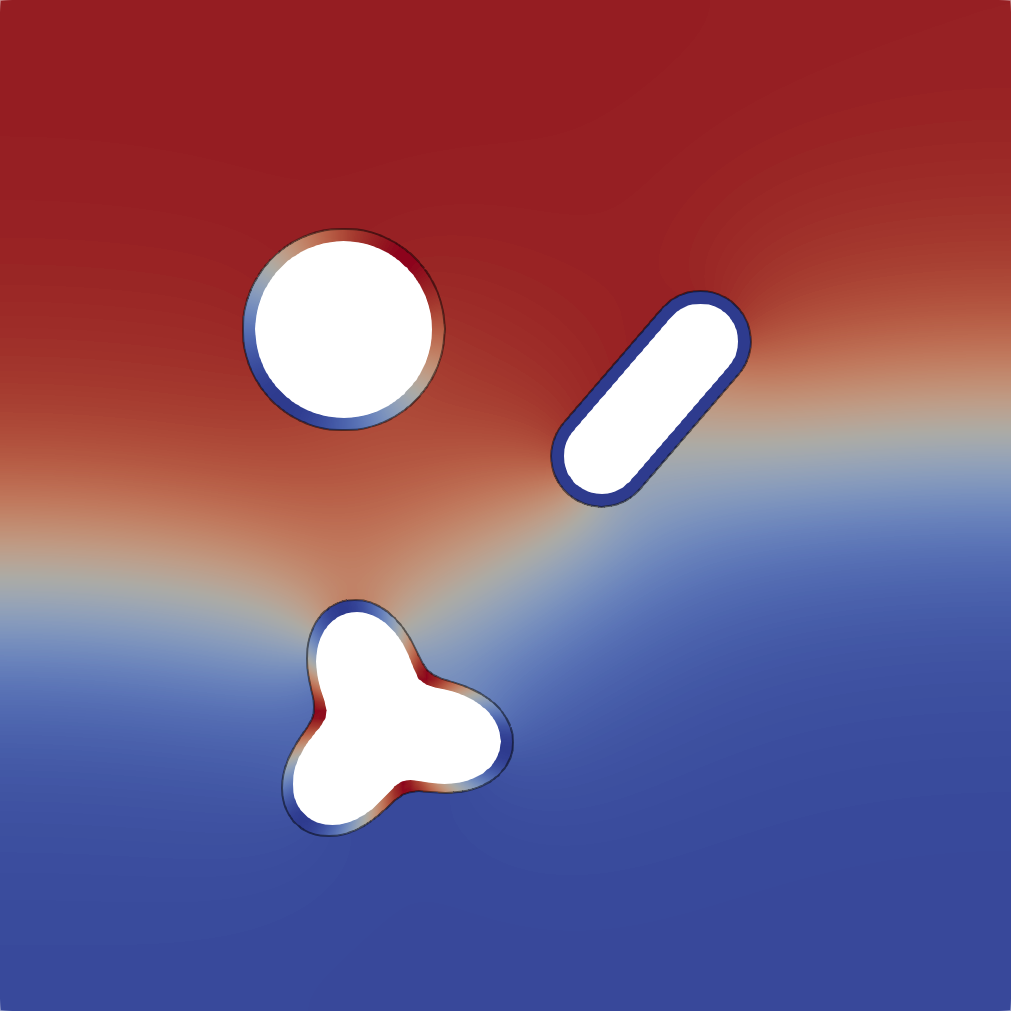}
\caption{Visualization of imposed motion in second example from section \ref{sec:results:free:test}.}
\label{fig:results:nano:test:visual}
\end{figure}

\begin{figure}[!h]
\centering
\includegraphics[width=.99\textwidth]{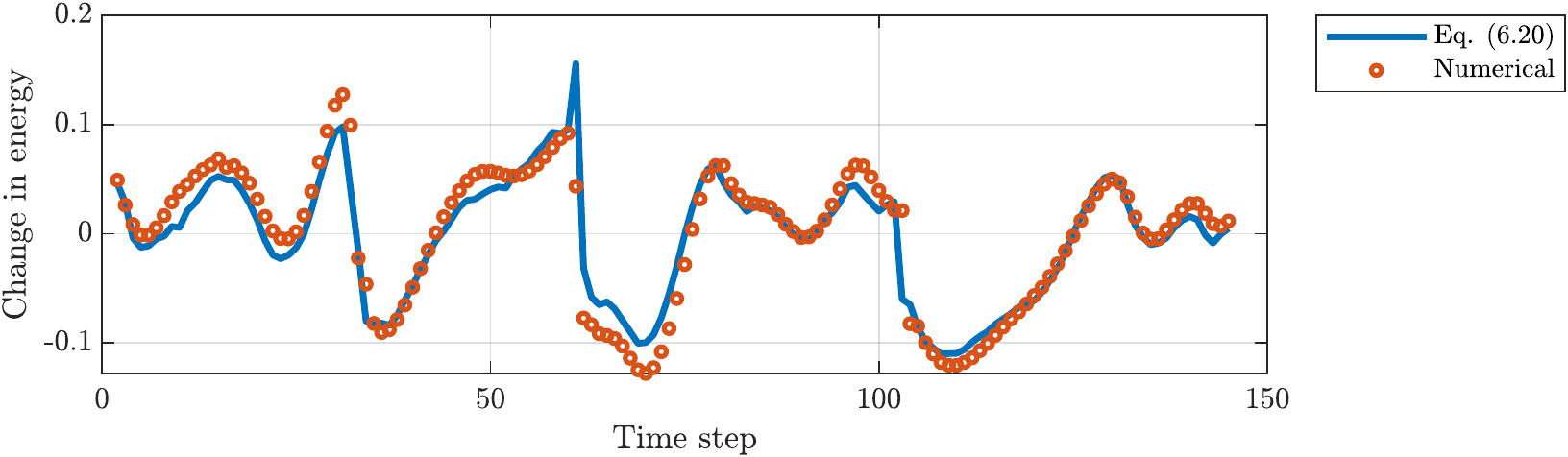}
\caption{Comparison between changes in energy computed using \eqref{eq:dhdt} and numerically for second example in section \ref{sec:results:free:test}.}
\label{fig:results:nano:test:compare}
\end{figure}

Similarly, to validate the energy sensitivity to particles' positions and orientations we consider a synthetic example with imposed velocity. Specifically, we consider three particles -- a disk, a rod, and a star -- rotating in a circular trajectory and spinning around their center of mass as illustrated in figure \ref{fig:results:nano:test:visual}. Note that the rod has a uniform surface energy, while the disk and the star have variable surface energies along their boundaries. Figure \ref{fig:results:nano:test:compare} compares the predicted and actual values of the energy change for each iteration.

Again, the predicted values match the numerically computed values well everywhere, expect for steps where the topology of polymer morphology undergoes a change. The small deviations between predicted and actual values are attributed to the errors introduced by numerical discretizations.

\subsection{Substrate-supported BCP droplets}
To demonstrate the capabilities of the proposed framework we consider several representative examples, specifically:
\begin{enumerate}
\item A droplet of lamella-forming diblock copolymer ($\XN{AB} =30$, $f=0.5$) on a substrate with lamellae oriented parallel to the substrate;
\item A droplet of lamella-forming diblock copolymer ($\XN{AB} =30$, $f=0.5$) on a substrate with lamellae oriented perpendicular to the substrate;
\item A droplet of cylinder-forming diblock copolymer ($\XN{AB} =30$, $f=0.3$) on a substrate;
\item A droplet of cylinder-forming diblock copolymer ($\XN{AB} =30$, $f=0.3$) in a groove.
\end{enumerate}
\begin{figure}[!h]
\centering
\begin{subfigure}{.32\textwidth}
\centering
\includegraphics[width=.8\textwidth]{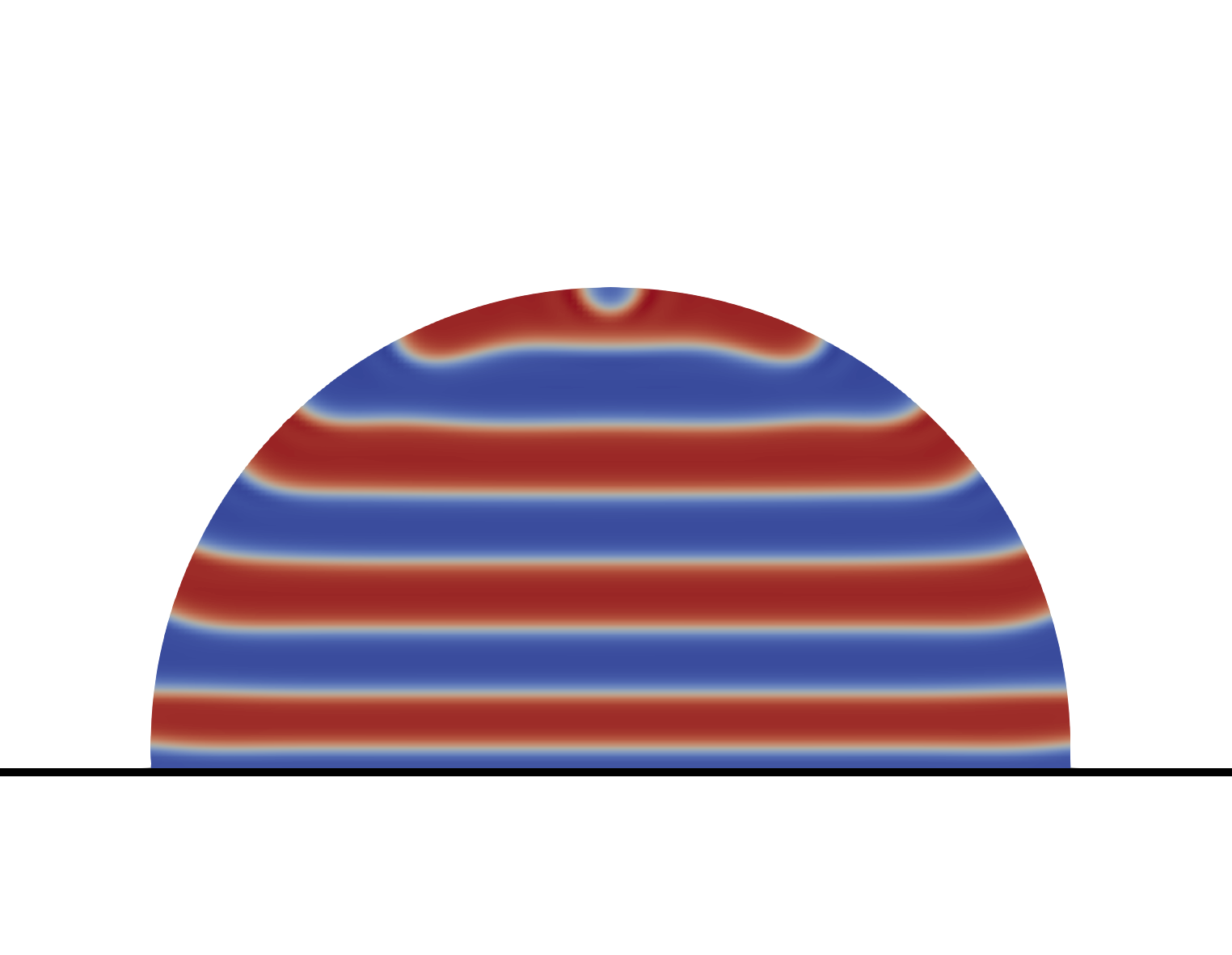}
\caption{}
\end{subfigure}
\begin{subfigure}{.32\textwidth}
\centering
\includegraphics[width=.8\textwidth]{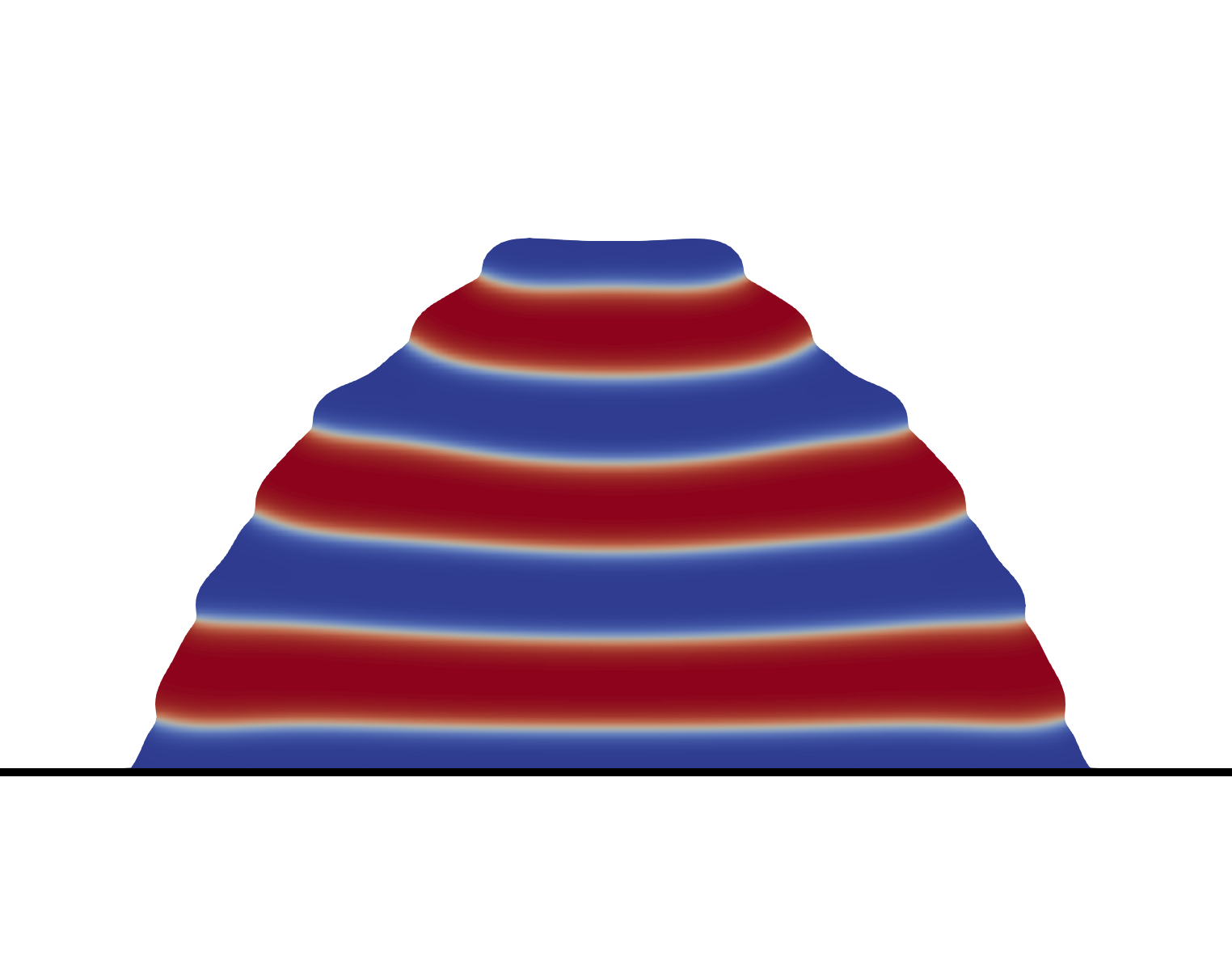}
\caption{}
\end{subfigure}
\begin{subfigure}{.32\textwidth}
\centering
\includegraphics[width=.8\textwidth]{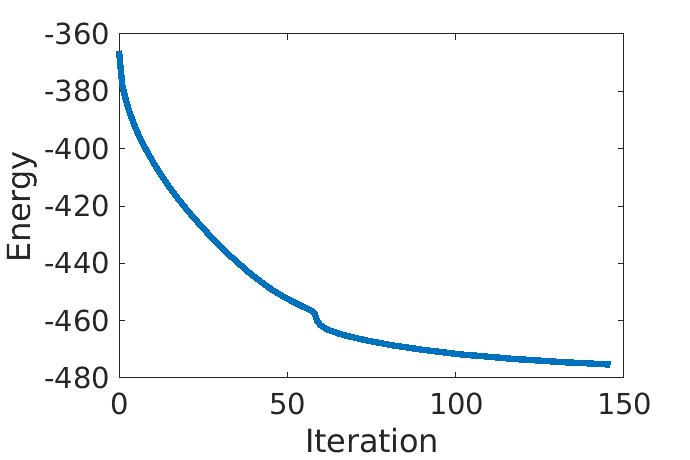}
\caption{}
\end{subfigure}
\caption{Equilibration of a droplet of a lamella-forming BCP on a substrate with lamellae oriented horizontally: (a) initial state, (b) final state, (c) energy evolution.}
\label{fig:results:free:case2}
\end{figure}
\begin{figure}[!h]
\centering
\begin{subfigure}{.32\textwidth}
\centering
\includegraphics[width=.8\textwidth]{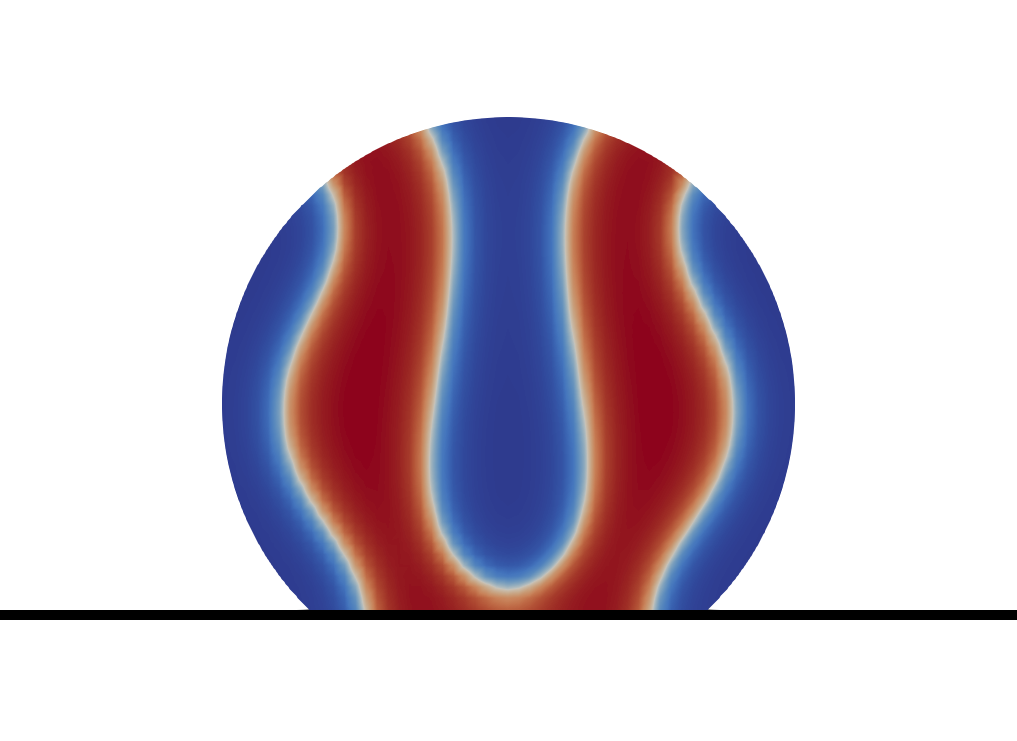}
\caption{}
\end{subfigure}
\begin{subfigure}{.32\textwidth}
\centering
\includegraphics[width=.8\textwidth]{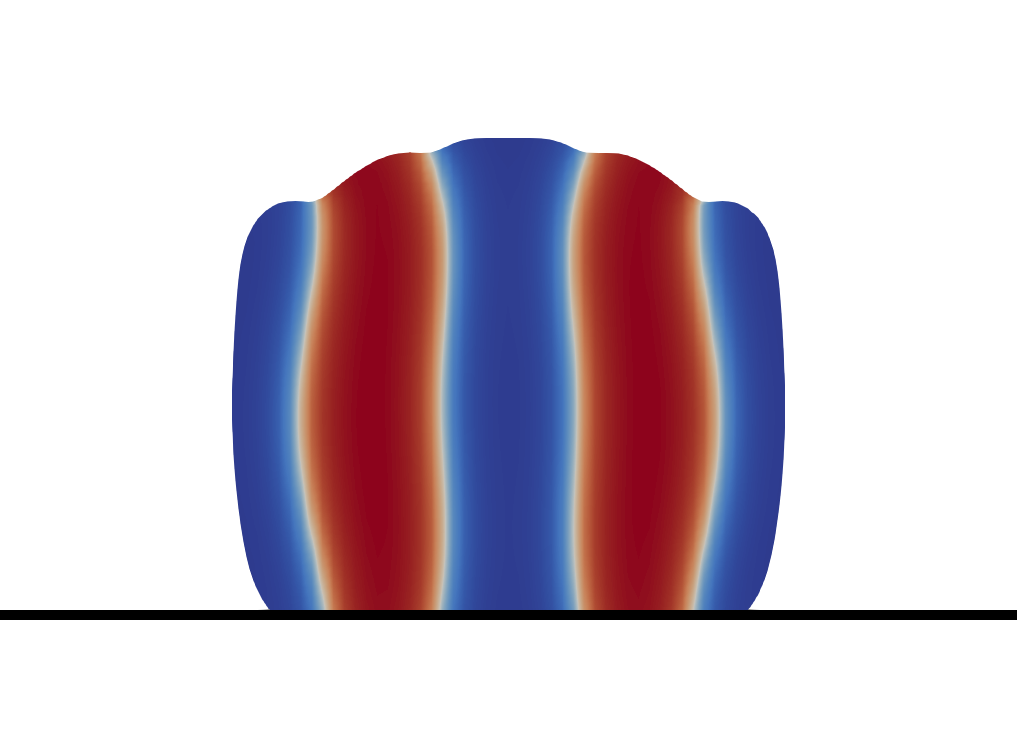}
\caption{}
\end{subfigure}
\begin{subfigure}{.32\textwidth}
\centering
\includegraphics[width=.8\textwidth]{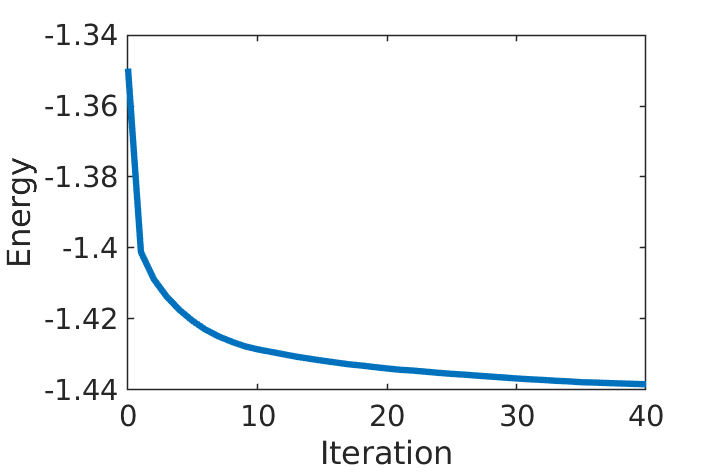}
\caption{}
\end{subfigure}
\caption{Equilibration of a droplet of a lamella-forming BCP on a substrate with lamellae oriented vertically: (a) initial state, (b) final state, (c) energy evolution.}
\label{fig:results:free:case3}
\end{figure}
\begin{figure}[!h]
\centering
\begin{subfigure}{.32\textwidth}
\centering
\includegraphics[width=.8\textwidth]{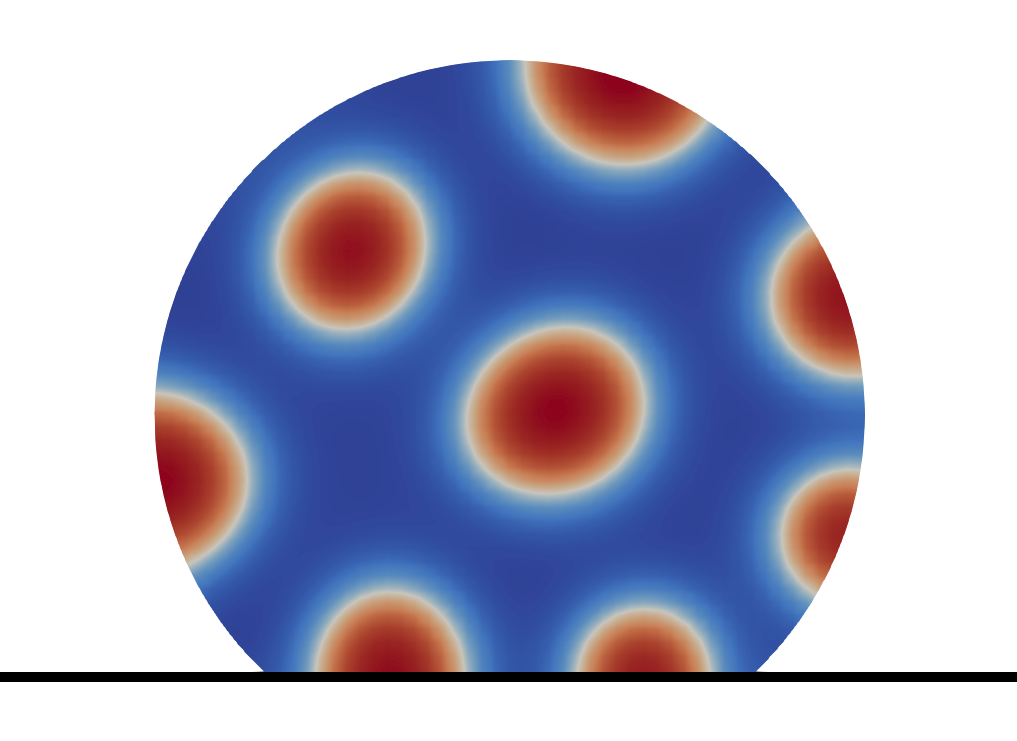}
\caption{}
\end{subfigure}
\begin{subfigure}{.32\textwidth}
\centering
\includegraphics[width=.8\textwidth]{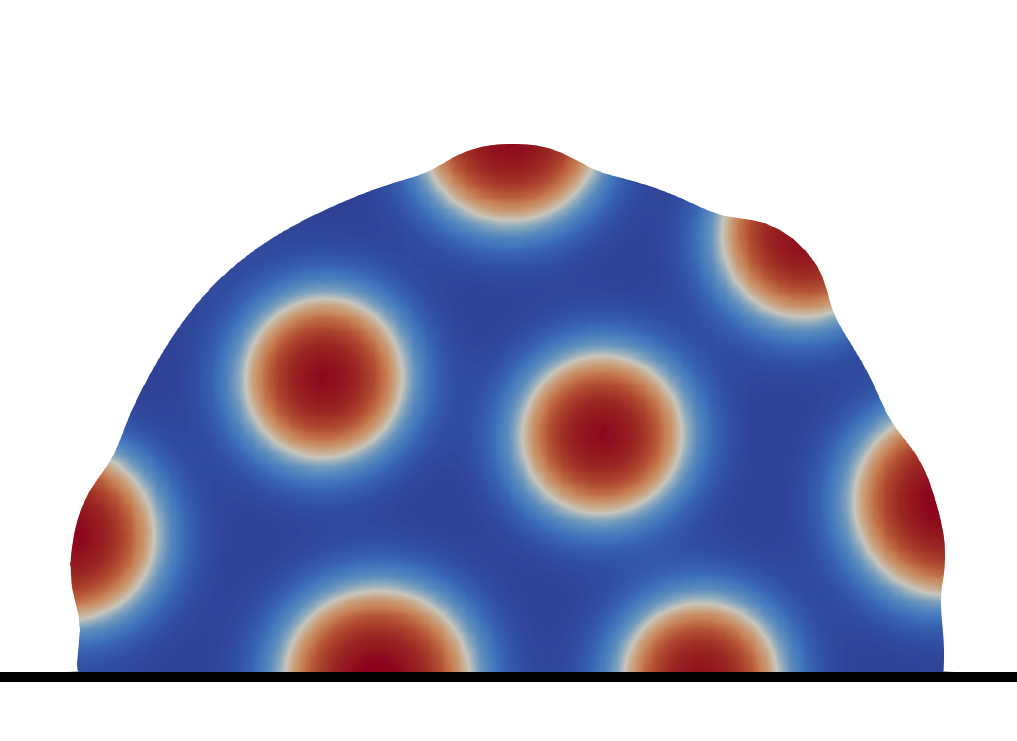}
\caption{}
\end{subfigure}
\begin{subfigure}{.32\textwidth}
\centering
\includegraphics[width=.8\textwidth]{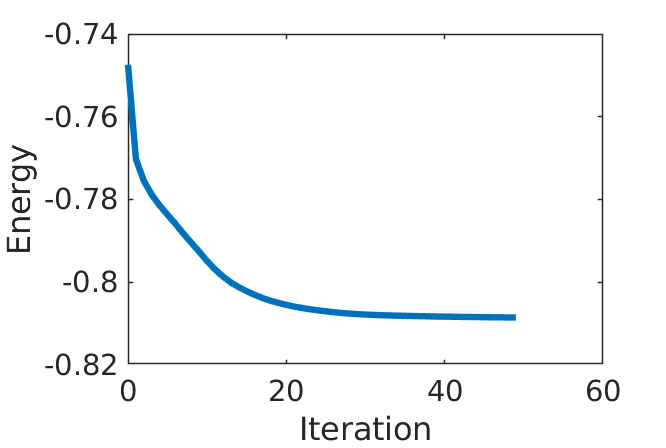}
\caption{}
\end{subfigure}
\caption{Equilibration of a droplet of a cylinder-forming BCP on a substrate: (a) initial state, (b) final state, (c) energy evolution.}
\label{fig:results:free:case1}
\end{figure}
\begin{figure}[!h]
\centering
\begin{subfigure}{.32\textwidth}
\centering
\includegraphics[width=.8\textwidth]{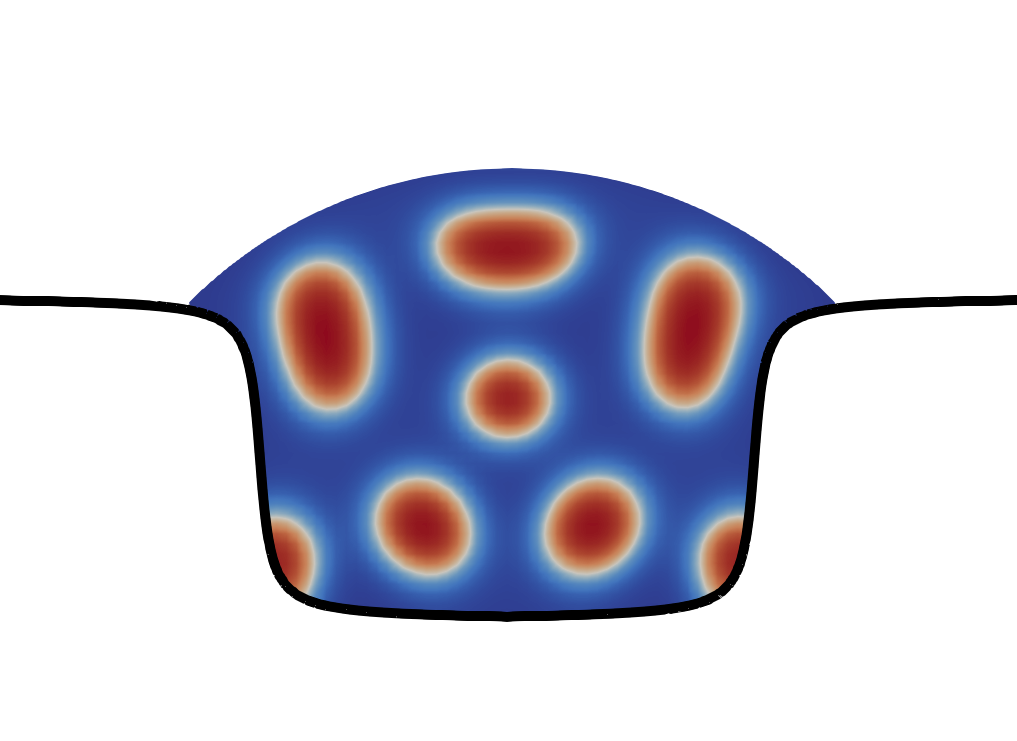}
\caption{}
\end{subfigure}
\begin{subfigure}{.32\textwidth}
\centering
\includegraphics[width=.8\textwidth]{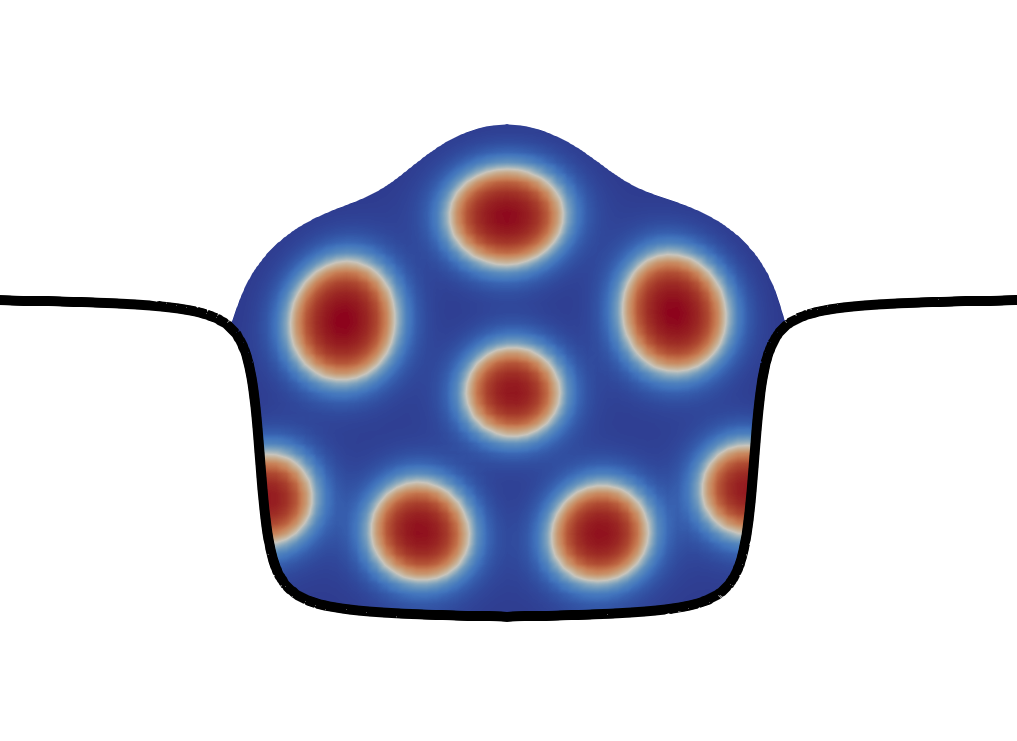}
\caption{}
\end{subfigure}
\begin{subfigure}{.32\textwidth}
\centering
\includegraphics[width=.8\textwidth]{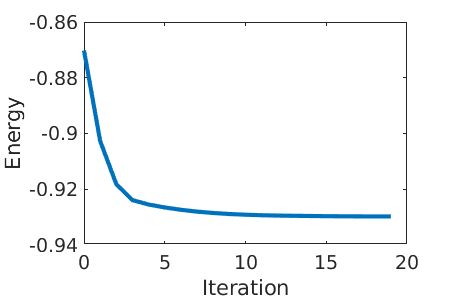}
\caption{}
\end{subfigure}
\caption{Equilibration of a droplet of a cylinder-forming BCP in a groove: (a) initial state, (b) final state, (c) energy evolution.}
\label{fig:results:free:case4}
\end{figure}
The simulation results for these cases are presented in figures \ref{fig:results:free:case1},  \ref{fig:results:free:case2}, \ref{fig:results:free:case3}, and \ref{fig:results:free:case4}. Specifically, they demonstrate the initial shape of a polymer droplet, the final shape, and the energy evolution. As one can see, in all cases the simulated system reaches a local energetic minimum.

\subsection{Meniscus formation in graphoepitaxy applications}

Next, we apply the presented method to study the meniscus formation in graphoepitaxy applications of block copolymers \cite{jeong2009soft,jeong2013directed}. Specifically, we consider vertically oriented lamellar-forming and horizontally oriented cylindrical-forming block copolymers in grooves and analyze the influence of the polymer-air surface tension as well as the value of the contact angle on the resulting self-assembling structures of polymer material. The parameters of the problem are chosen in the following way: the width of the groove is $20 R_g$, the height of the unperturbed film is $5 R_g$, the interaction strength between $A$ and $B$ polymer species is $\XN{AB} = 30$, the fraction of the $A$ component is $f=0.5$ and $f=0.3$ in the case of lamellae- and cylinder-forming diblock copolymers, respectively. For simplicity, the surface tensions between the polymer species and the groove's walls are assumed to be equal (i.e., no preferential attraction). 

\begin{figure}[!h]
\centering  
\begin{tabular}{ c  c  c  c }
$\chi_{ap} N$ &
$120^o$ &
$90^o$ &
$60^o$  \\
$1$ &
\includegraphics[scale=0.09]{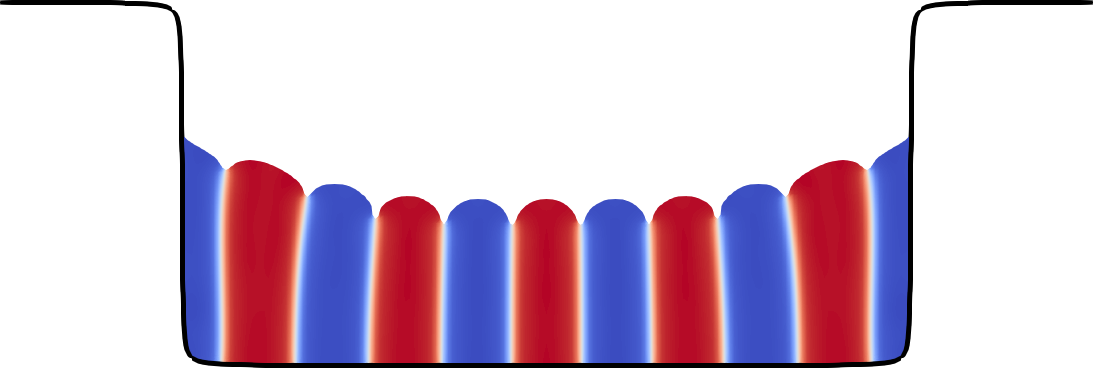} &
\includegraphics[scale=0.09]{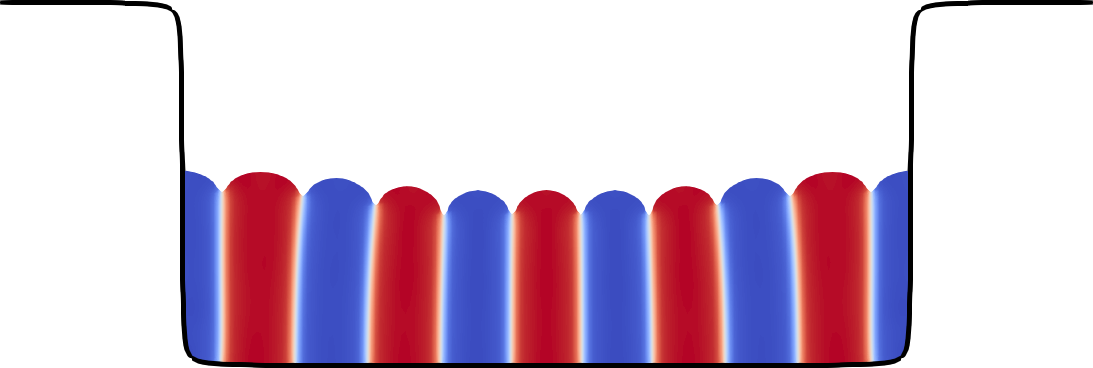} &
\includegraphics[scale=0.09]{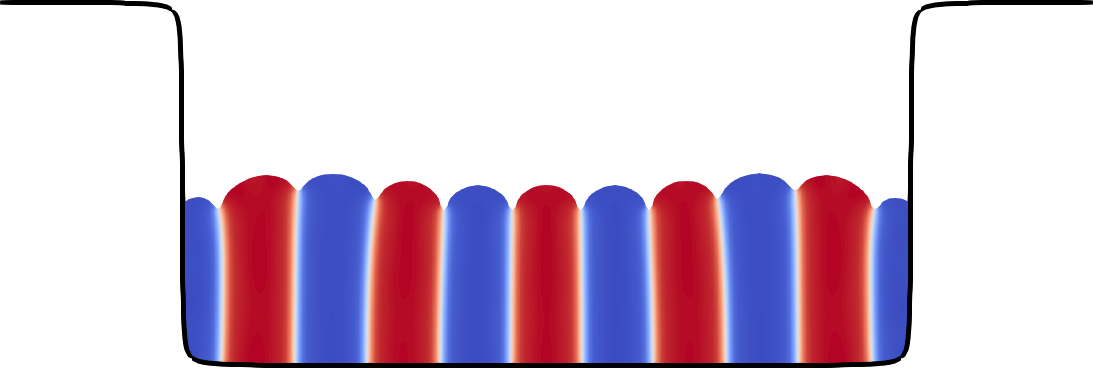} \\
$9$ &
\includegraphics[scale=0.09]{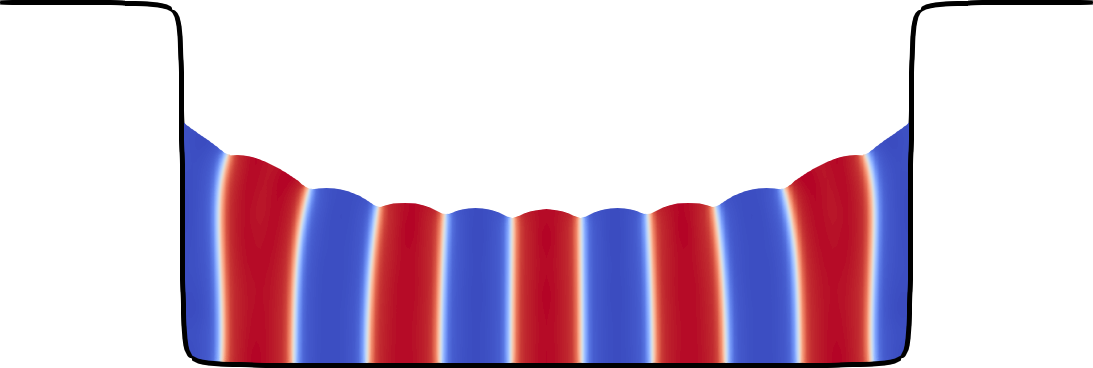} &
\includegraphics[scale=0.09]{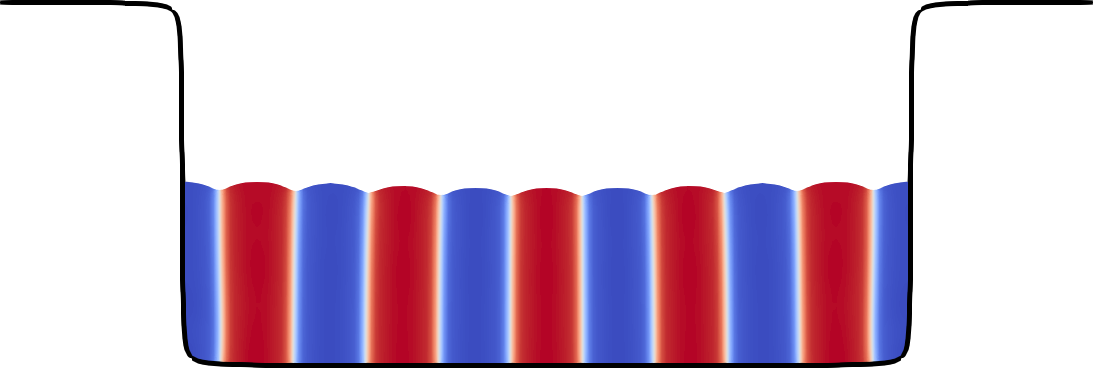} &
\includegraphics[scale=0.09]{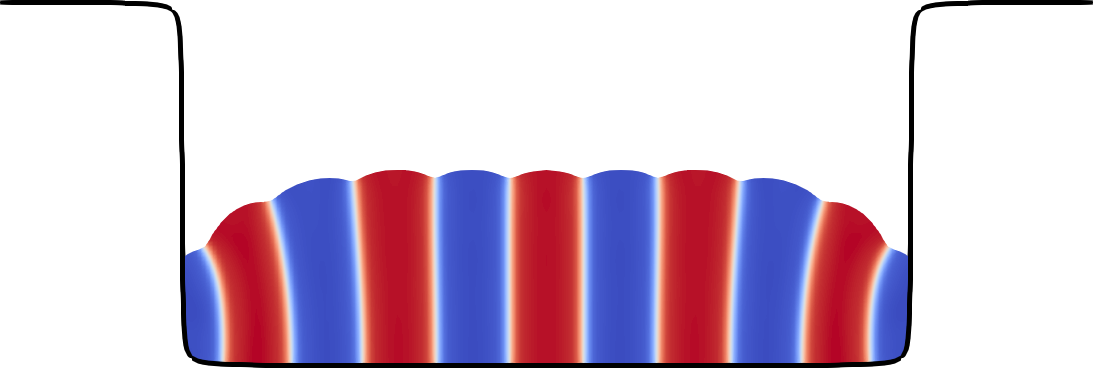} \\
$25$ &
\includegraphics[scale=0.09]{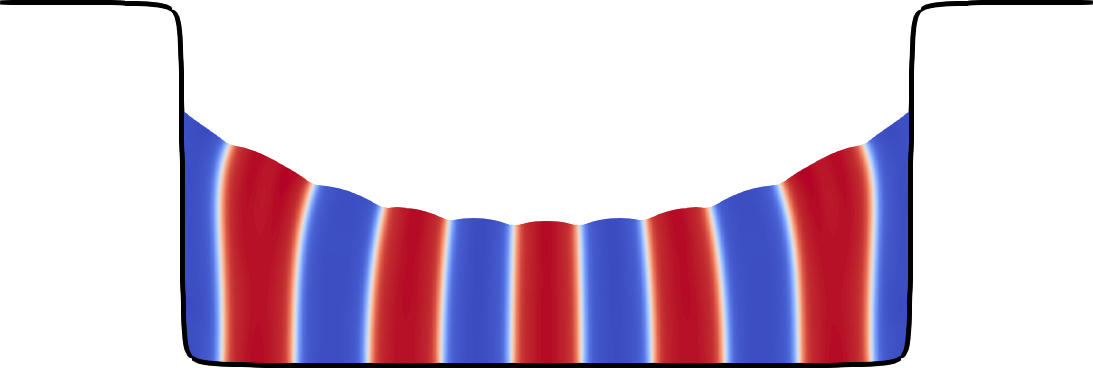} &
\includegraphics[scale=0.09]{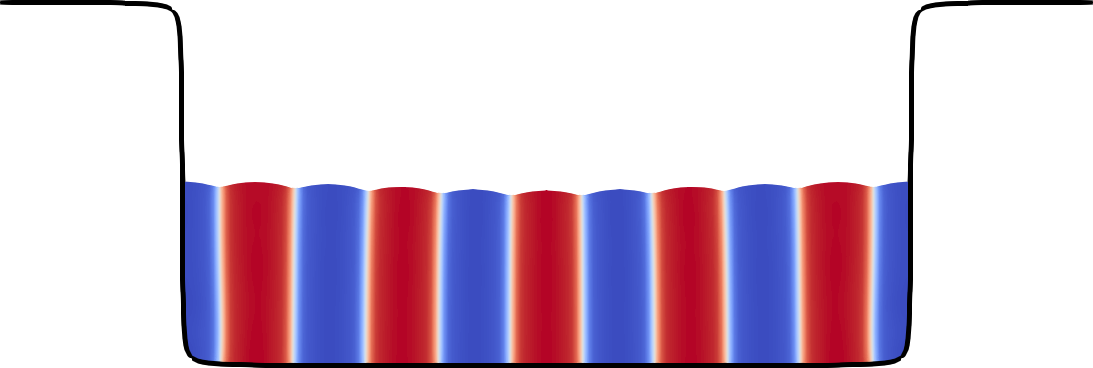} &
\includegraphics[scale=0.09]{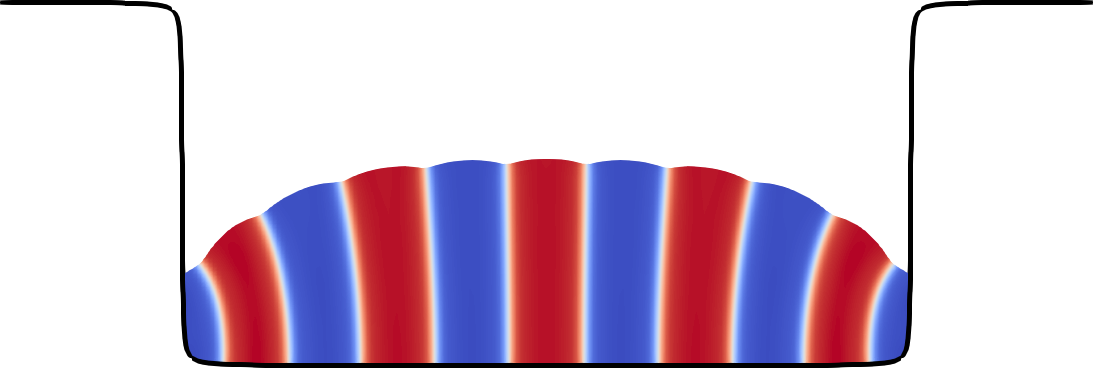} \\
$100$ &
\includegraphics[scale=0.09]{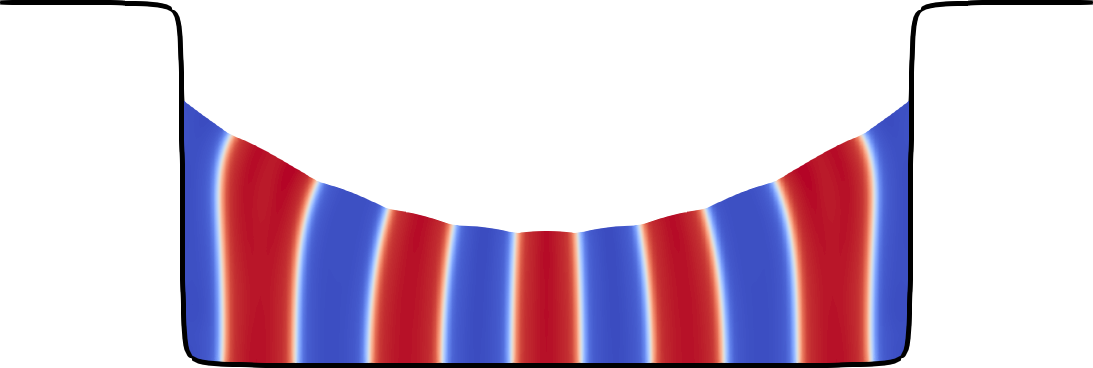} &
\includegraphics[scale=0.09]{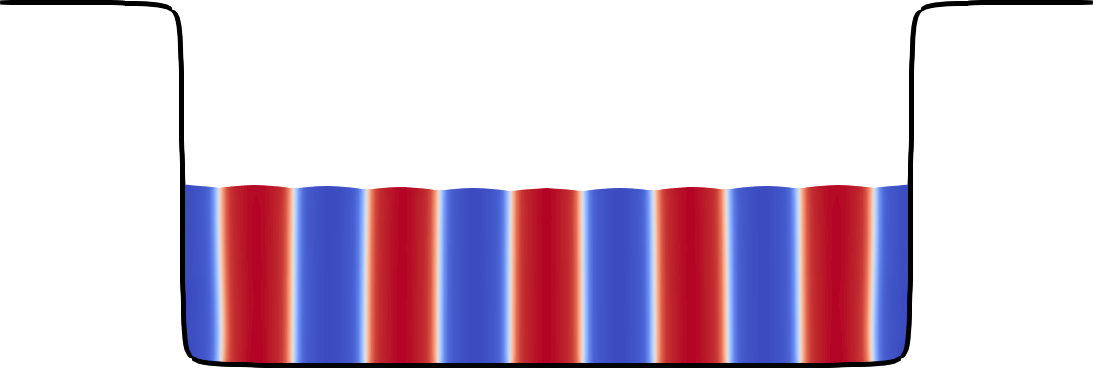} &
\includegraphics[scale=0.09]{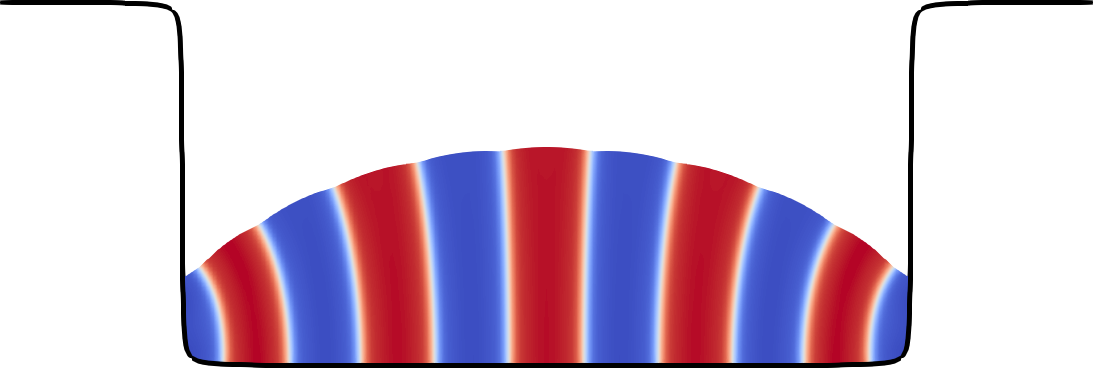} 
\end{tabular}
\caption{Equilibrium shapes of a lamellae-forming diblock copolymer film of height $4R_g$ in a groove of width $20R_g$ for several values of the polymer-air surface tension and contact angle.}
\label{fig:results:free:film:lamellar}
\end{figure}

\begin{figure}[!h]
\centering  
\begin{tabular}{ c  c  c  c }
$\chi_{ap} N$ &
$120^o$ &
$90^o$ &
$60^o$  \\
$1$ &
\includegraphics[scale=0.09]{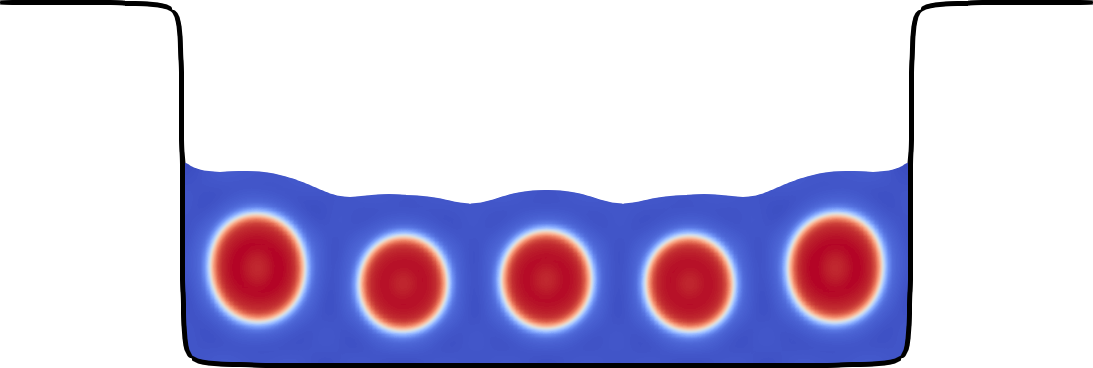} &
\includegraphics[scale=0.09]{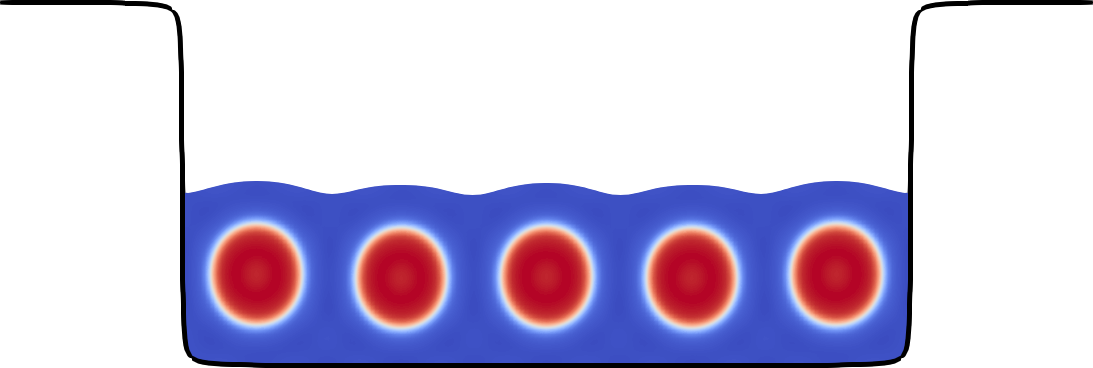} &
\includegraphics[scale=0.09]{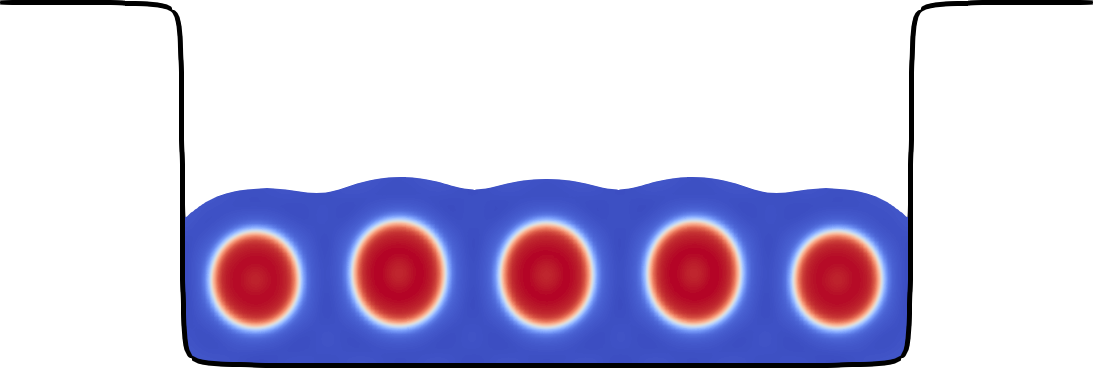} \\
$9$ &
\includegraphics[scale=0.09]{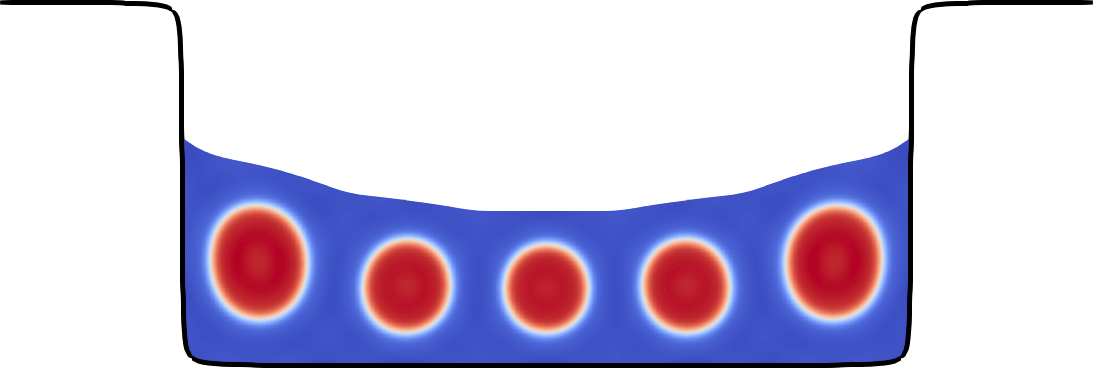} &
\includegraphics[scale=0.09]{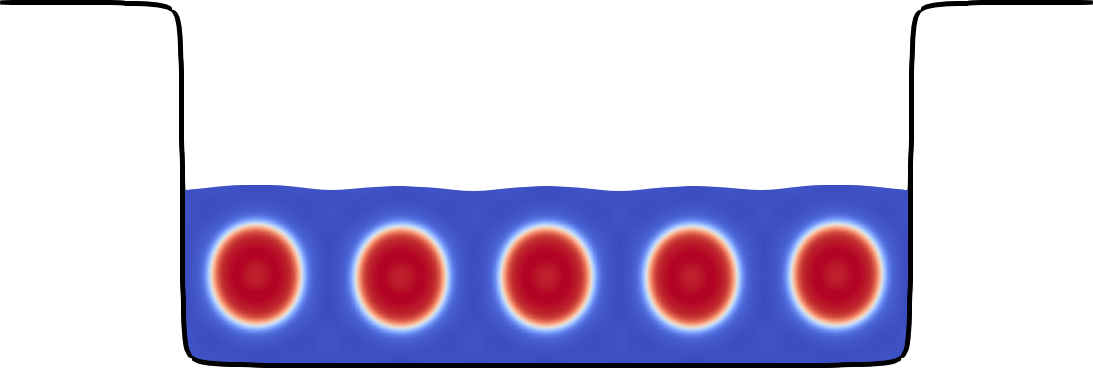} &
\includegraphics[scale=0.09]{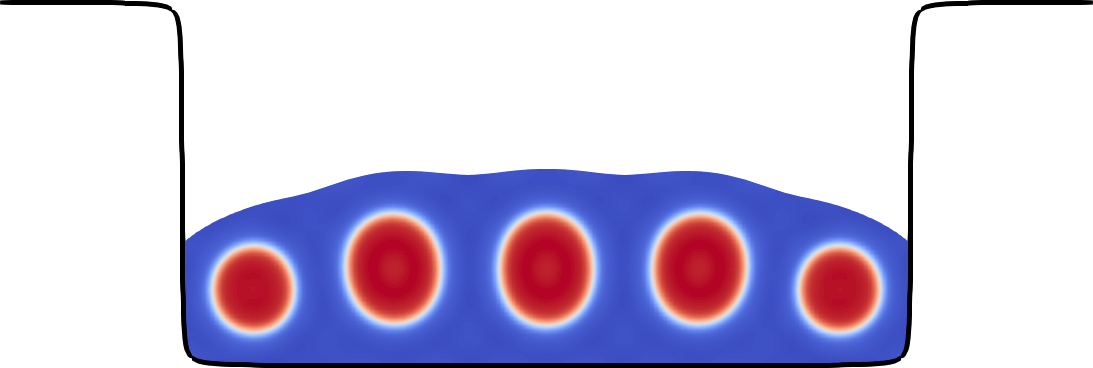} \\
$25$ &
\includegraphics[scale=0.09]{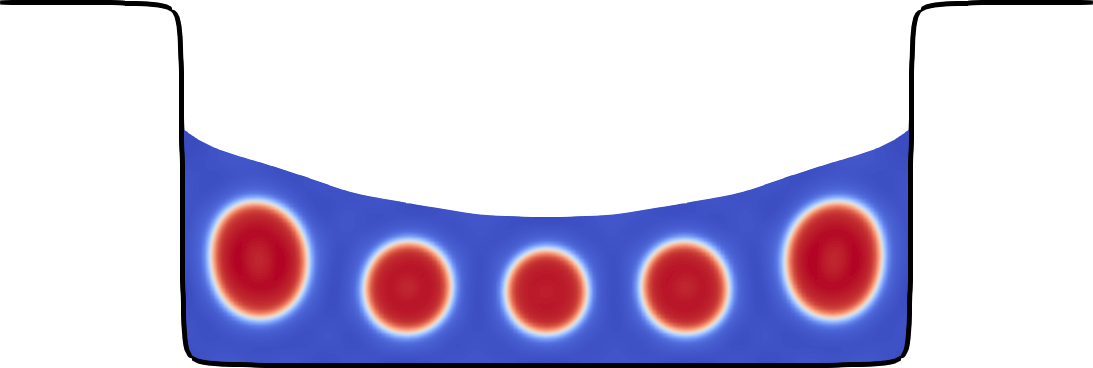} &
\includegraphics[scale=0.09]{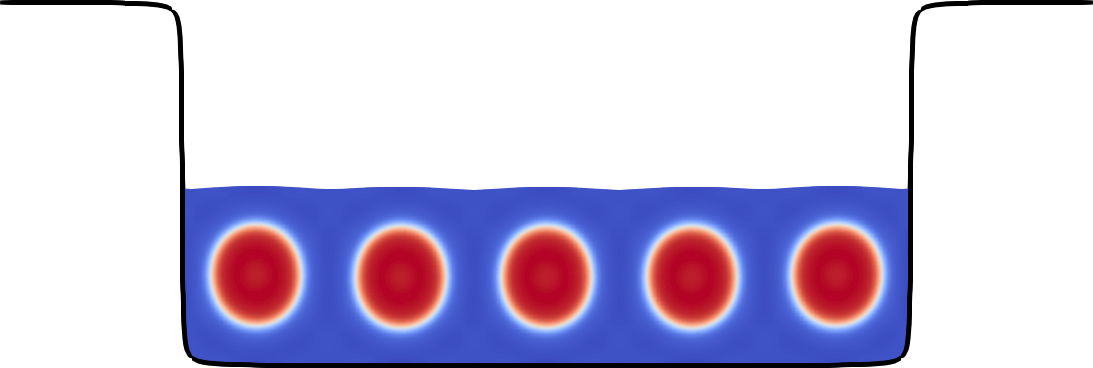} &
\includegraphics[scale=0.09]{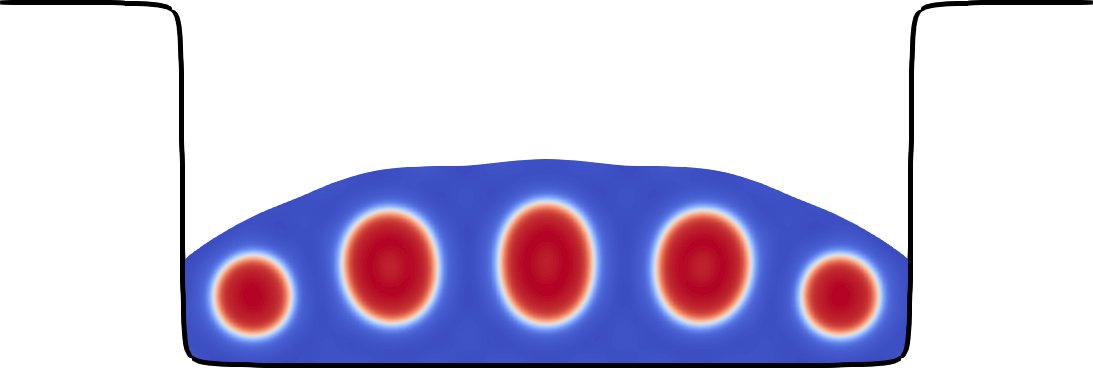} \\
$100$ &
\includegraphics[scale=0.09]{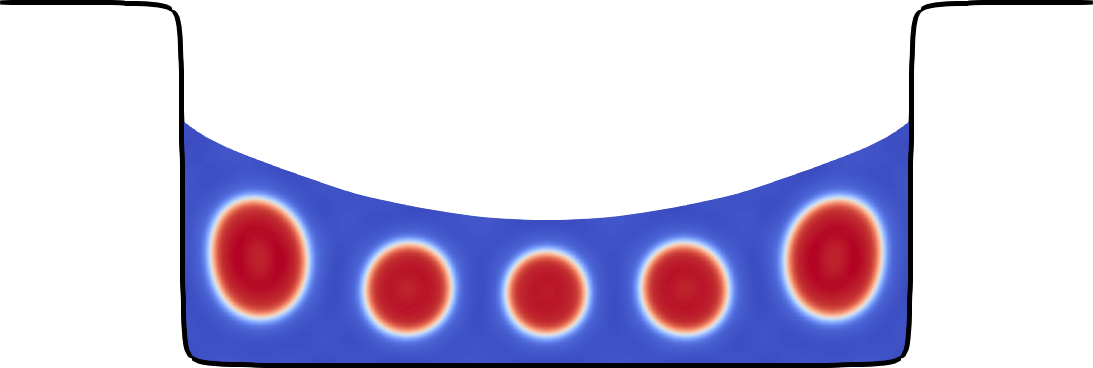} &
\includegraphics[scale=0.09]{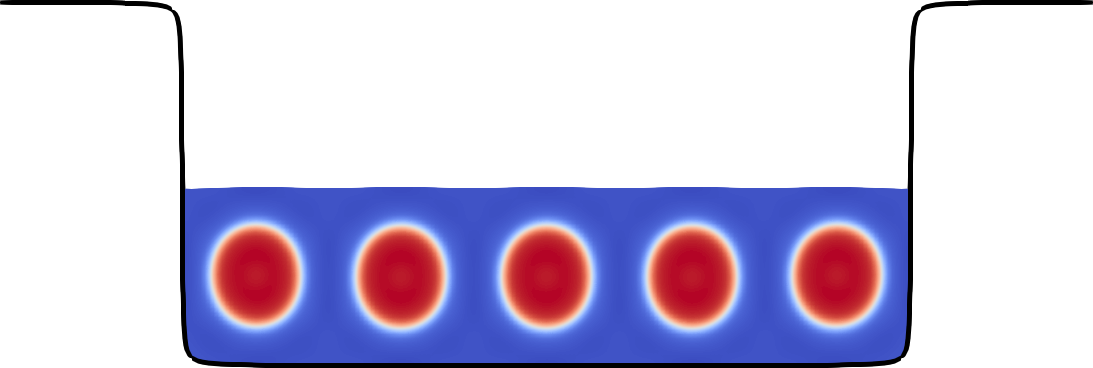} &
\includegraphics[scale=0.09]{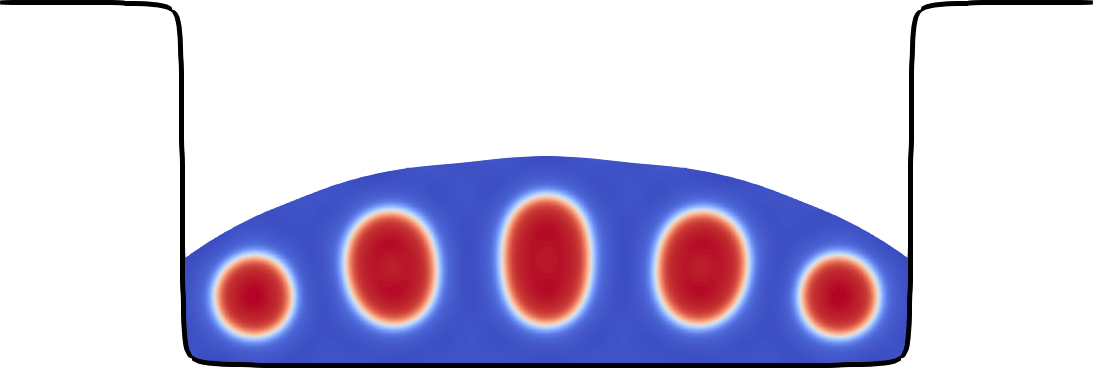} 
\end{tabular}
\caption{Equilibrium shapes of a cylinder-forming diblock copolymer film of height $4R_g$ in a groove of width $20R_g$ for several values of the polymer-air surface tension and contact angle.}
\label{fig:results:free:film:cylindrical}
\end{figure}

Figures \ref{fig:results:free:film:lamellar} and \ref{fig:results:free:film:cylindrical} demonstrate the stable shapes of the polymer films  obtained in cases when the contact angle formed by the polymer material with the groove's wall is $60^o$, $90^o$, or $120^o$ and for values of polymer-air surface tensions corresponding to $\chi_{ap} N=1$, $9$, $25$, and $100$. 

These equilibrium shapes can be qualitatively explained in terms of the balance between the internal polymer stiffness and the polymer-air surface tension. In cases when the polymer-air surface tension $\chi_{ap} N = 1$ is much lower than the interaction strength between the polymer components $\XN{AB} = 30$, the overall film shape remains almost flat with the curvature caused by the meniscus propagating only a very small distance away from the walls. Note that the low surface tension also allows a small-scale corrugation of polymer-air interface due to the internal film morphology. In the other limit when the polymer-air surface tension $\XN{ap} = 100$ is much higher than the polymer interaction strength $\XN{AB} = 30$, the overall film shapes are very close to the spherical ones expected for simple liquids with only a slight small-scale corrugations. Note that in this case, the polymer domains next to groove's walls also undergo significant deformations. Finally, in intermediate cases $\XN{ap} = 9$ and  $\XN{ap} = 25$, which are comparable to $\XN{AB} = 30$, the curvature due to the meniscus formation propagates a significant distance away from the walls, while the center region of the film is still flat.

\begin{figure}[!h]
\centering  
\begin{tabular}{ c  c  c  c }
$\chi_{ap} N$ &
$120^o$ &
$90^o$ &
$60^o$  \\
$1$ &
\includegraphics[scale=0.09]{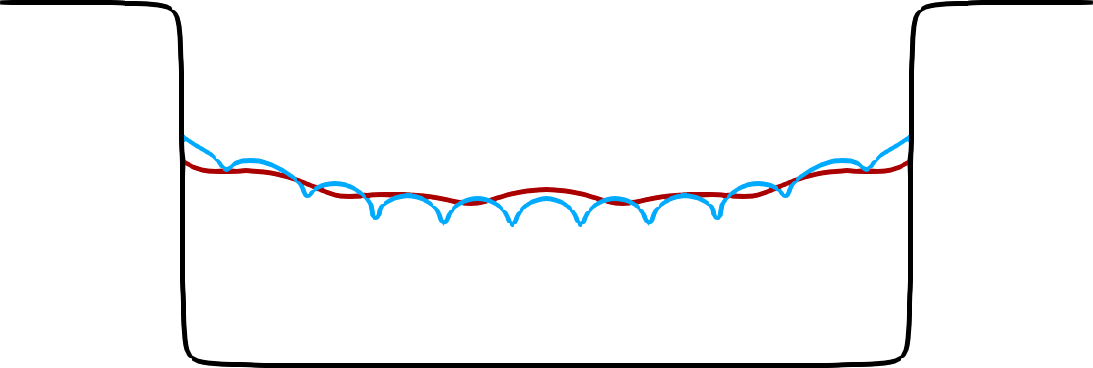} &
\includegraphics[scale=0.09]{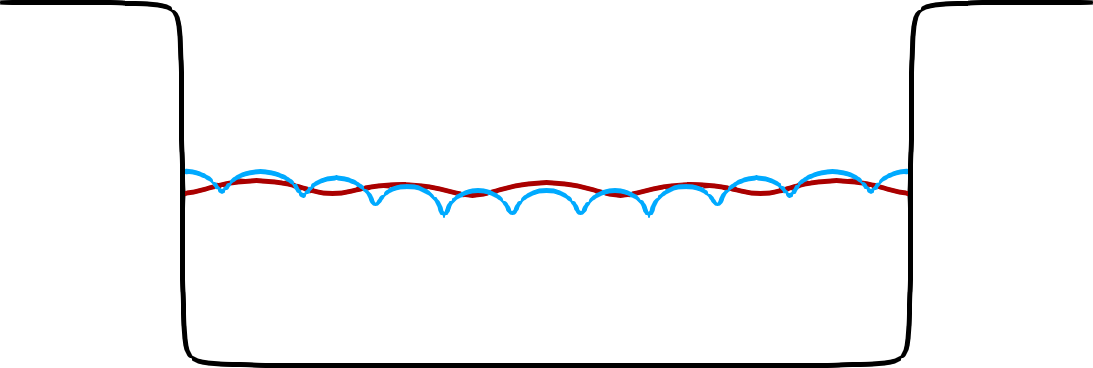} &
\includegraphics[scale=0.09]{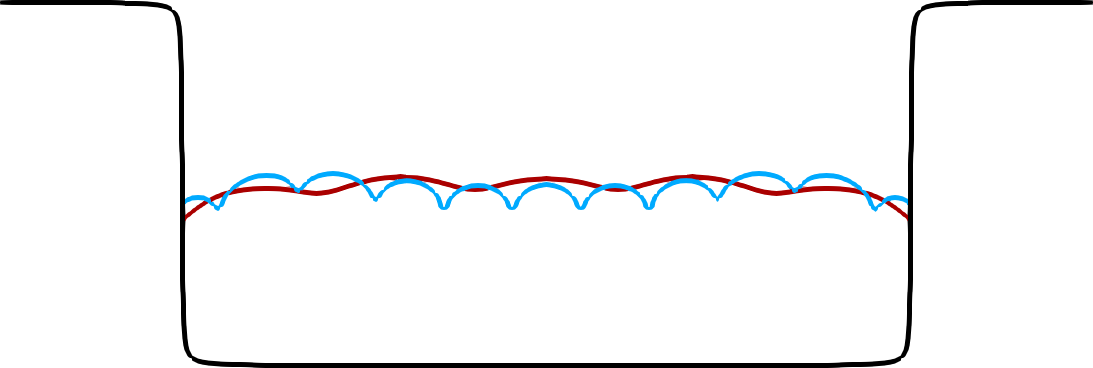} \\
$9$ &
\includegraphics[scale=0.09]{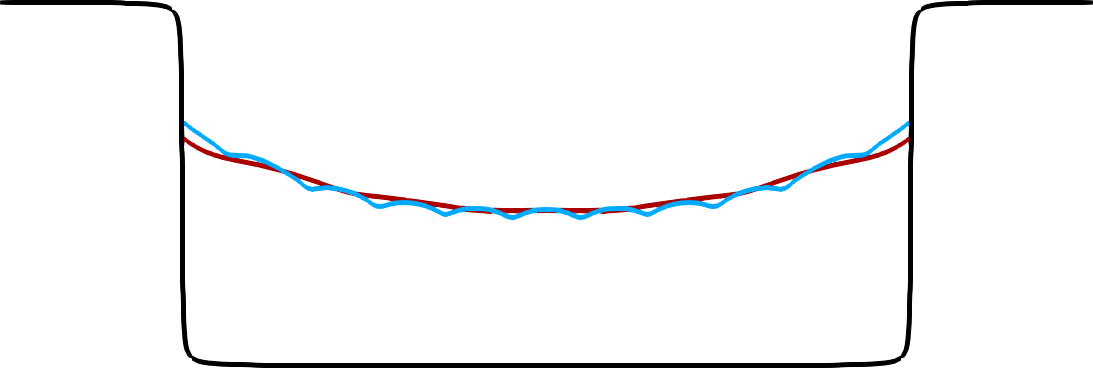} &
\includegraphics[scale=0.09]{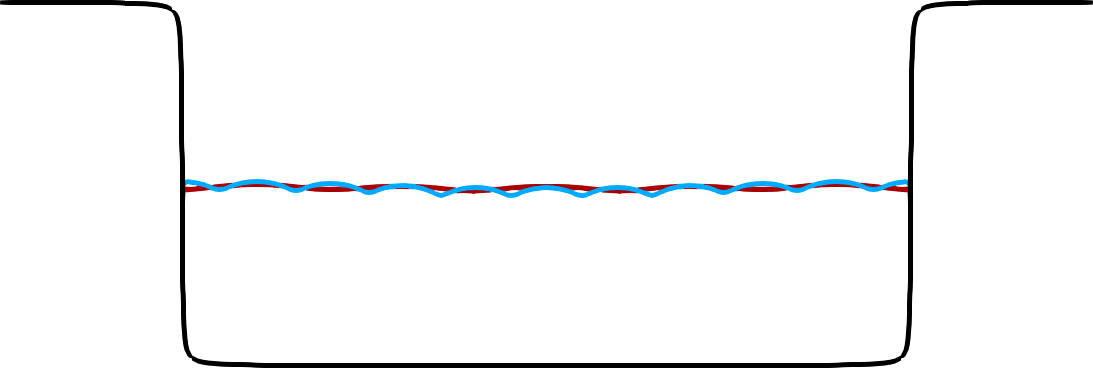} &
\includegraphics[scale=0.09]{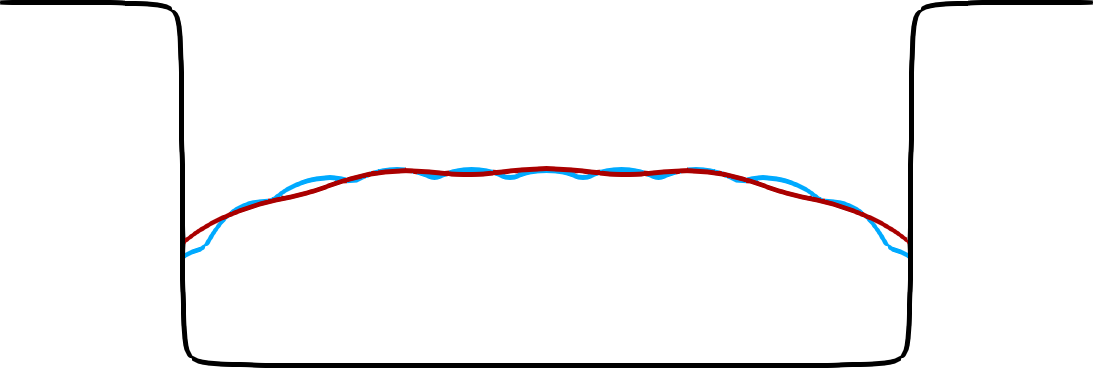} \\
$25$ &
\includegraphics[scale=0.09]{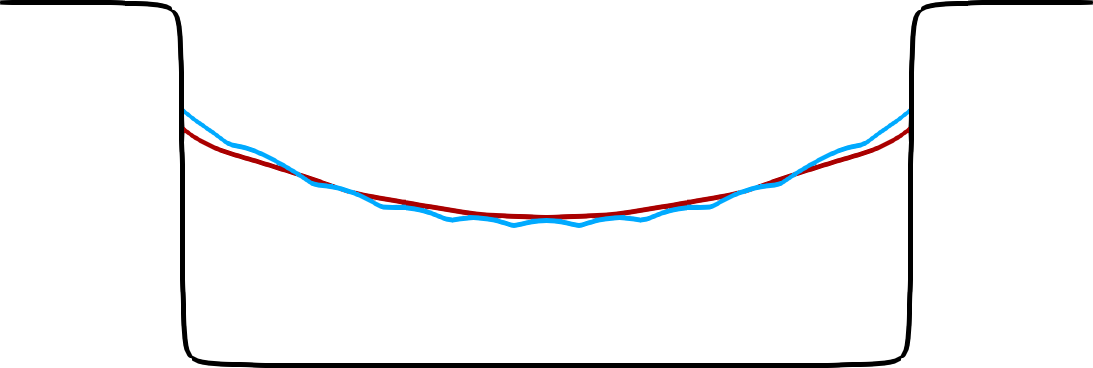} &
\includegraphics[scale=0.09]{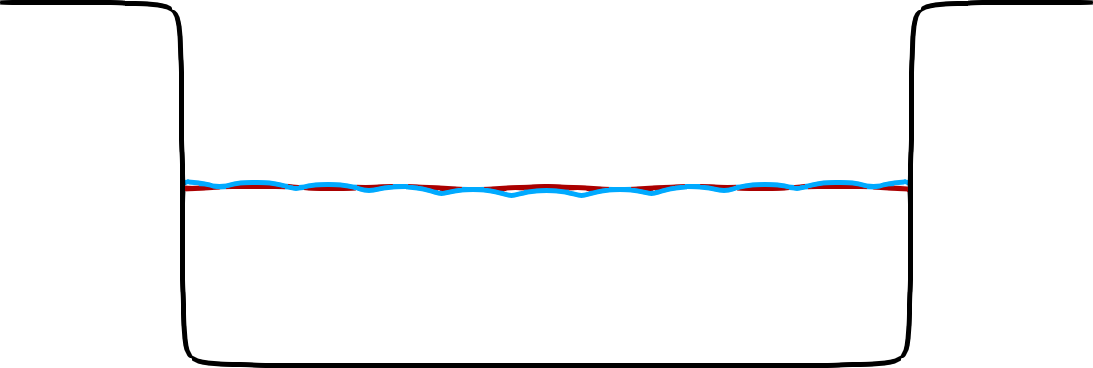} &
\includegraphics[scale=0.09]{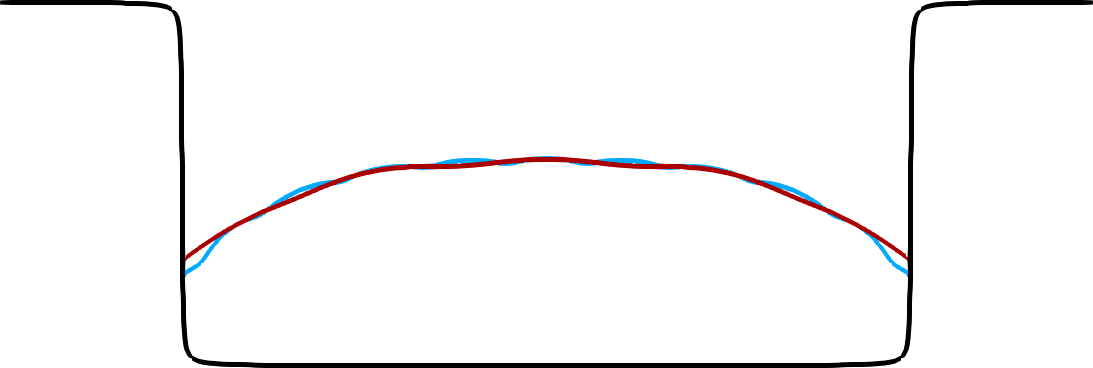} \\
$100$ &
\includegraphics[scale=0.09]{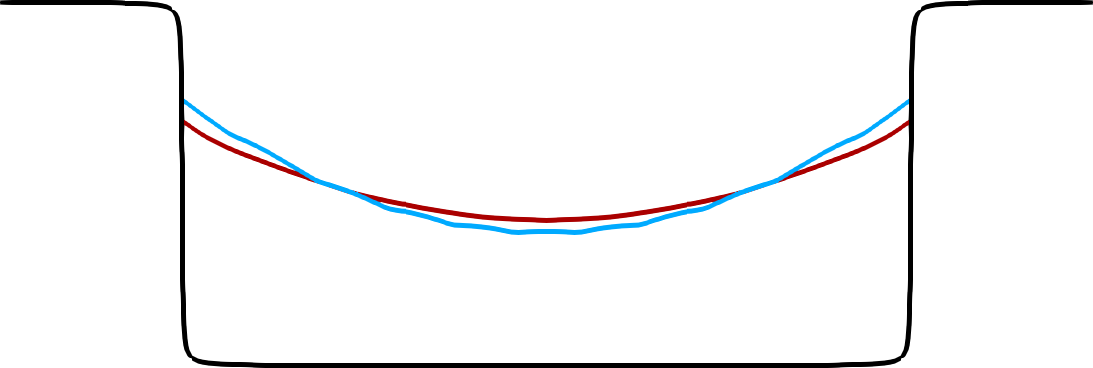} &
\includegraphics[scale=0.09]{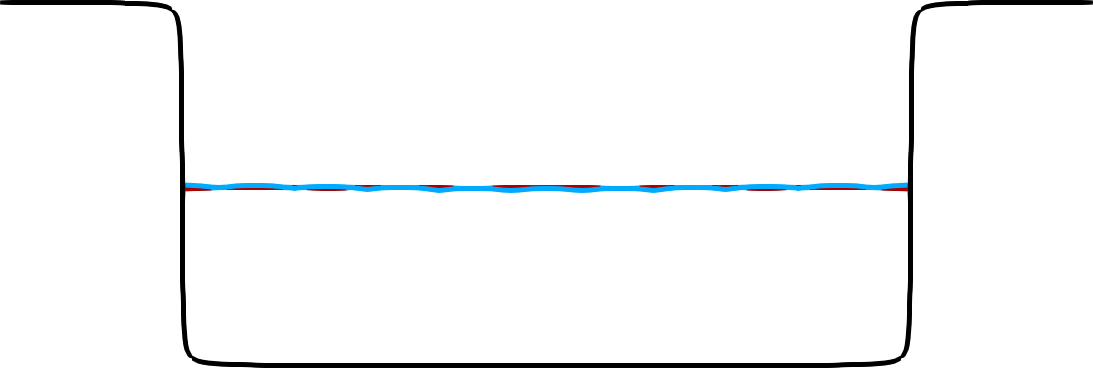} &
\includegraphics[scale=0.09]{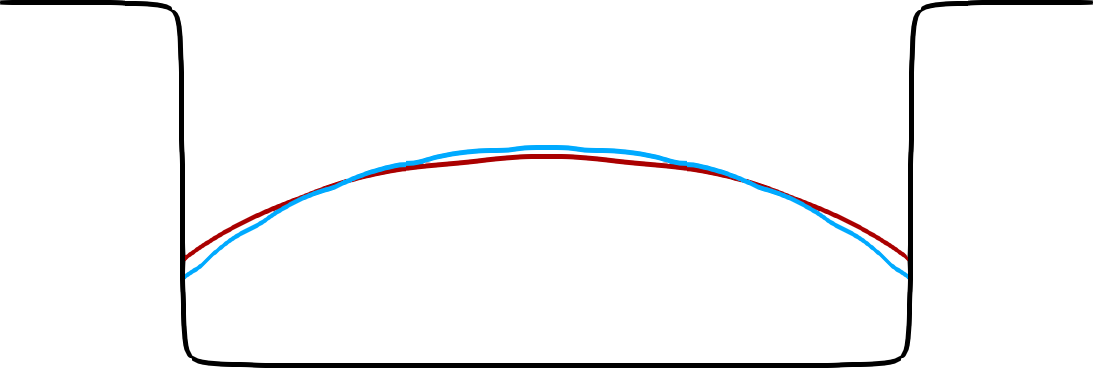} 
\end{tabular}
\caption{Comparison of the equilibrium shapes of lamellae-forming (blue line) and cylinder-forming (red line) diblock copolymer films of height $4R_g$ in a groove of width $20R_g$ for several values of the polymer-air surface tension and contact angle.}
\label{fig:results:free:film:compare}
\end{figure}

It is also interesting to see that the degree of the film deformation is slightly higher in the case of lamellae-forming diblock copolymer as demonstrated in figure \ref{fig:results:free:film:compare}. This indicates that the stiffness of the cylindrical morphology is slightly higher and the deformation of a cylindrical arrangement of polymer chains causes a higher energetic penalty compared to the deformation of the lamellar arrangements. 

To summarize, the result presented in this example demonstrate that the overall equilibrium shapes of copolymer films in guiding grooves are defined by the balance between the internal stiffness of polymer morphologies and the polymer-air surface tension. These results indicate that in graphoepitaxy applications of block copolymer, the surface energies of guiding templates must be carefully selected to prevent significant deformations of the self-assembling films.

\subsection{Placement of Janus rods in lamellae-forming BCP}
We now apply the proposed computational approach for investigating the co-assembly of lamellar diblock copolymer ($\XN{AB}=30$, $f=0.5$) with rod-like Janus particles, that is, particles that are attracted to the $A$ polymer component on the one end and to the the $B$ polymer component on the other end. On the one hand, the elongated shape of such particles promotes the orientation parallel to the $AB$ interface to reduce the system's energy, while the selective attraction on opposite sides of such particles favors the perpendicular orientation. As the result, one can expect the resulting placement of the nanoparticle to be a function of its length and the magnitude of its surface energy. Another factor expected to affect the particle placement is the particles density because the crowding may result in insufficient space for  nanoparticles to assume their equilibrium orientation.

In numerical simulations, we consider particles of capsule-like shapes with radius of curvature $r_0 = 0.1R_g$ and nonuniform surface energies given by:
\begin{myalign*}
\gamma_{Ap} &= \frac{1}{\sigma} \hf \sqrt{\XN{ap}} \left(1+ \cos \of{\theta} \right),\\
\gamma_{Bp} &= \frac{1}{\sigma} \hf \sqrt{\XN{ap}} \left(1- \cos \of{\theta} \right),
\end{myalign*} 
where $\theta$ is the angle of rotation from the particle's axis and $\XN{ap}$ characterizes the strength of preferential attraction between the nanoparticle and polymer species and, for convenience, is also referred to as \textit{the polarization strength}.

\begin{figure}[!h]
\centering
\includegraphics[width=.24\textwidth]{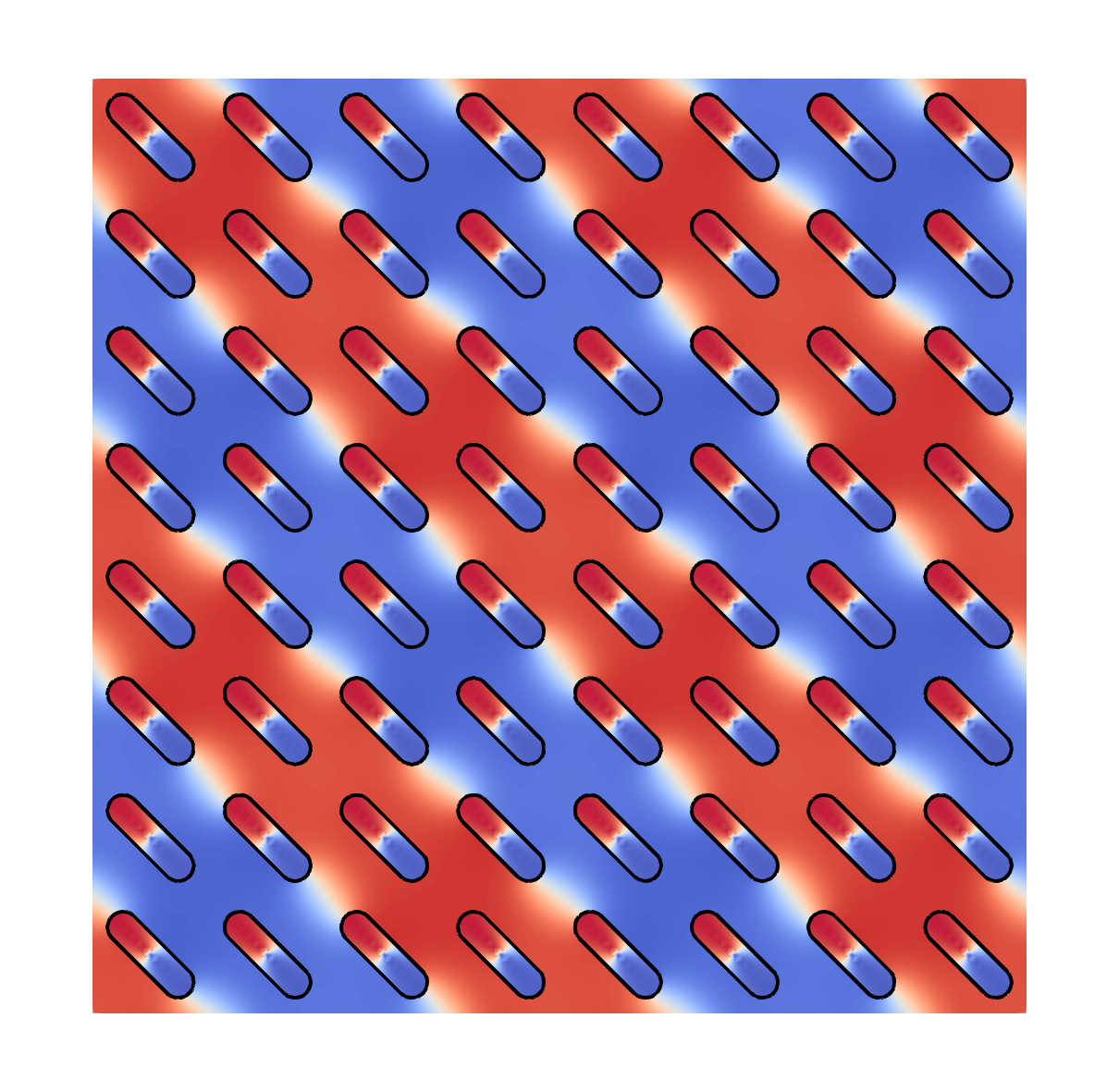}
\includegraphics[width=.24\textwidth]{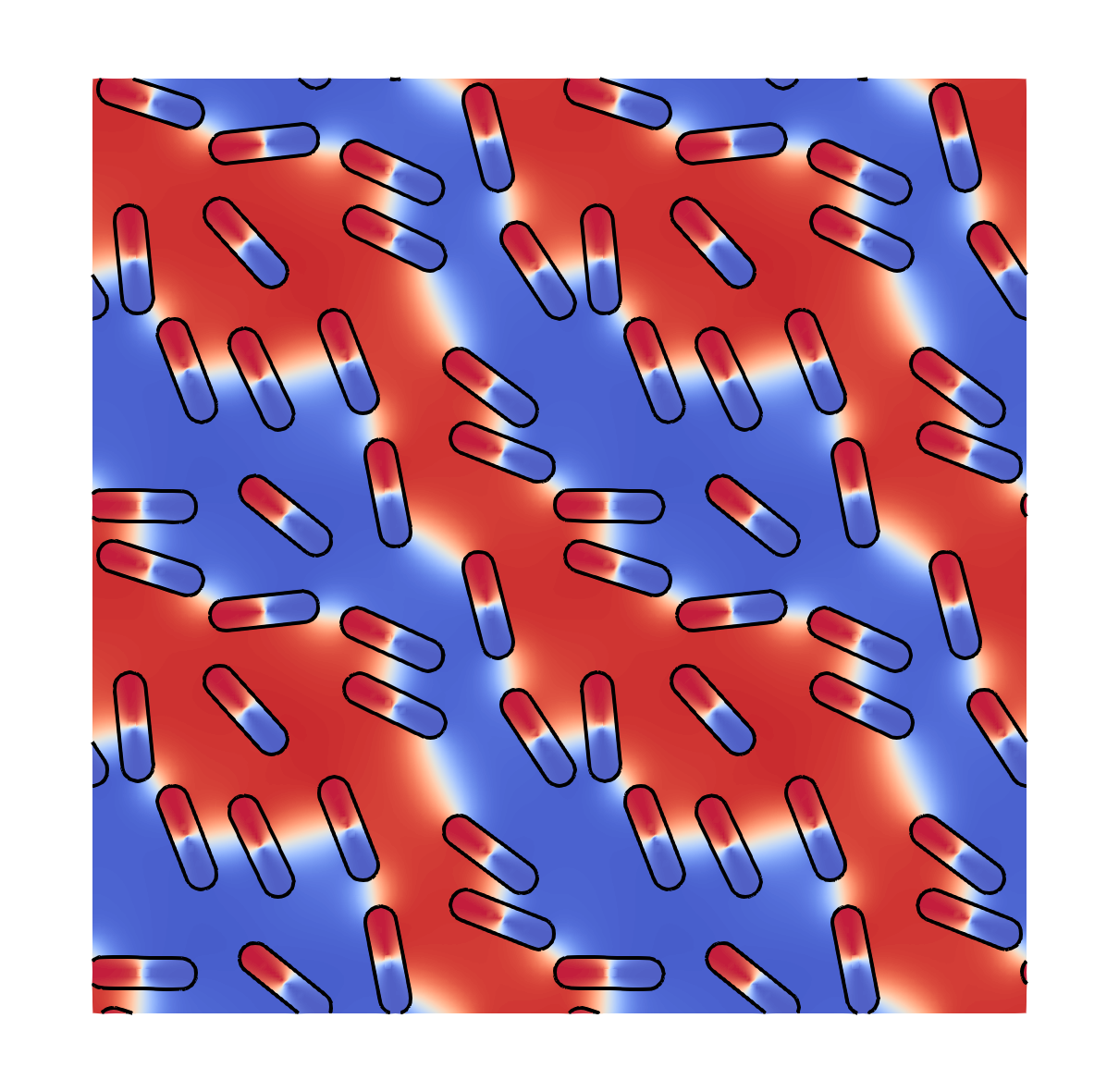}
\includegraphics[width=.24\textwidth]{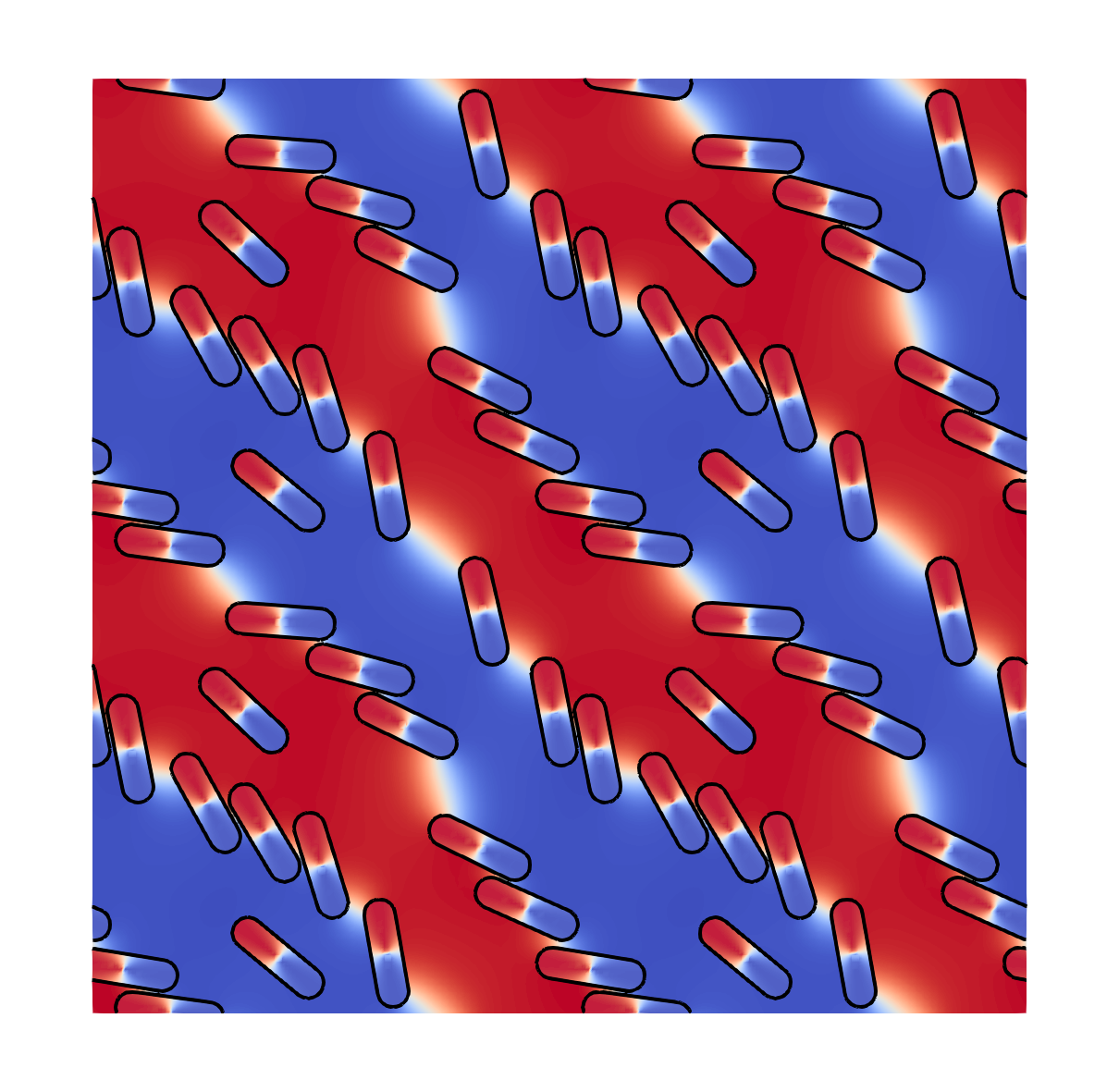}
\includegraphics[width=.24\textwidth]{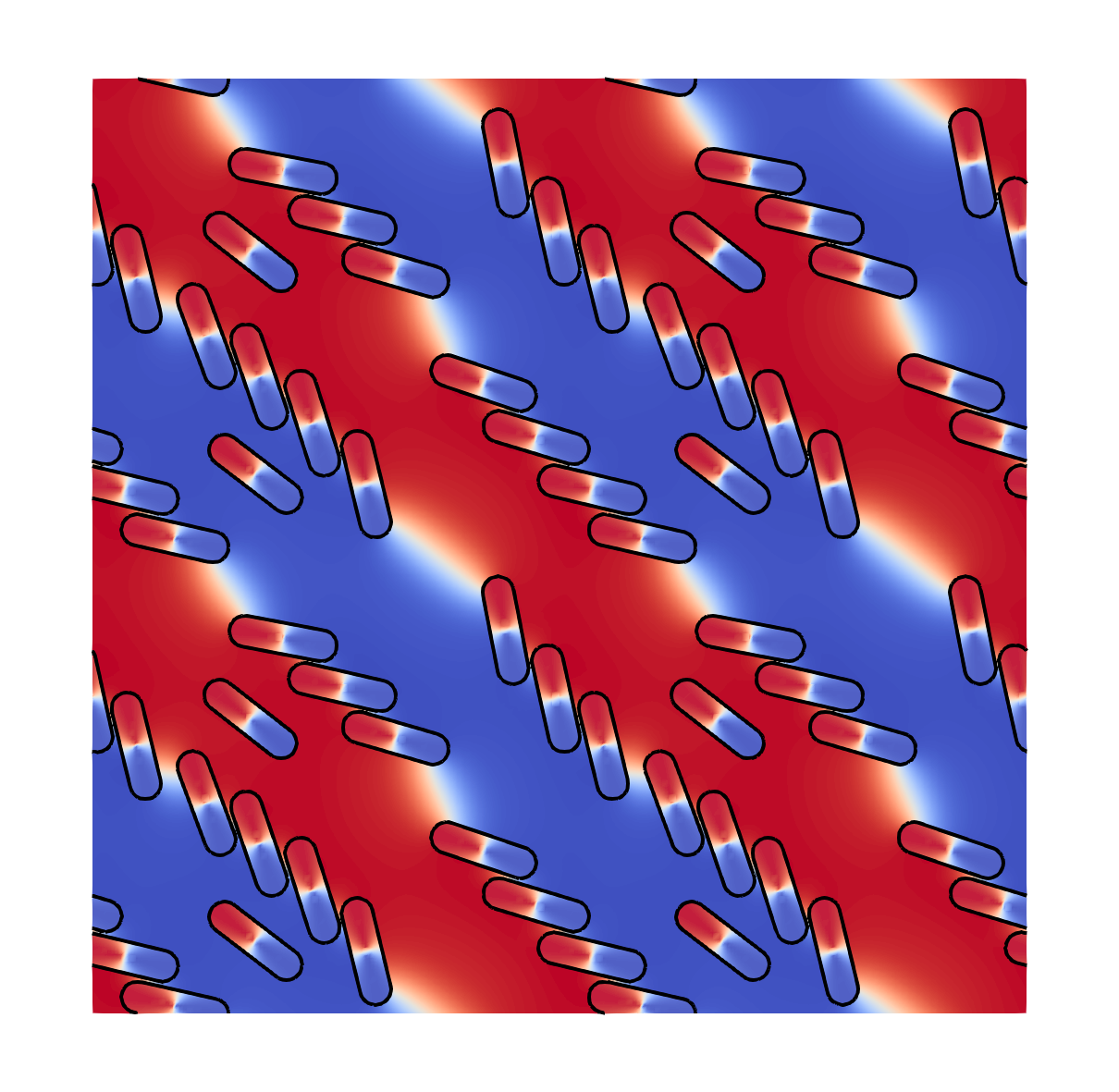}
\caption{Representative simulation results for co-assembly of a lamellar diblock copolymer ($\XN{AB} = 30$, $f=0.5$) and Janus nanorods of length $0.4 R_g$ (periodically tiled for visualization purposes).}
\label{fig:results:nano:many}
\end{figure}

\begin{figure}[!h]
\centering  
\setlength{\tabcolsep}{0.1em}
\renewcommand{\arraystretch}{0.5}
\begin{tabular}{ c  c  c  c  c c c }
$\chi_{p} N$ &
$0.2 R_g$ &
$0.3 R_g$ &
$0.4 R_g$ &
$0.5 R_g$ &
$0.6 R_g$ &
$0.7 R_g$ \\
$1$ 
&
\includegraphics[scale=0.05]{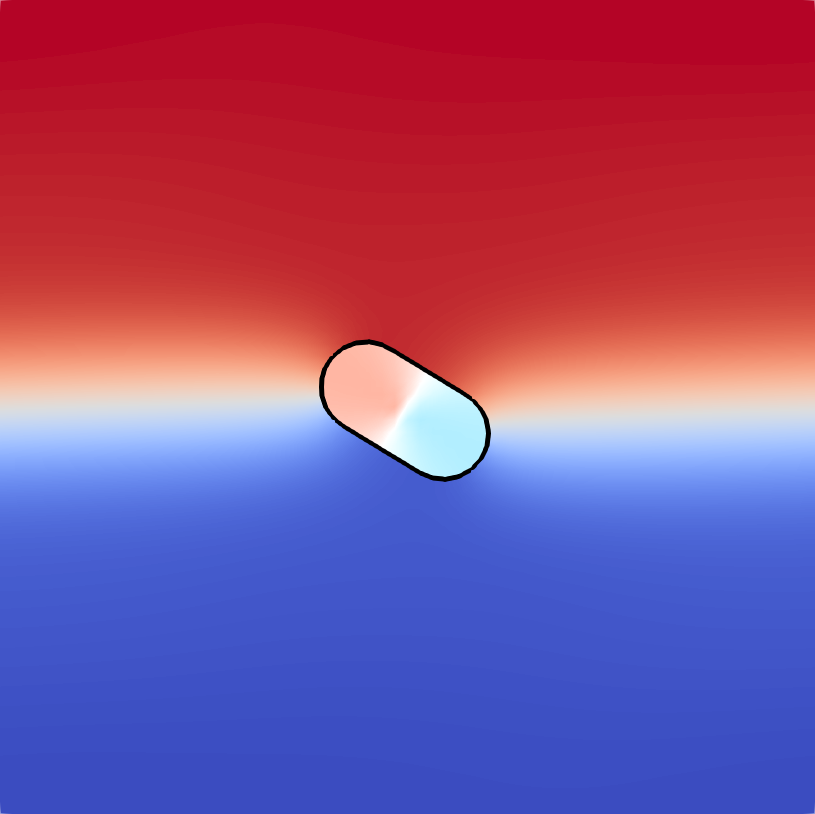} &
\includegraphics[scale=0.05]{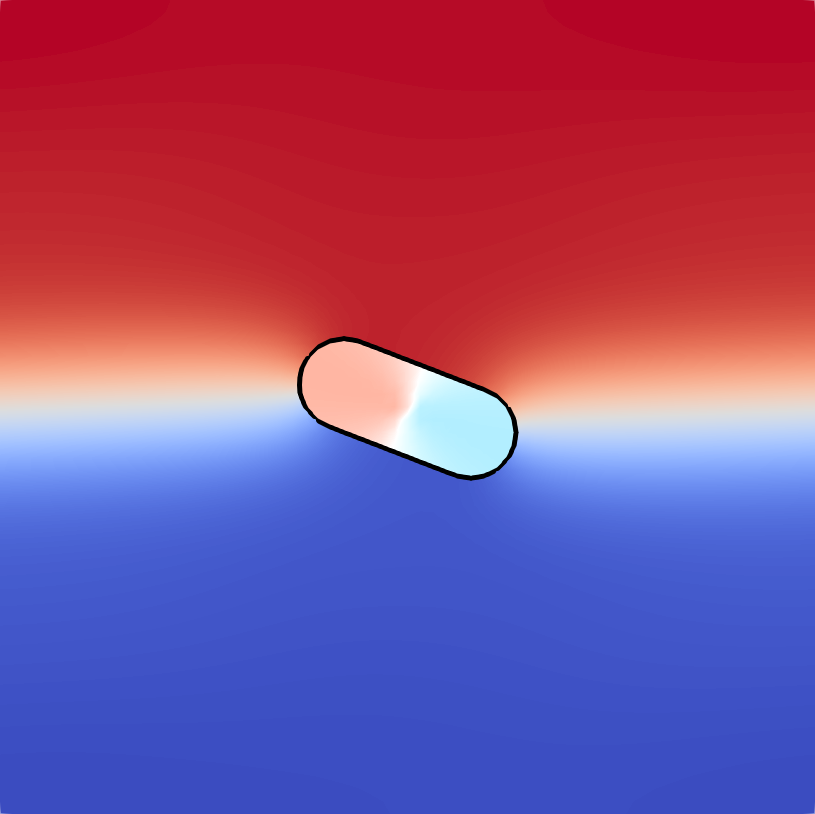} &
\includegraphics[scale=0.05]{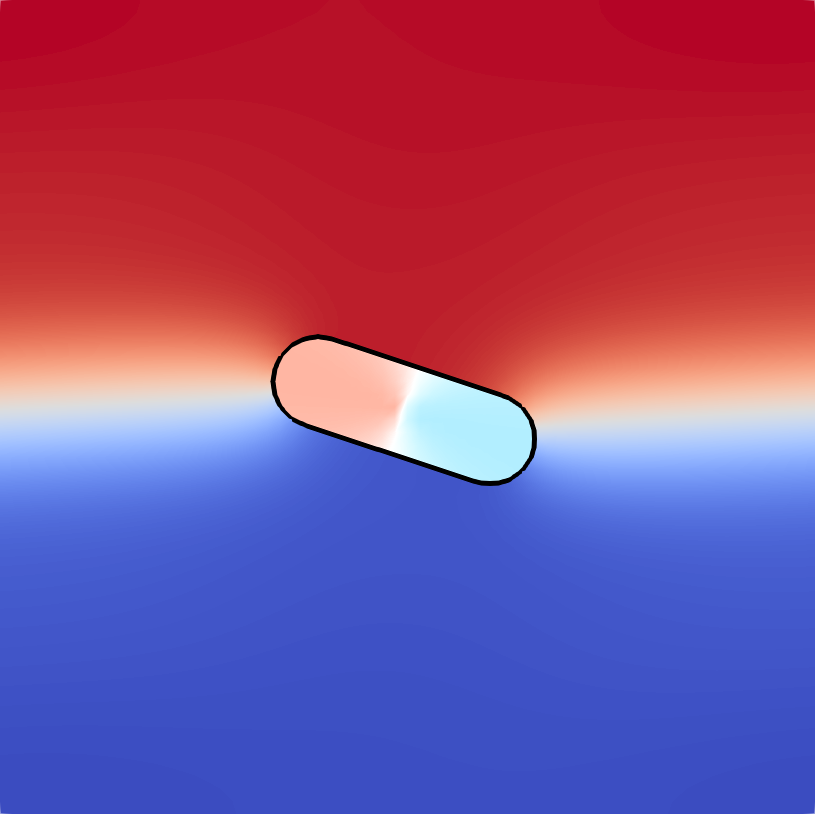} &
\includegraphics[scale=0.05]{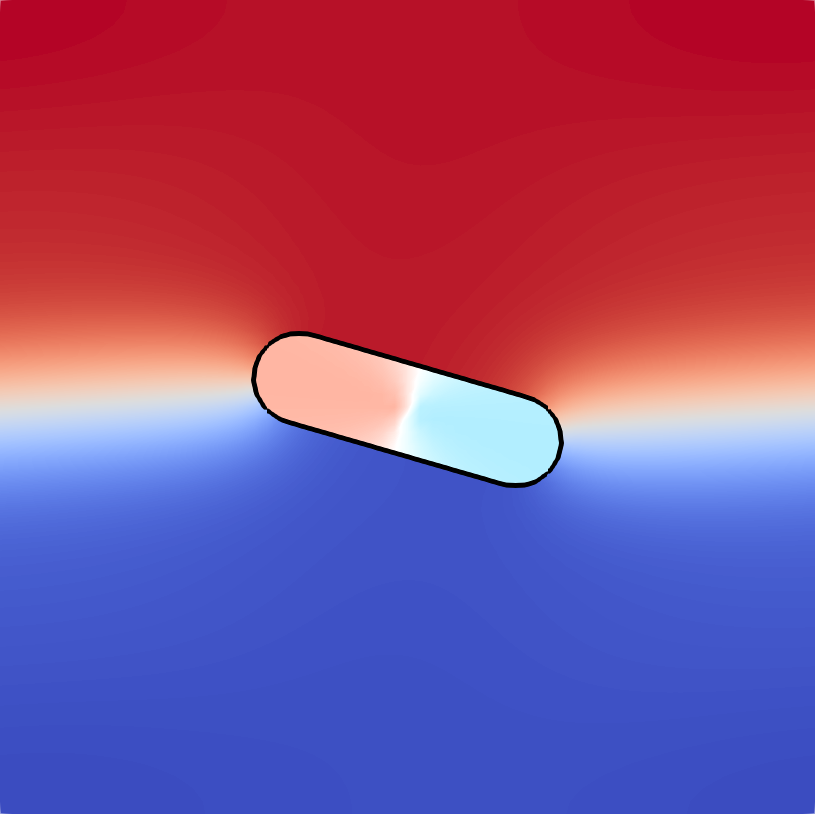} &
\includegraphics[scale=0.05]{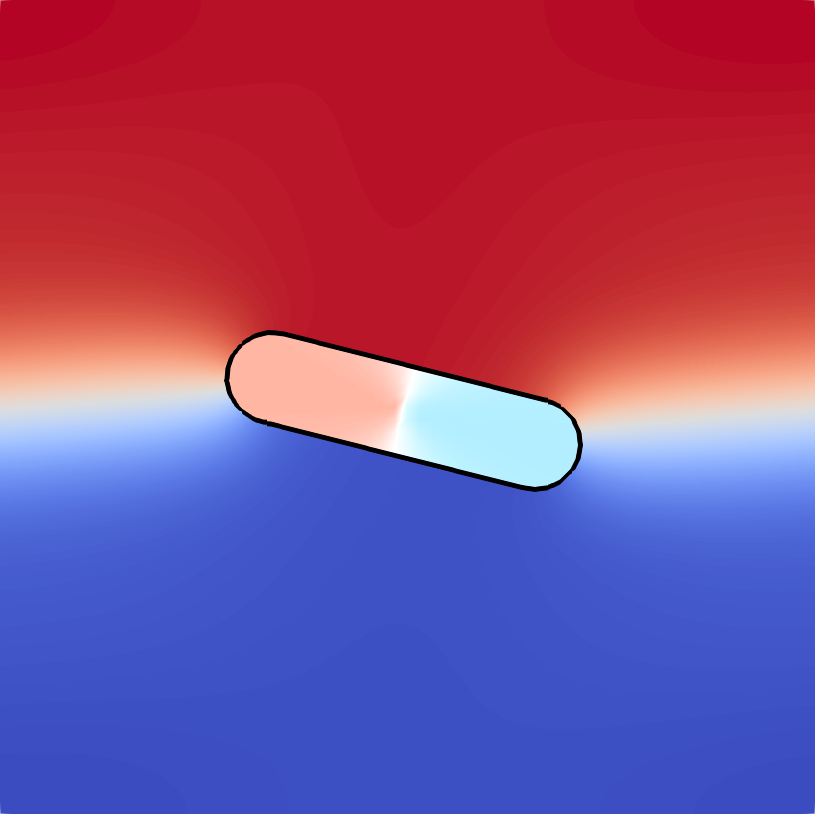} &
\includegraphics[scale=0.05]{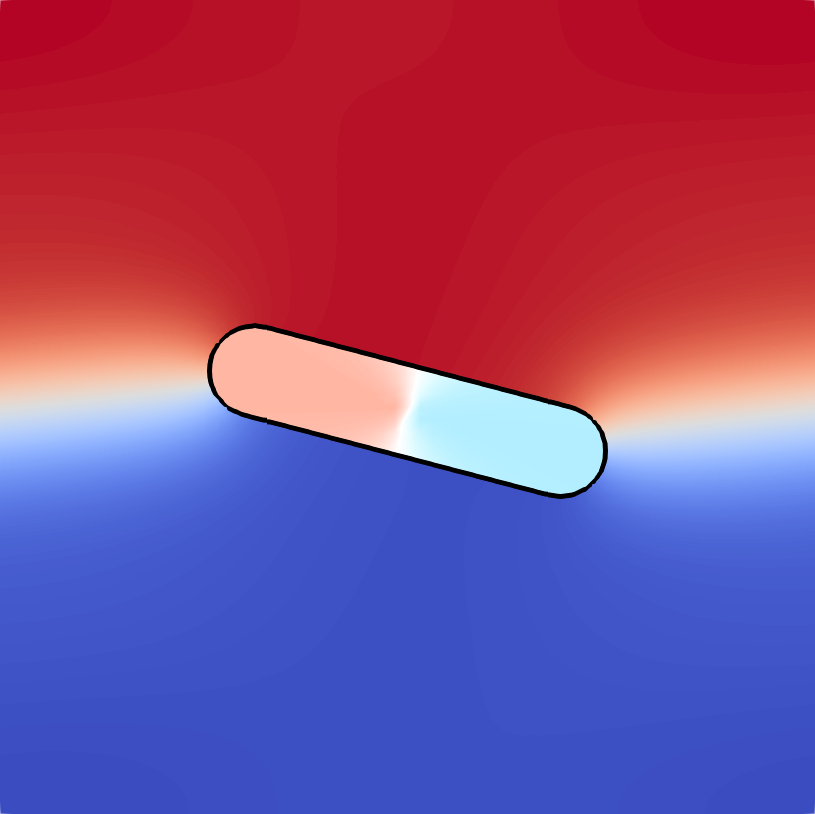} \\
$2$ &
\includegraphics[scale=0.05]{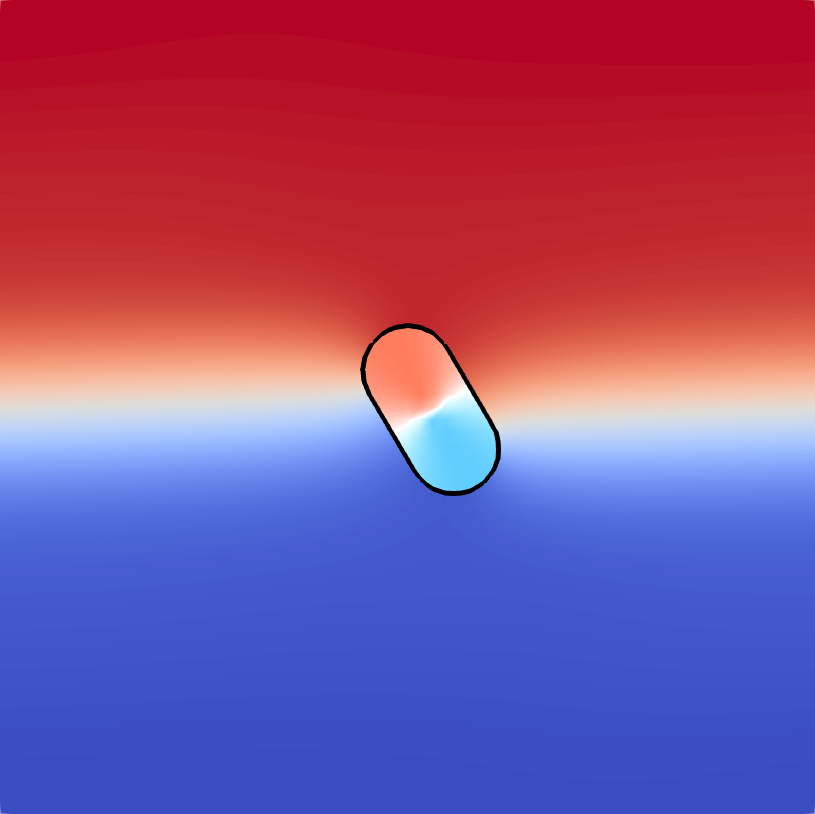} &
\includegraphics[scale=0.05]{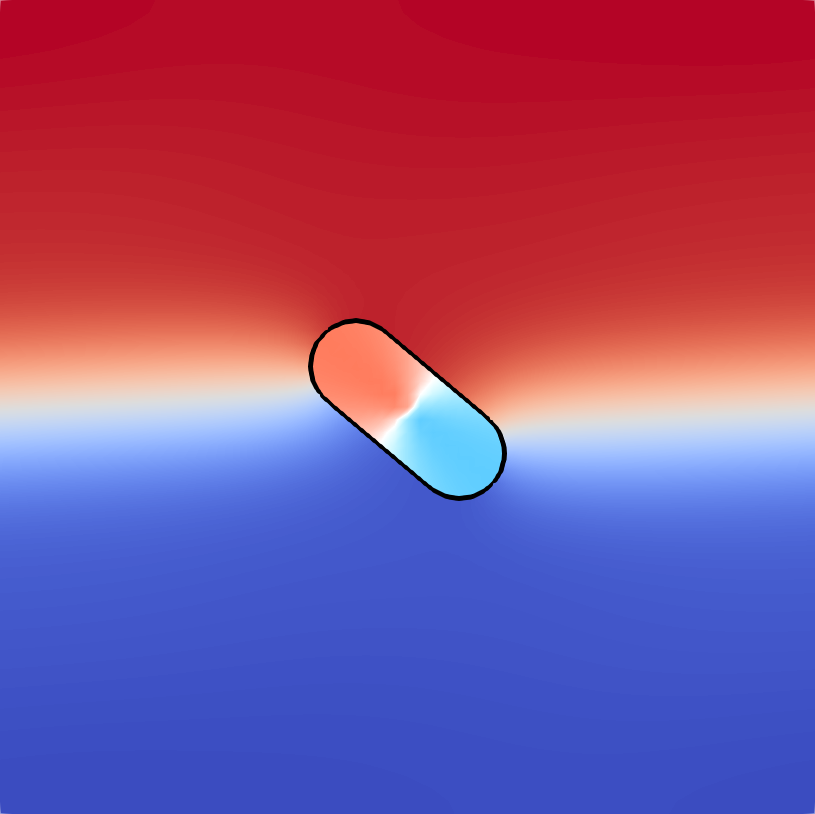} &
\includegraphics[scale=0.05]{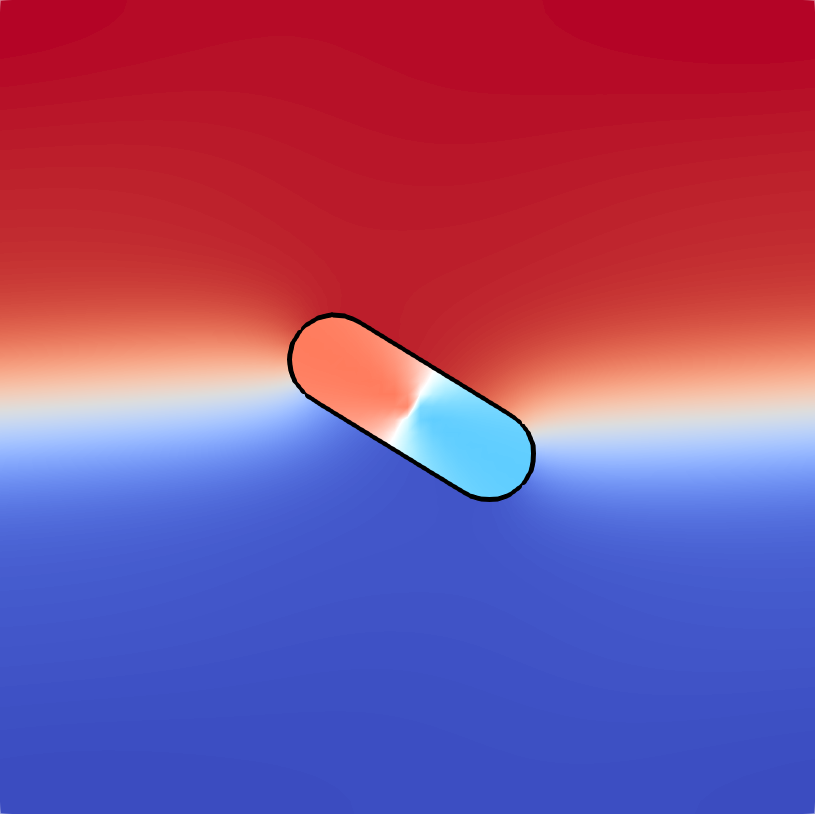} &
\includegraphics[scale=0.05]{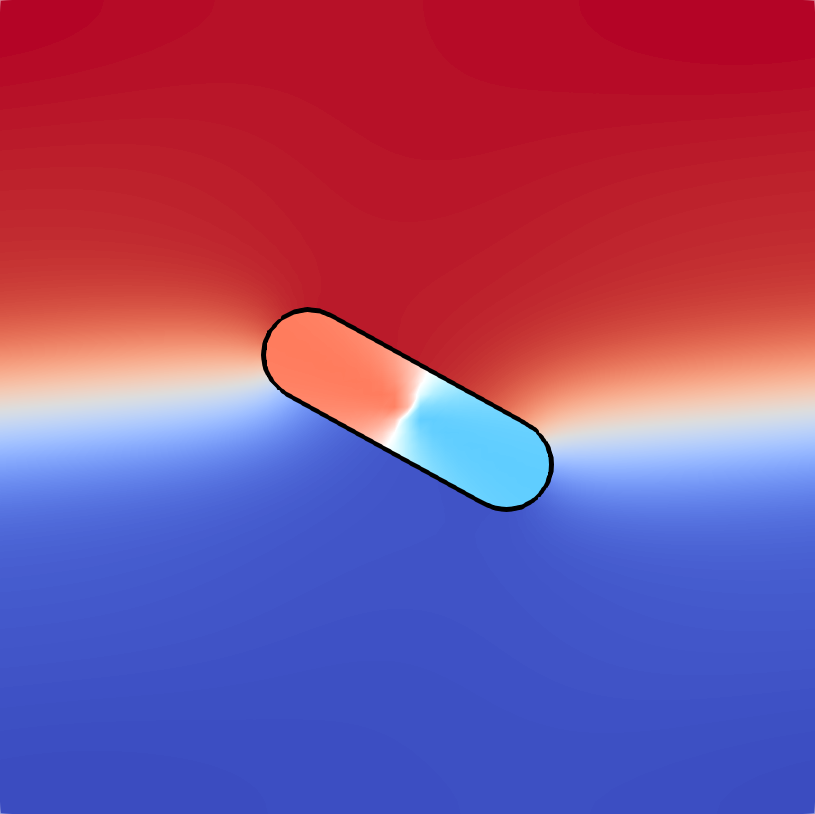} &
\includegraphics[scale=0.05]{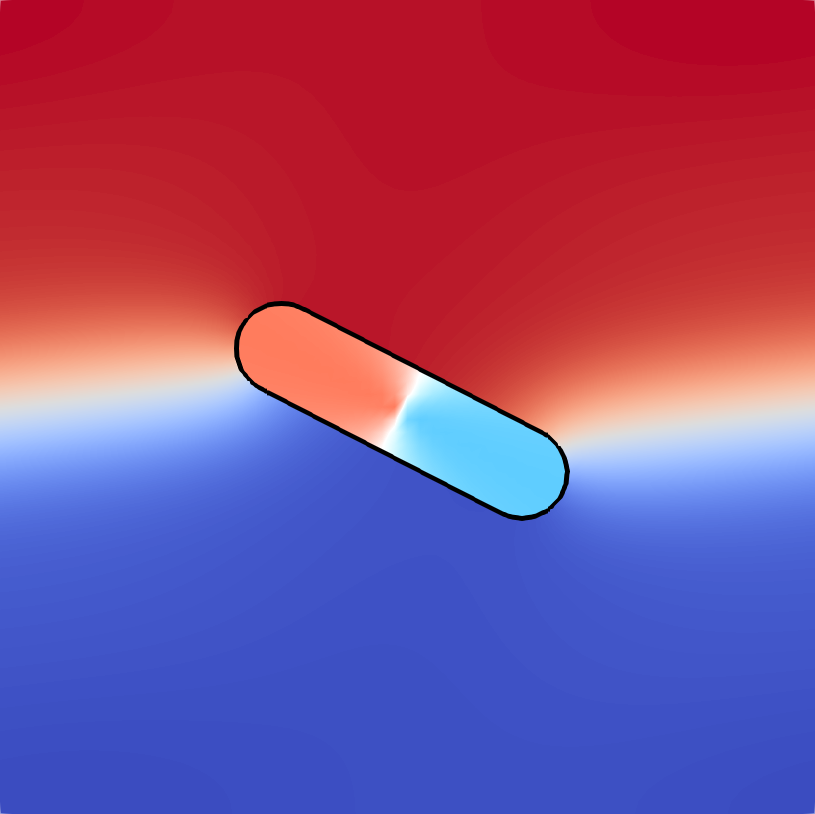} &
\includegraphics[scale=0.05]{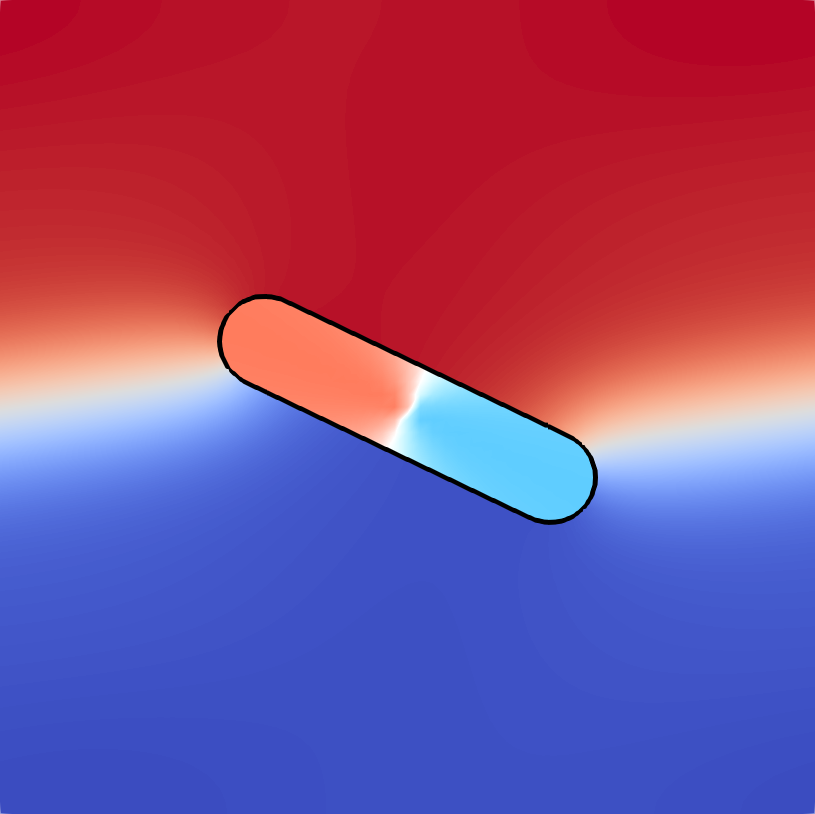} \\
$3$ &
\includegraphics[scale=0.05]{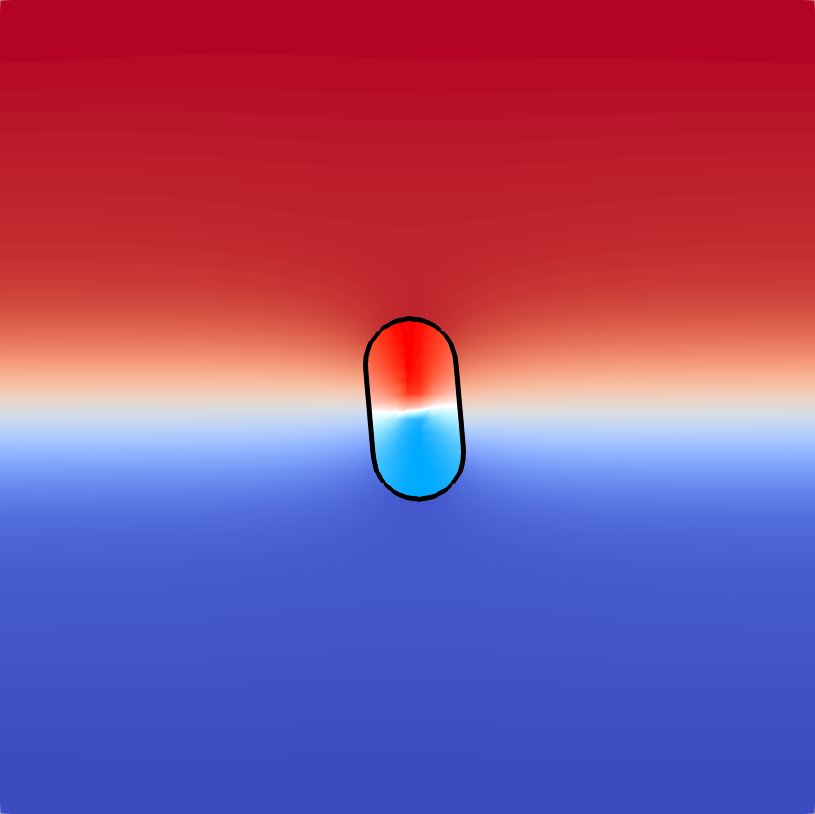} &
\includegraphics[scale=0.05]{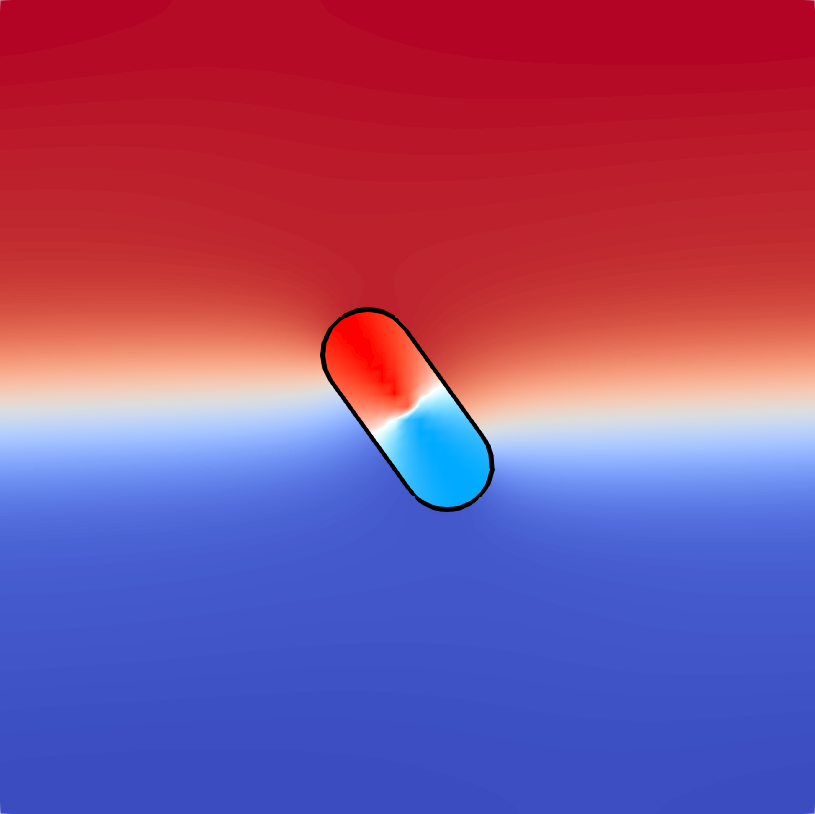} &
\includegraphics[scale=0.05]{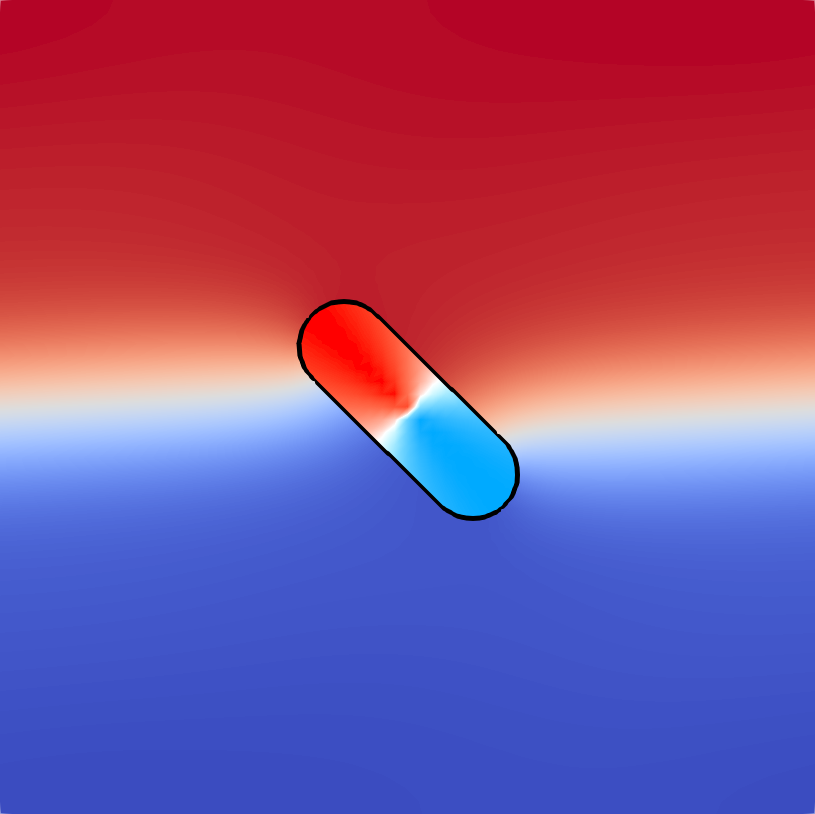} &
\includegraphics[scale=0.05]{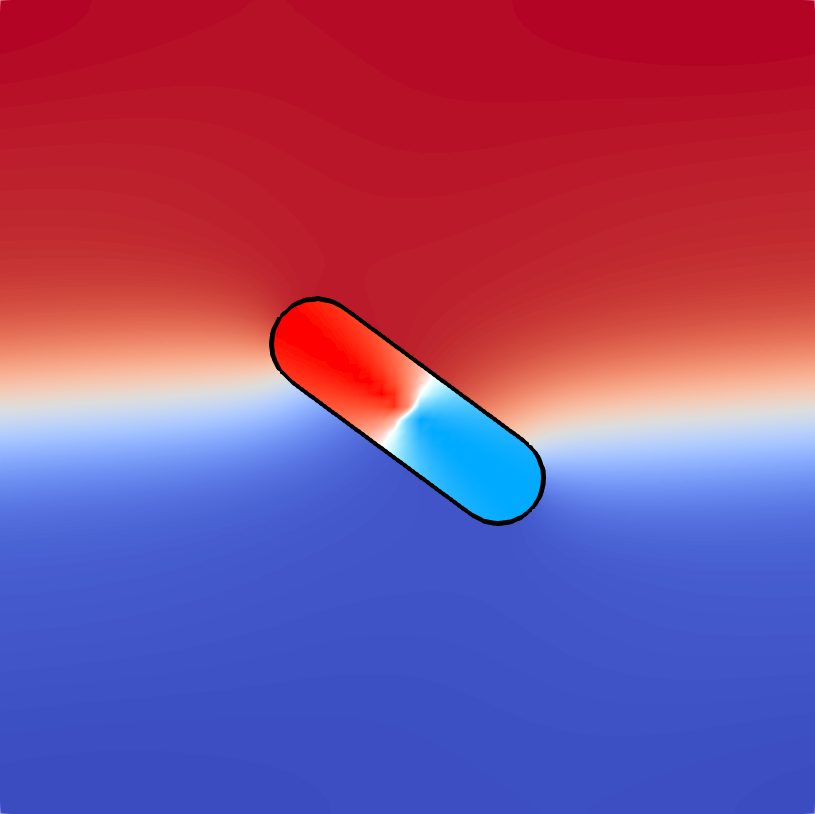} &
\includegraphics[scale=0.05]{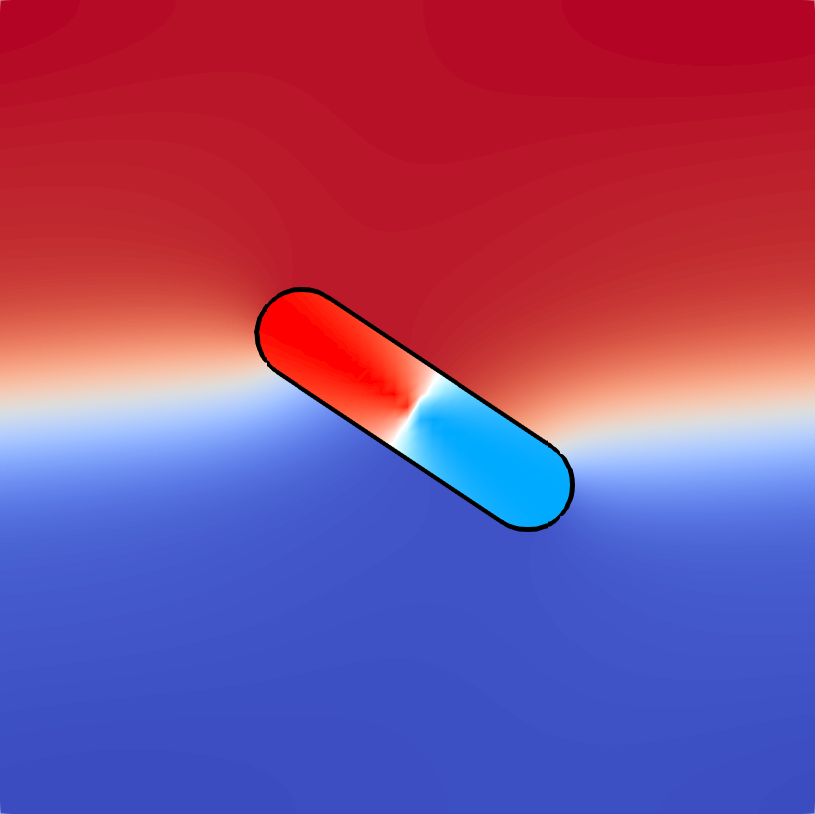} &
\includegraphics[scale=0.05]{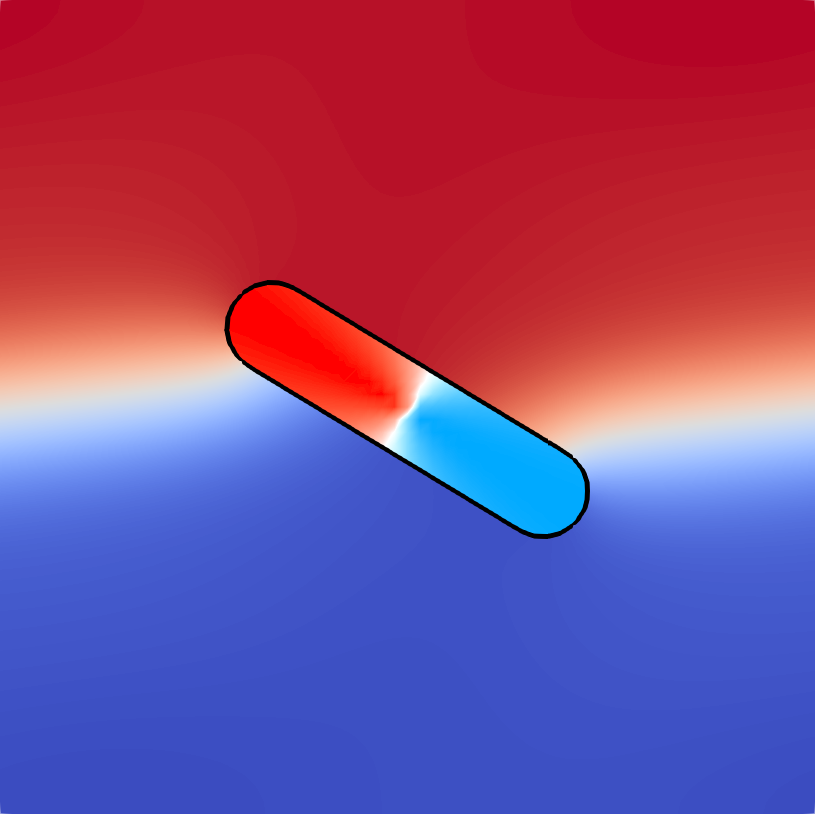} \\
$4$ &
\includegraphics[scale=0.05]{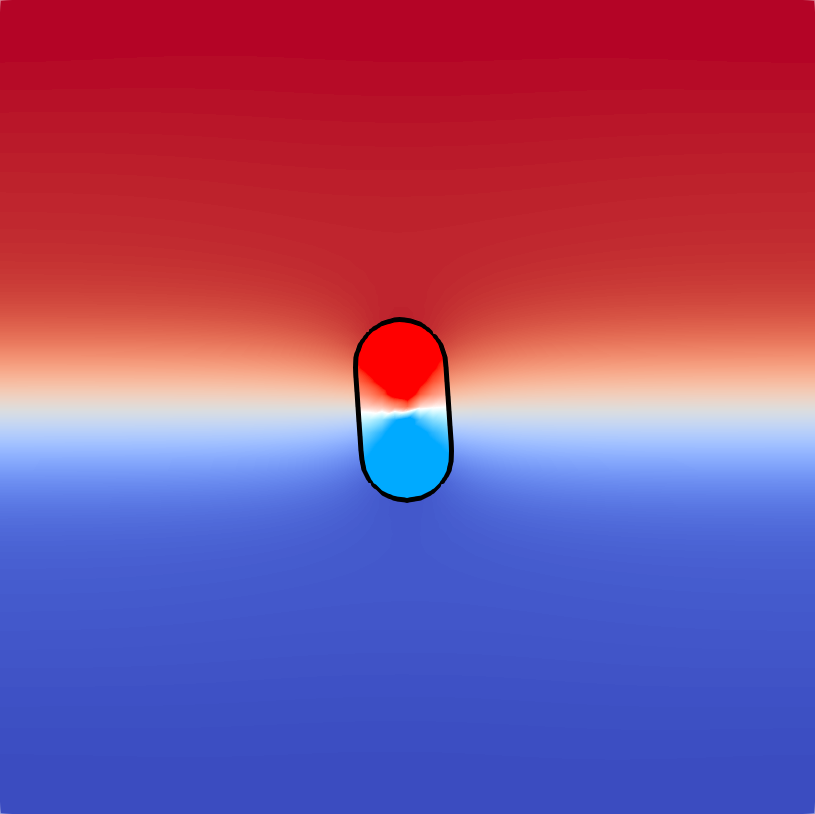} &
\includegraphics[scale=0.05]{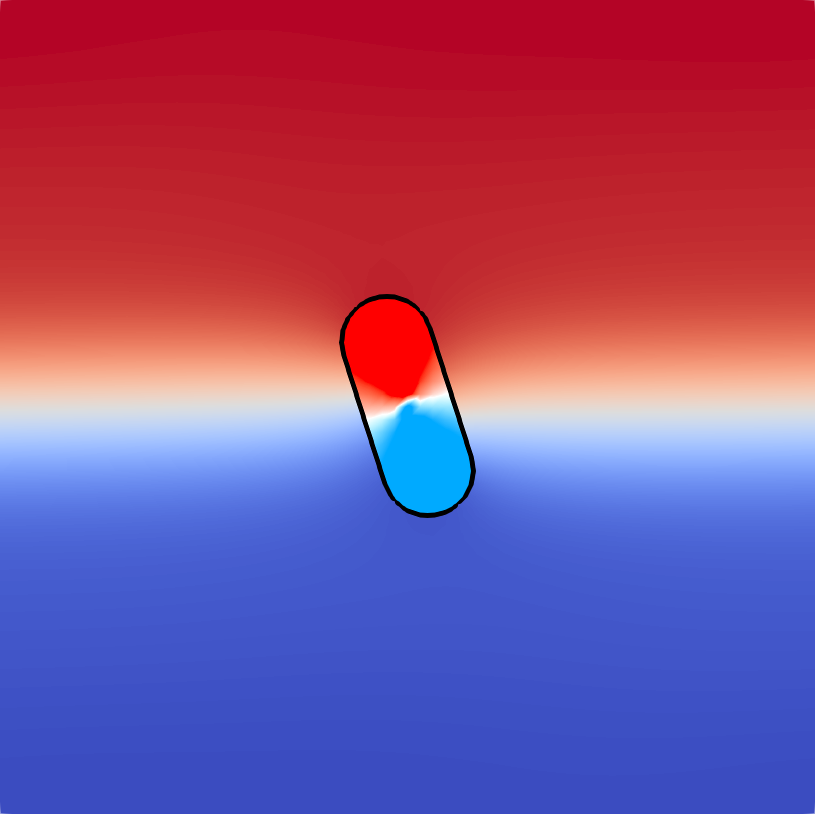} &
\includegraphics[scale=0.05]{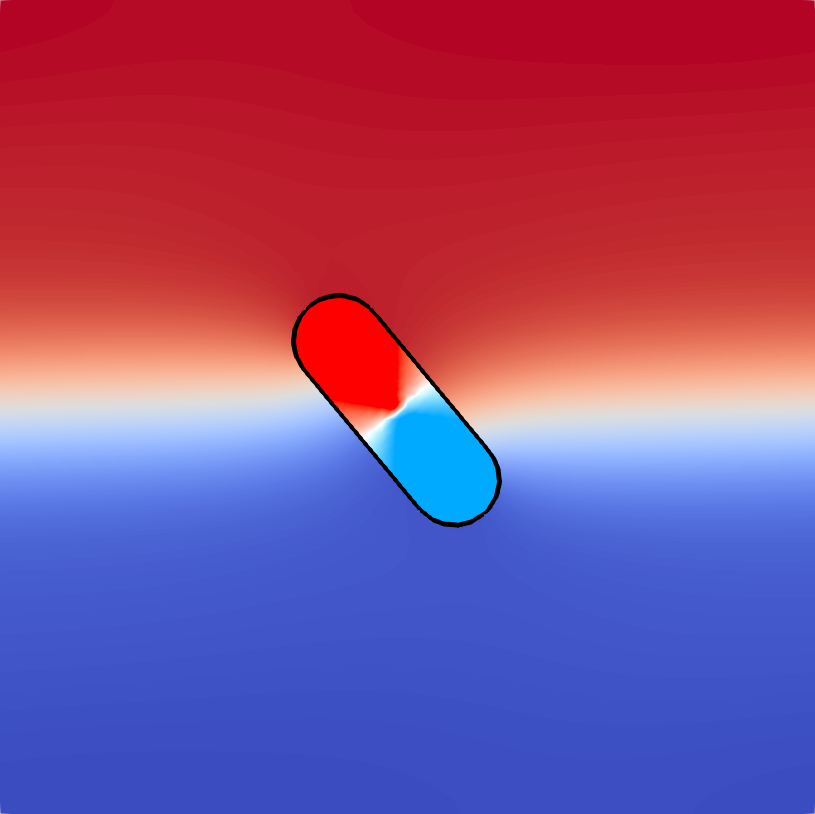} &
\includegraphics[scale=0.05]{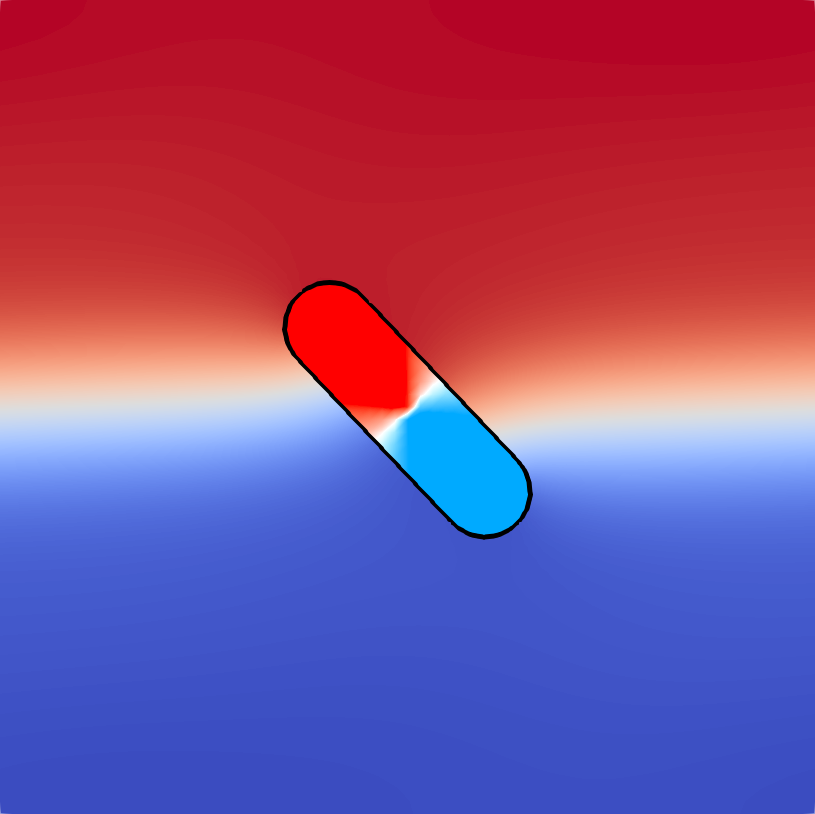} &
\includegraphics[scale=0.05]{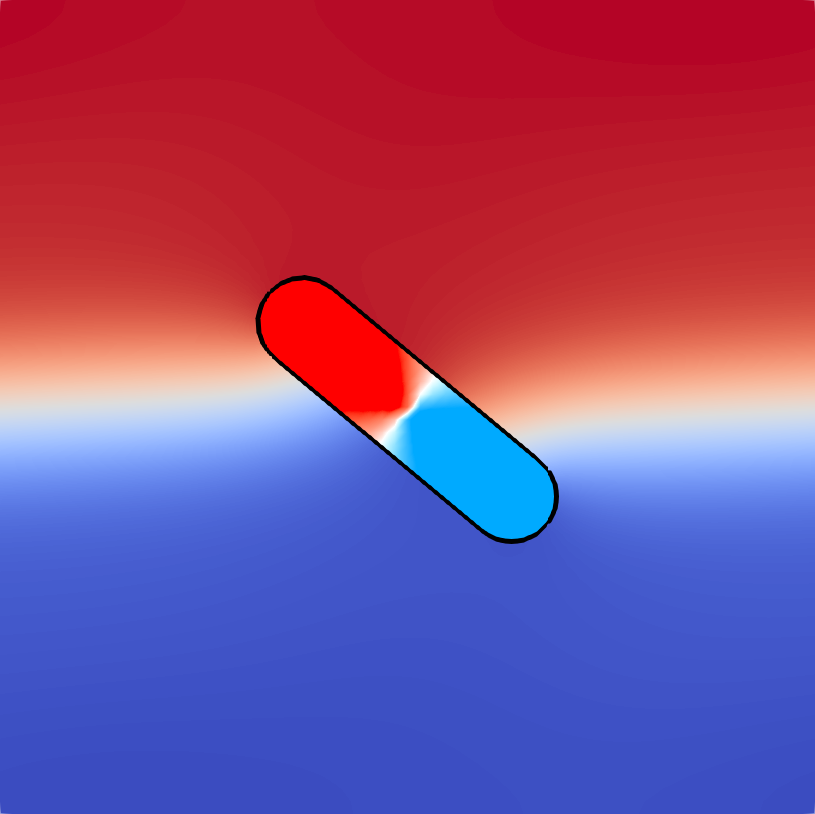} &
\includegraphics[scale=0.05]{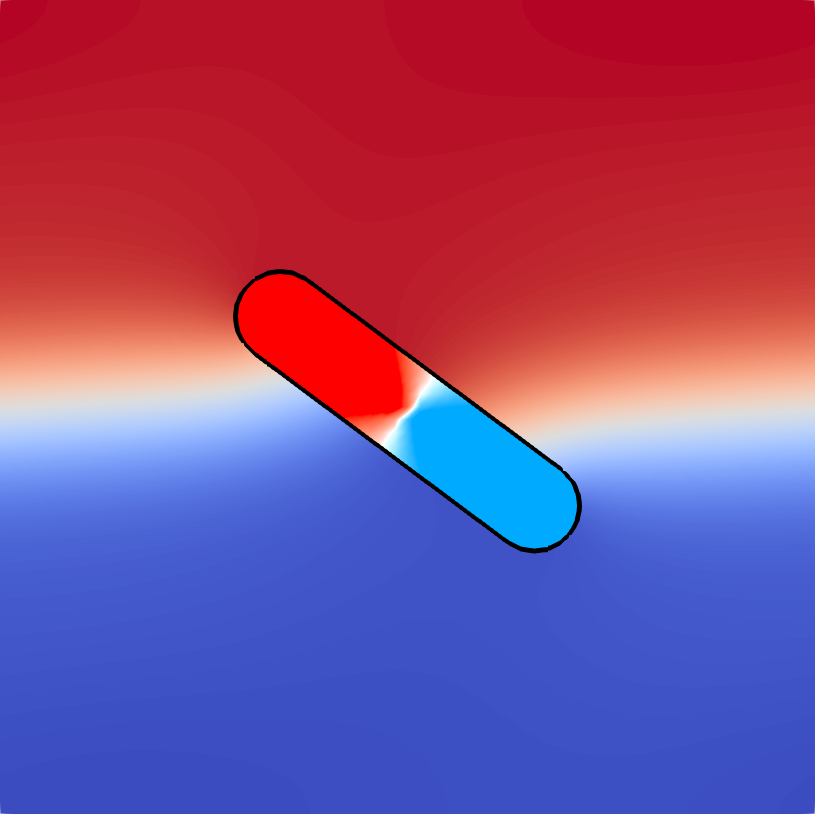} \\
$5$ &
\includegraphics[scale=0.05]{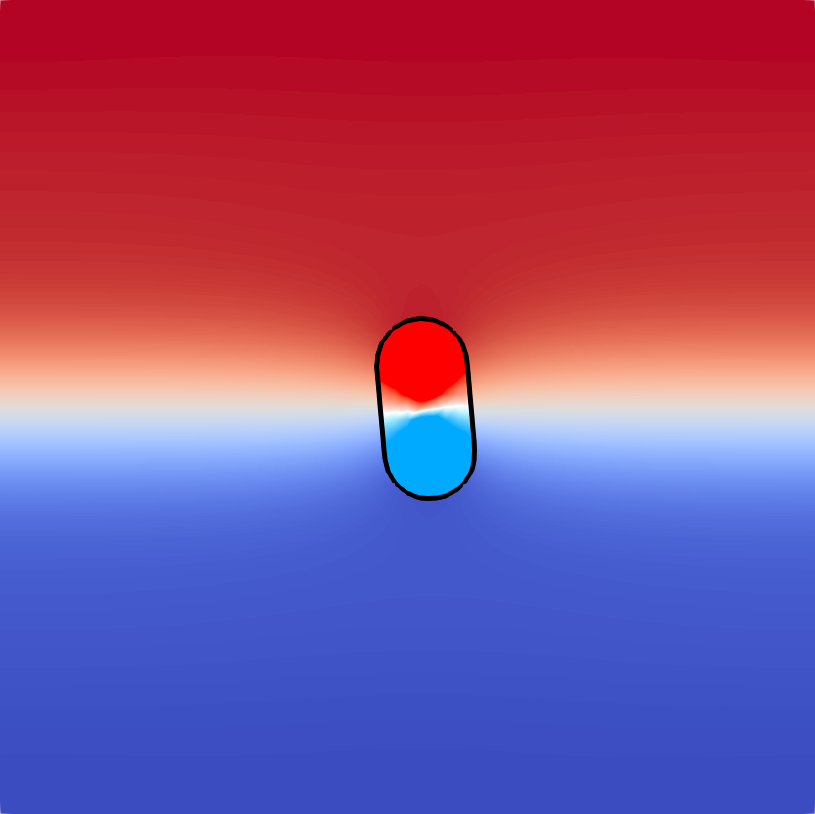} &
\includegraphics[scale=0.05]{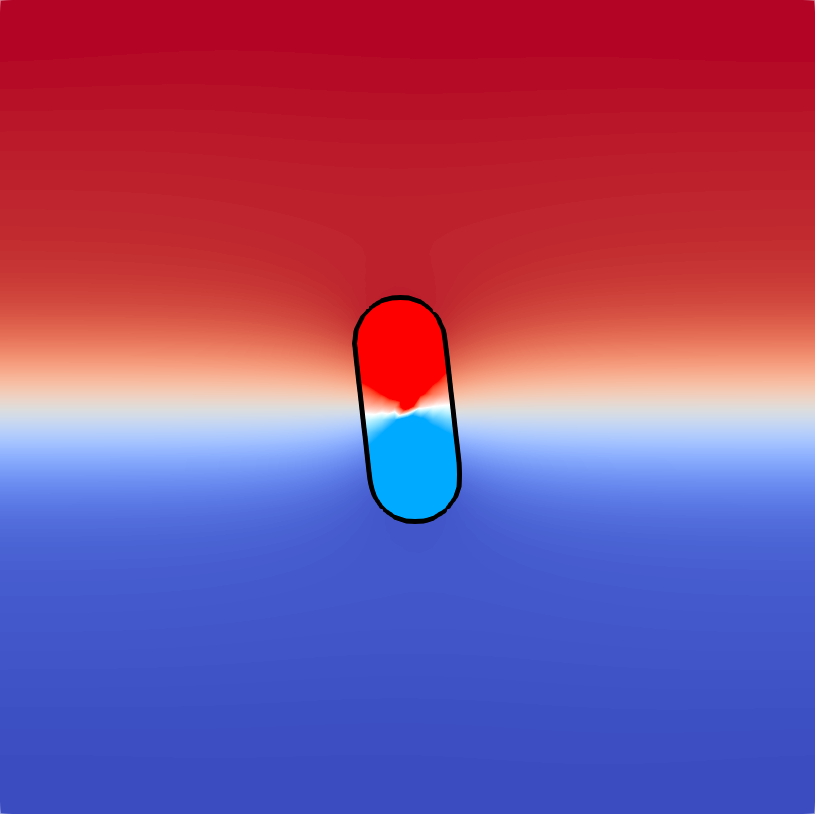} &
\includegraphics[scale=0.05]{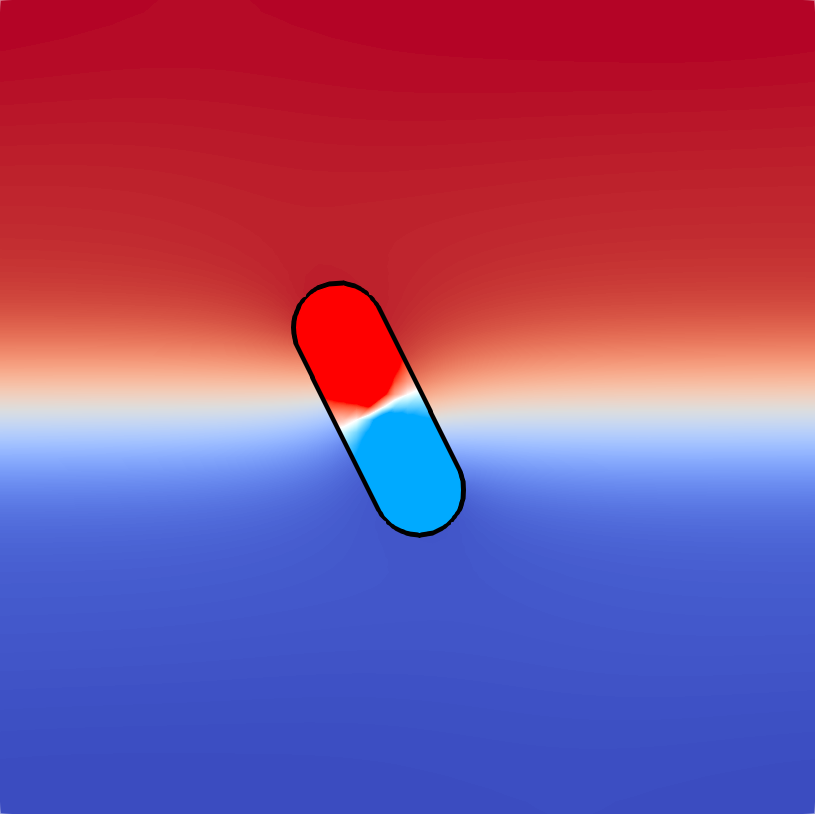} &
\includegraphics[scale=0.05]{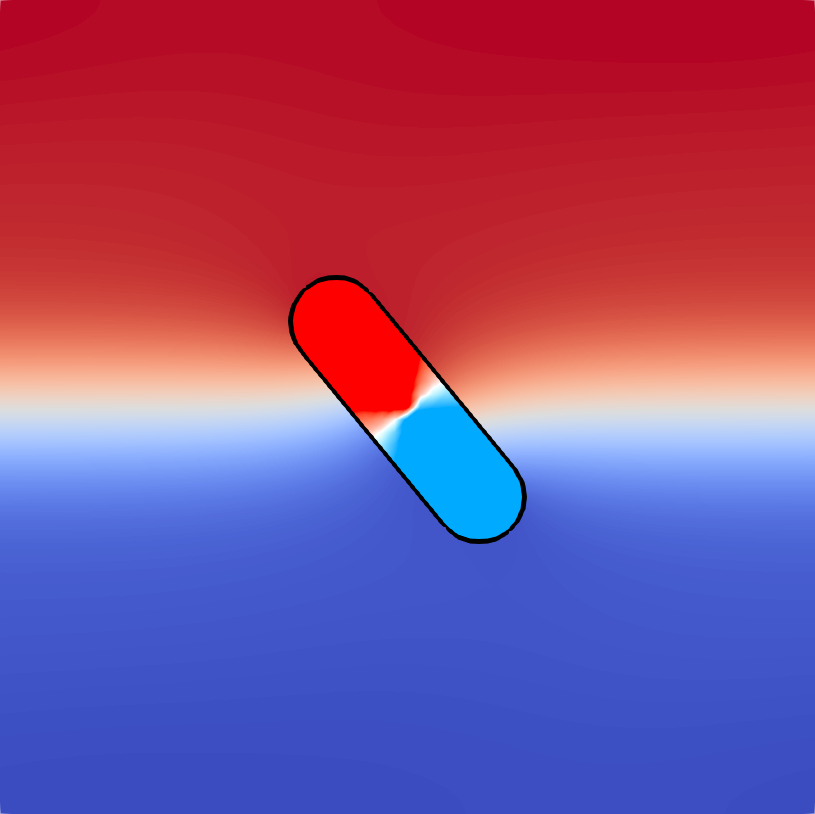} &
\includegraphics[scale=0.05]{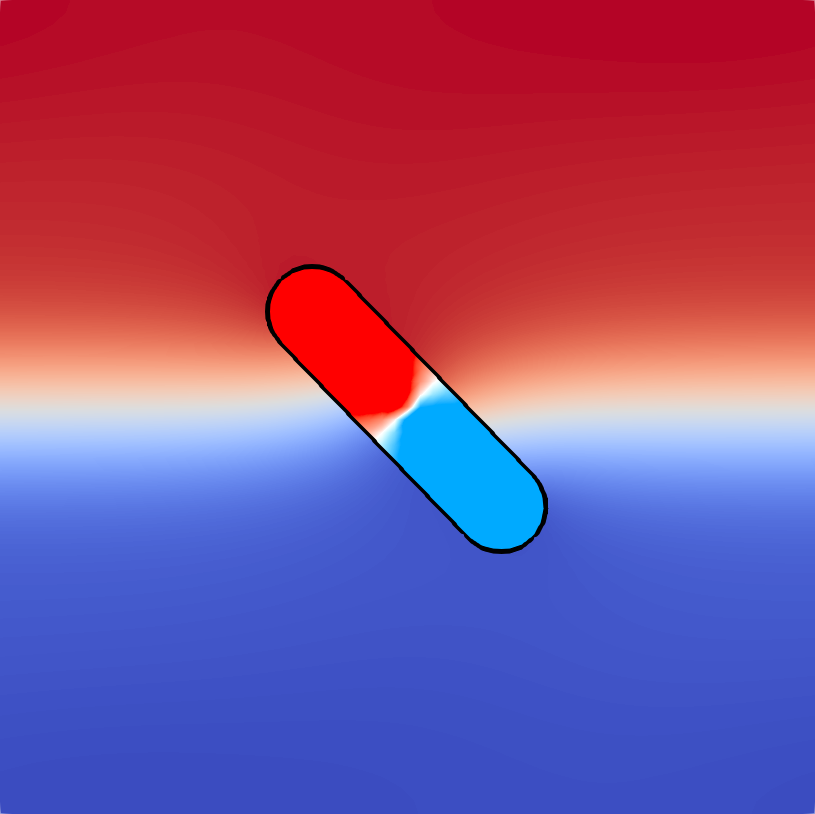} &
\includegraphics[scale=0.05]{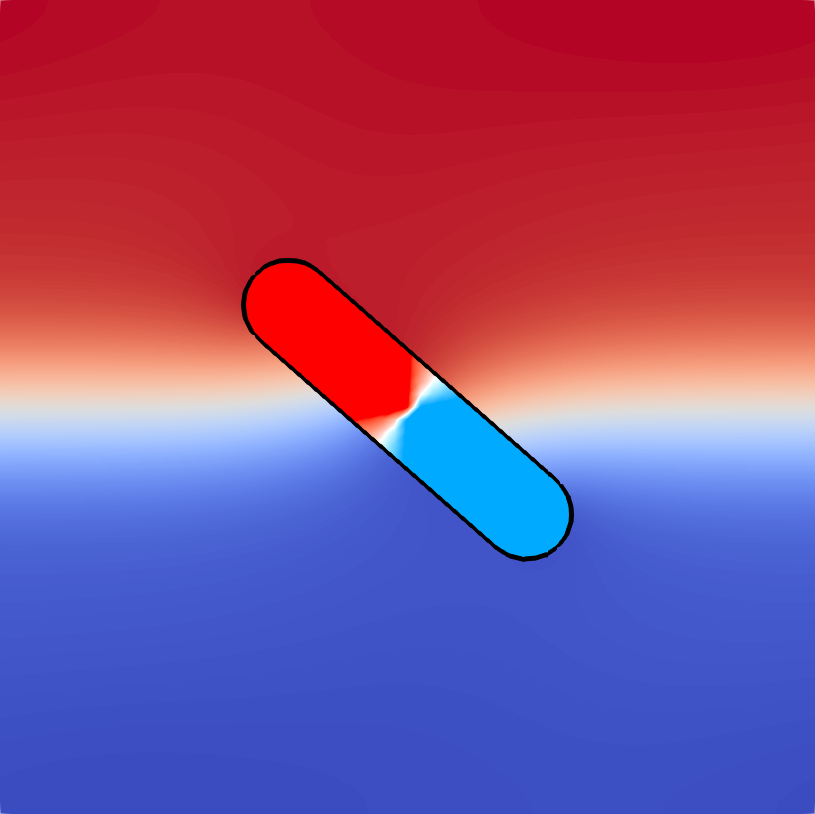} \\
$6$ &
\includegraphics[scale=0.05]{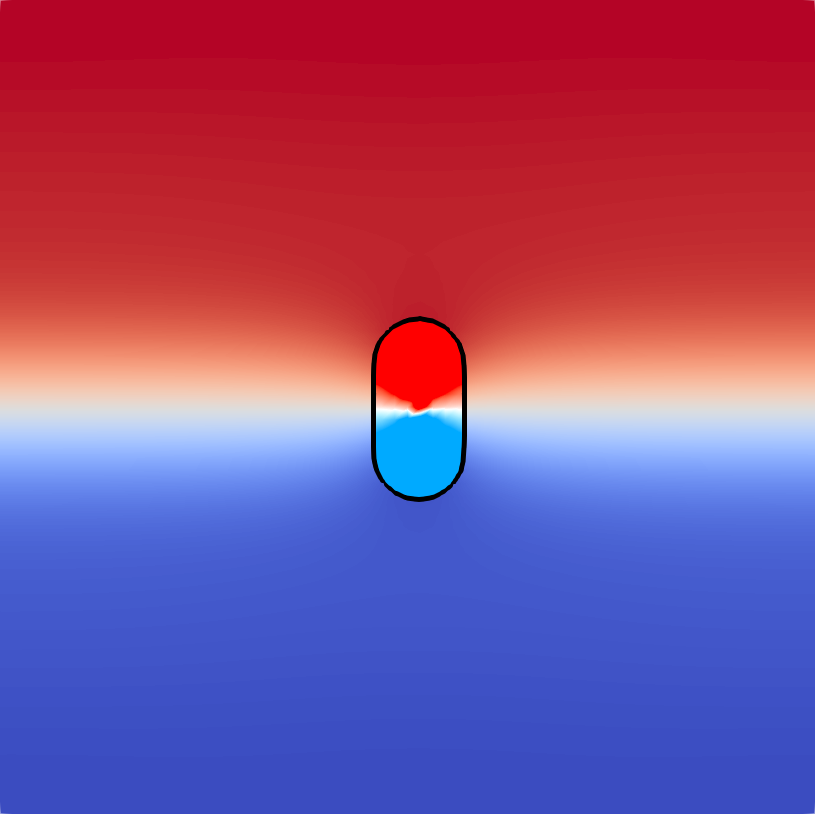} &
\includegraphics[scale=0.05]{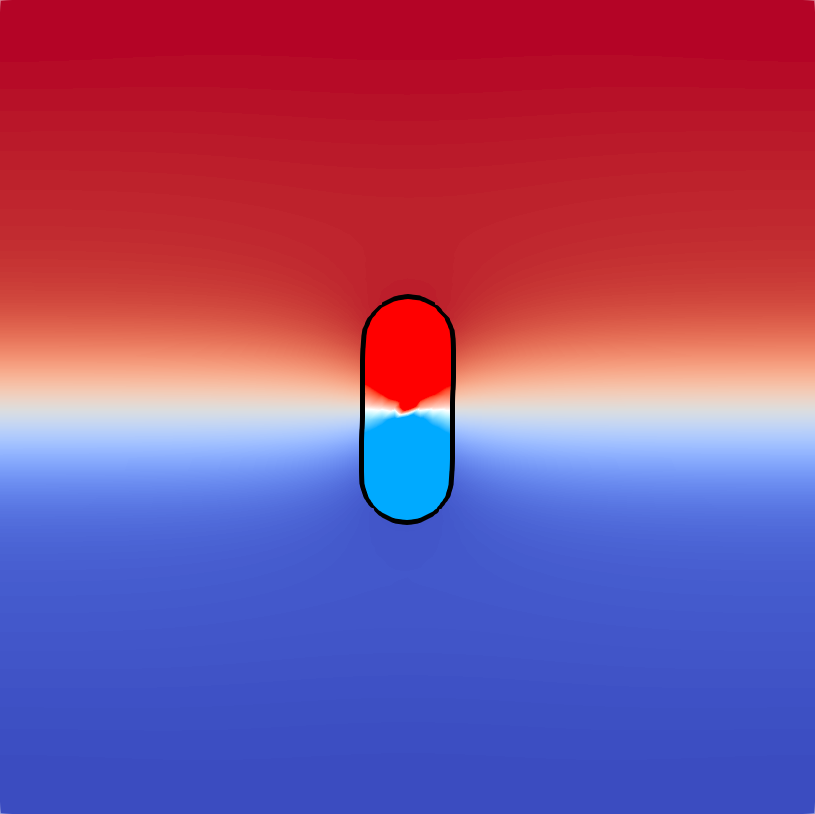} &
\includegraphics[scale=0.05]{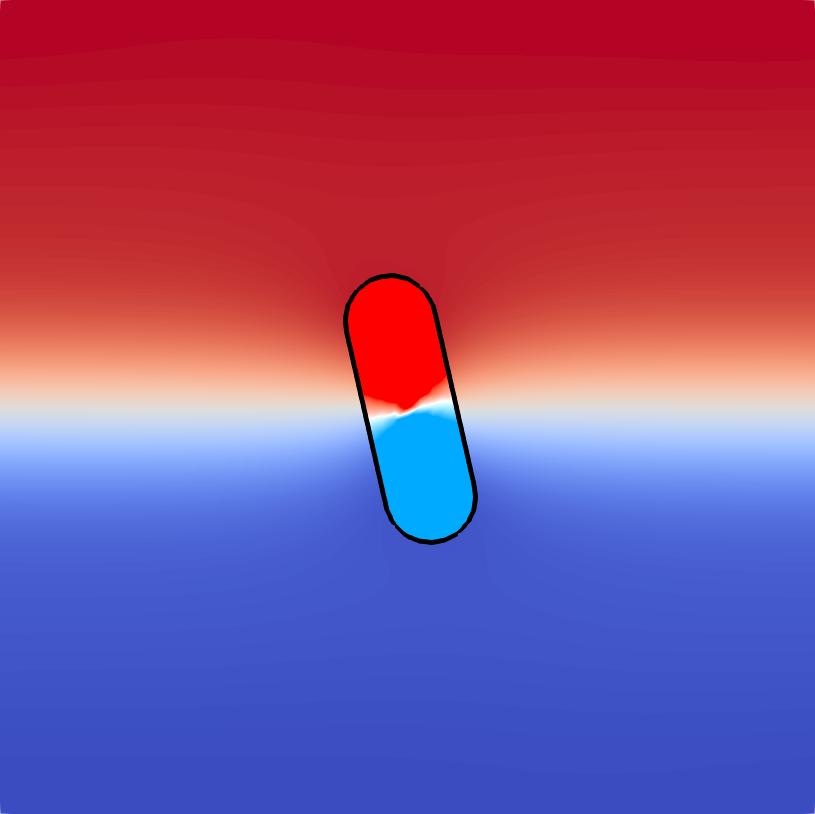} &
\includegraphics[scale=0.05]{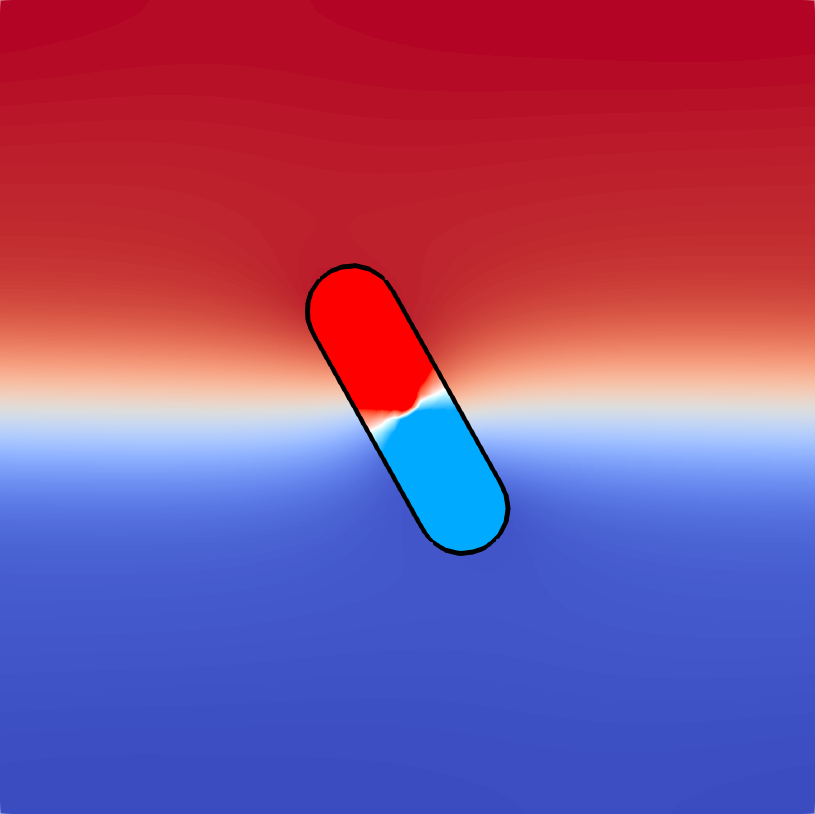} &
\includegraphics[scale=0.05]{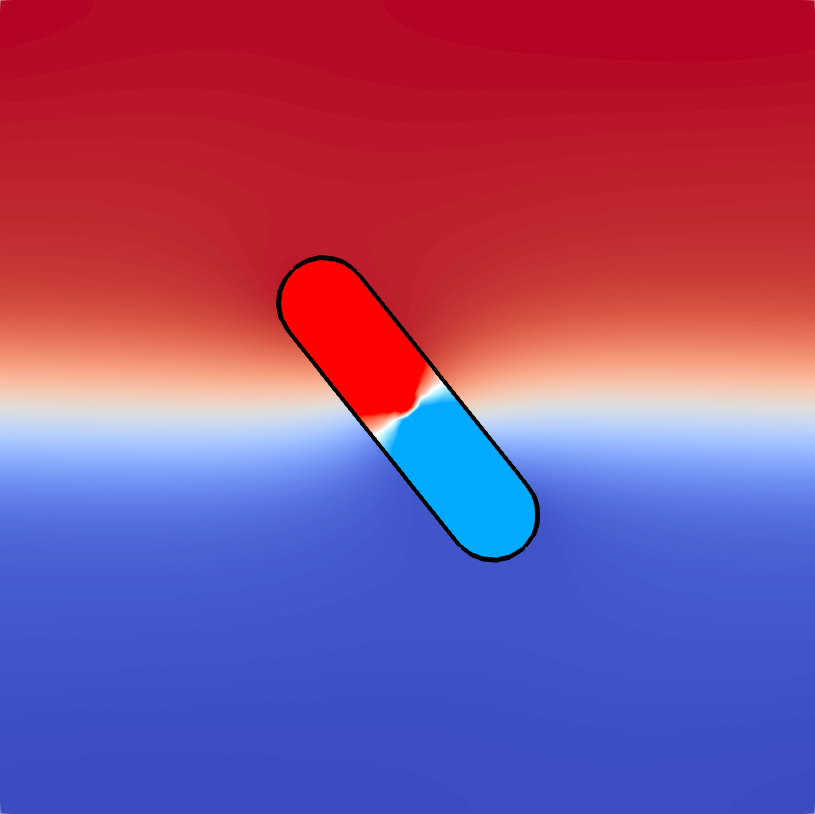} &
\includegraphics[scale=0.05]{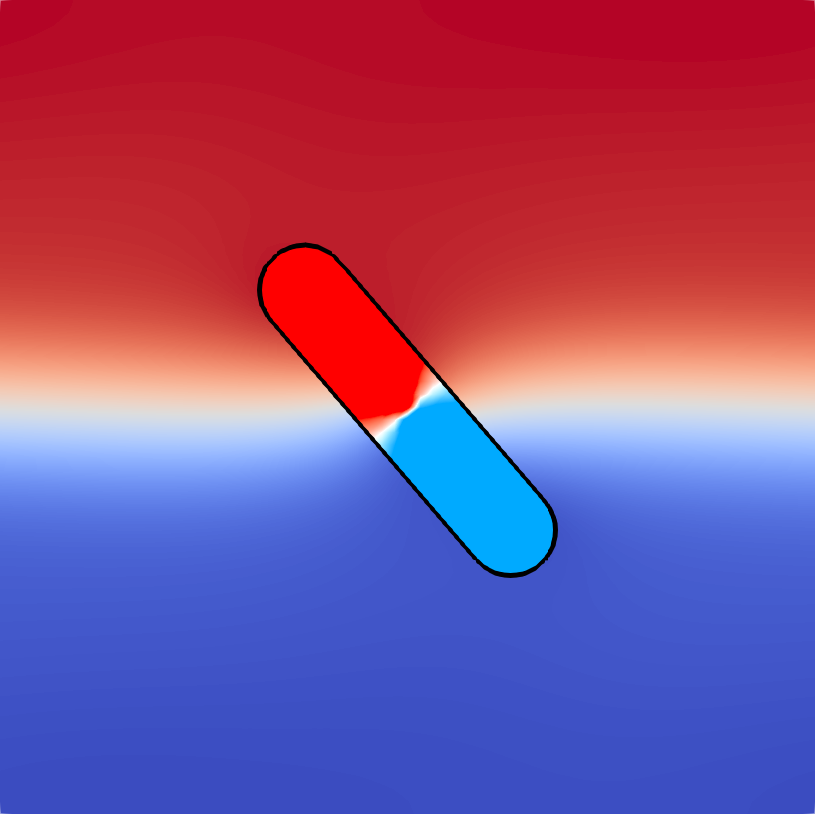}
\end{tabular}
\caption{Equilibrium orientation of a Janus nanorod at the interface between $A$ and $B$ blocks obtained for different values of the particle's lengths and of the polarization strength.}
\label{fig:results:nano:length}
\end{figure}

\begin{figure}[!h]
\centering  
\includegraphics[width=.6\textwidth]{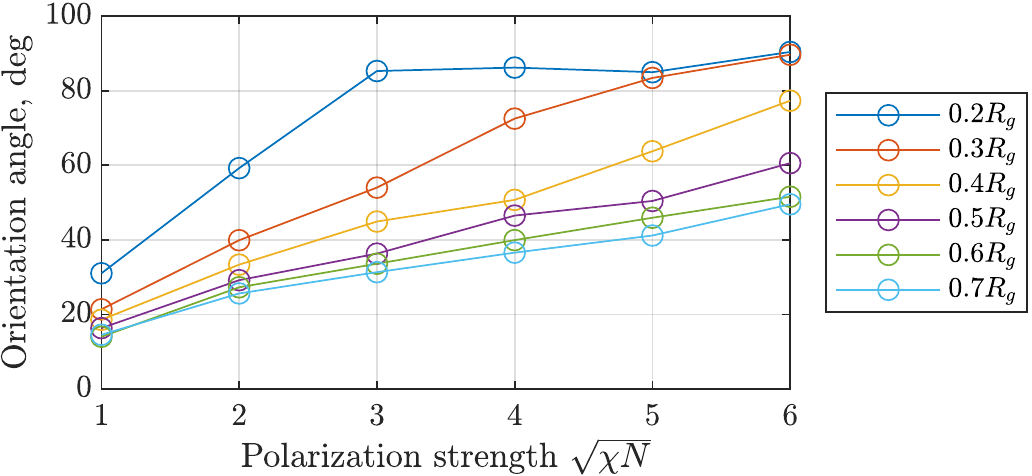}
\caption{Dependence of the Janus nanorod orientation on the polarization strength for several values of the particle's length.}
\label{fig:results:nano:length:plot}
\end{figure}

Figure \ref{fig:results:nano:many} depicts a representative simulation result for the co-assembly of lamellae-forming diblock copolymer ($\XN{AB} = 30$, $f=0.5$) with Janus nanorods of length $0.4 R_g$ in a computational domain of dimensions $3 R_g \times 3 R_g$. One observes that, while a small number of nanoparticles are trapped inside of $A$ or $B$ domains the majority of nanorods tend to aggregate at the interface between $A$ and $B$ polymer components and form a certain angle with the interface.
In order to investigate this co-assembly behavior in greater detail, we first consider a single nanorod in a smaller rectangular domain with dimension $1.8\times 1.8 R_g$, which approximately corresponds to the single diblock lamellar spacing. Periodic boundary conditions are imposed in the $x$-direction and homogeneous Neumann boundary conditions are imposed in the $y$-direction. Figure \ref{fig:results:nano:length} depicts the equilibrium placements of nanorods of different lengths and for several values of the polarization strength $\XN{ap}$, while figure \ref{fig:results:nano:length:plot} summarizes the quantitative values for the equilibrium angles. It is clear from these results that the stable orientation of an isolated Janus nanorod in a lamellar diblock copolymer with respect to the interface between polymer components is a smooth function of the particle's length and of the polarization strength.

\begin{figure}[!h]
\centering  
\setlength{\tabcolsep}{0.1em}
\renewcommand{\arraystretch}{0.5}
\begin{tabular}{ c  c  c  c  c c c  }
$\chi_{p} N$ &
$$ &
$$ &
$$ &
$$ \\
$1$ &
\includegraphics[scale=0.05]{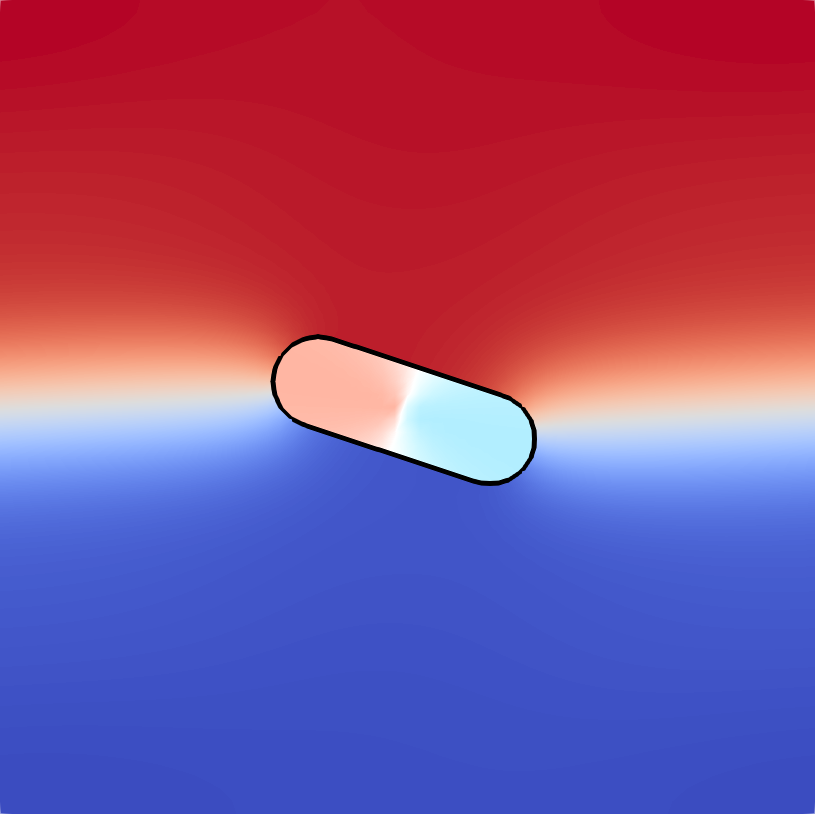} &
\includegraphics[scale=0.05]{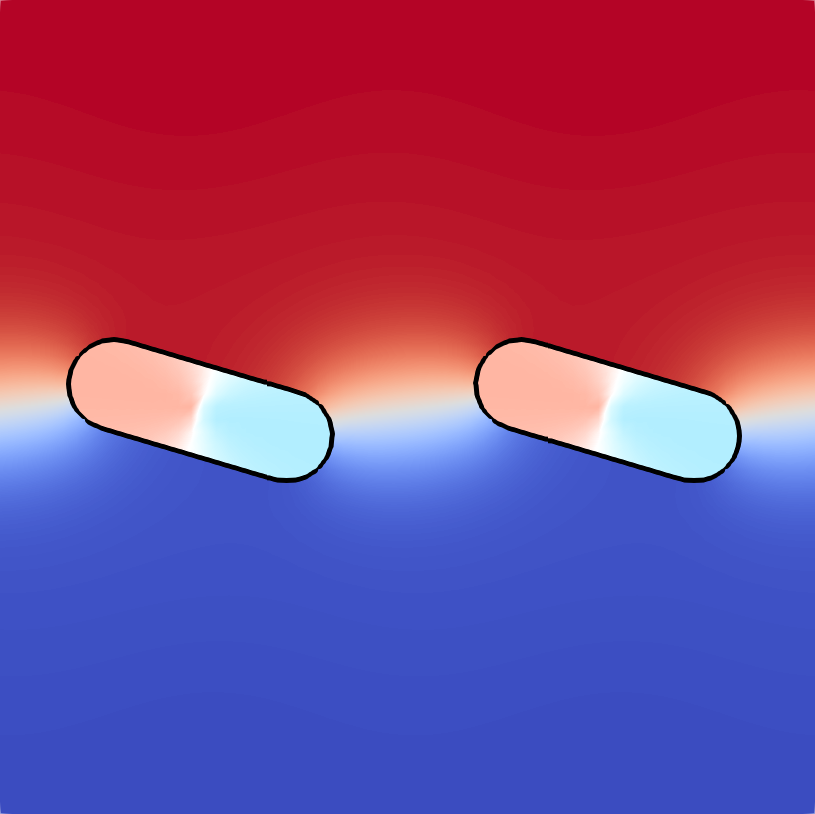} &
\includegraphics[scale=0.05]{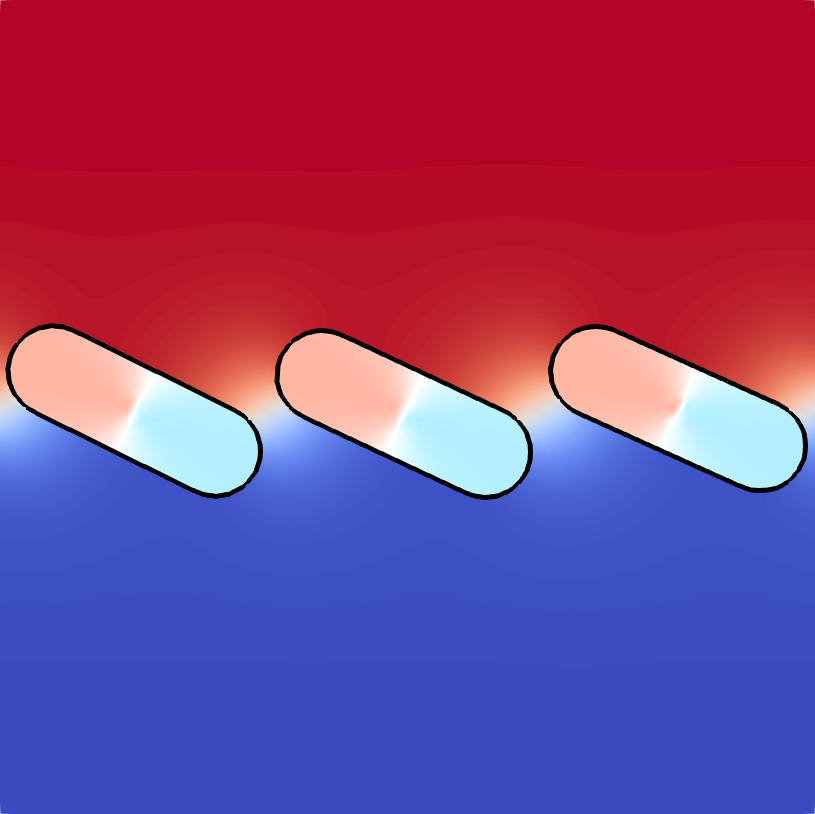} &
\includegraphics[scale=0.05]{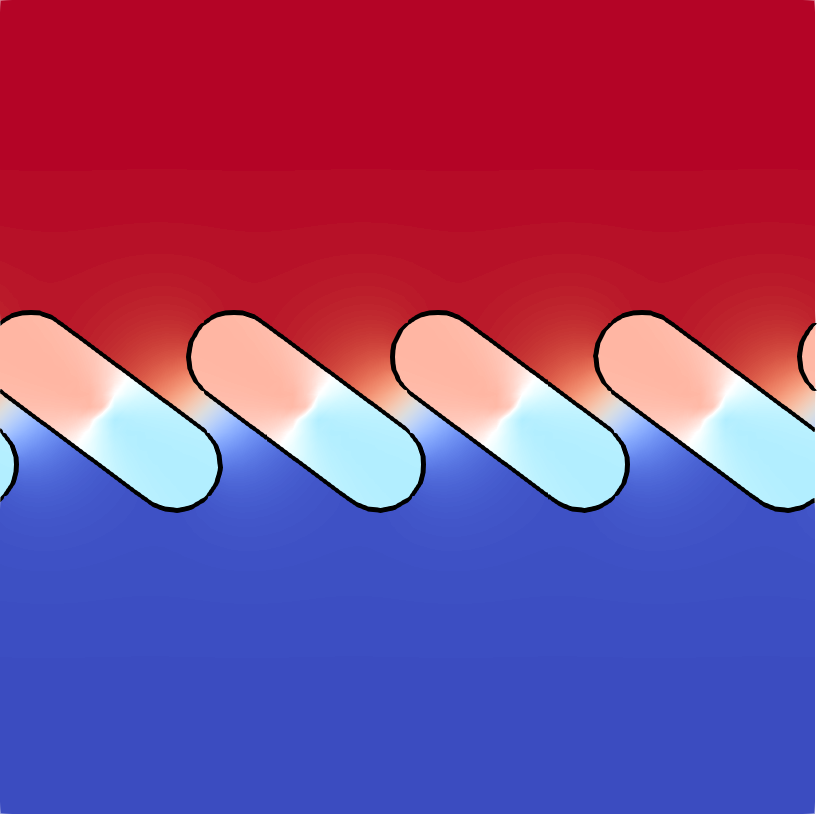} &
\includegraphics[scale=0.05]{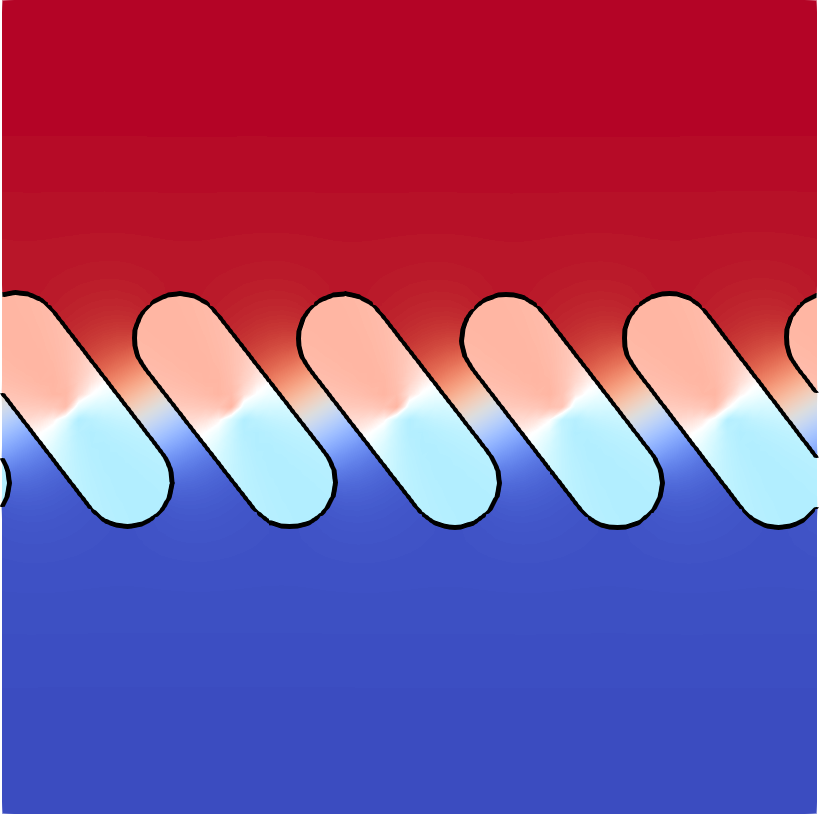} &
\includegraphics[scale=0.05]{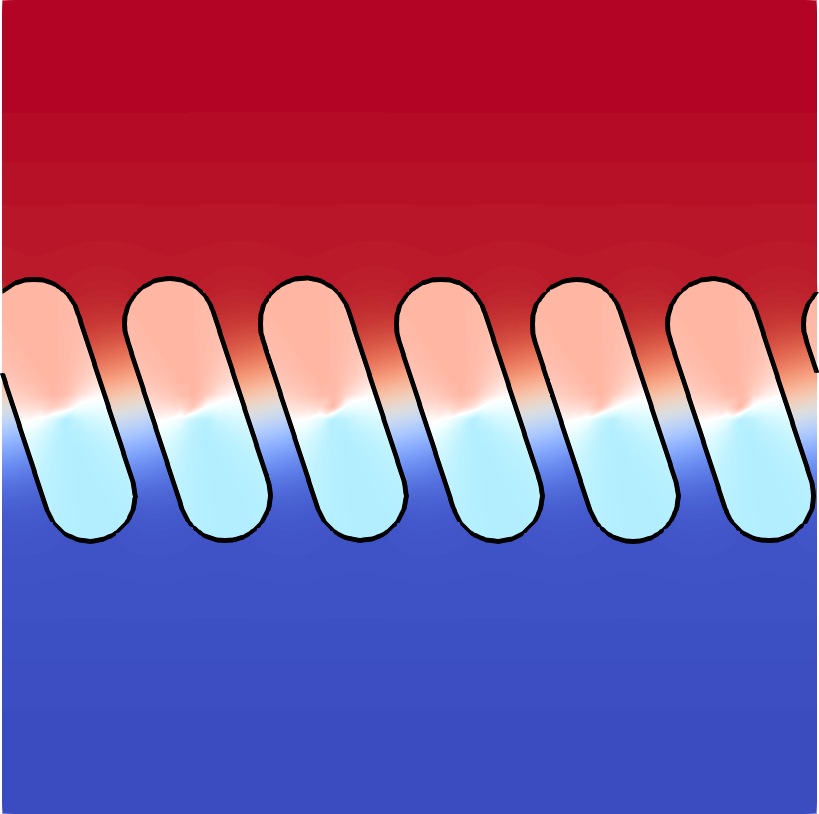} \\
$2$ &
\includegraphics[scale=0.05]{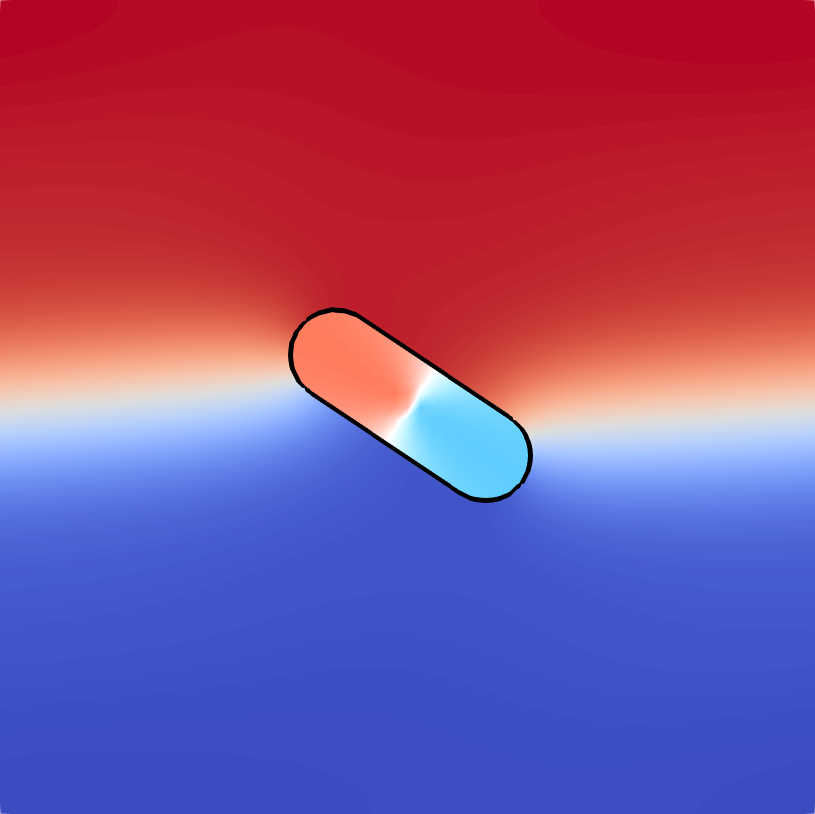} &
\includegraphics[scale=0.05]{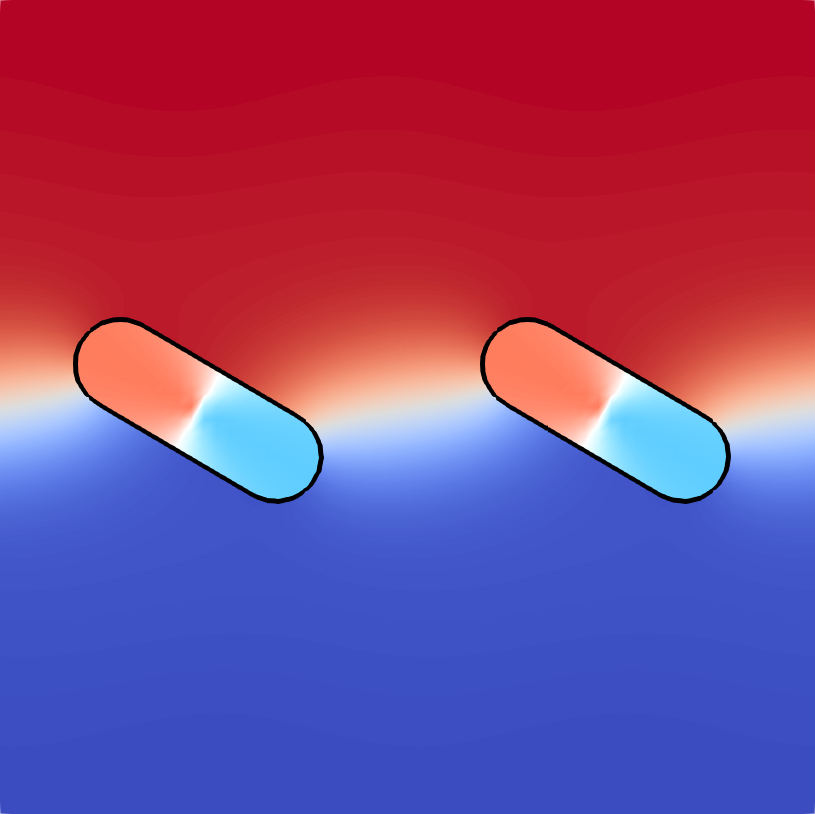} &
\includegraphics[scale=0.05]{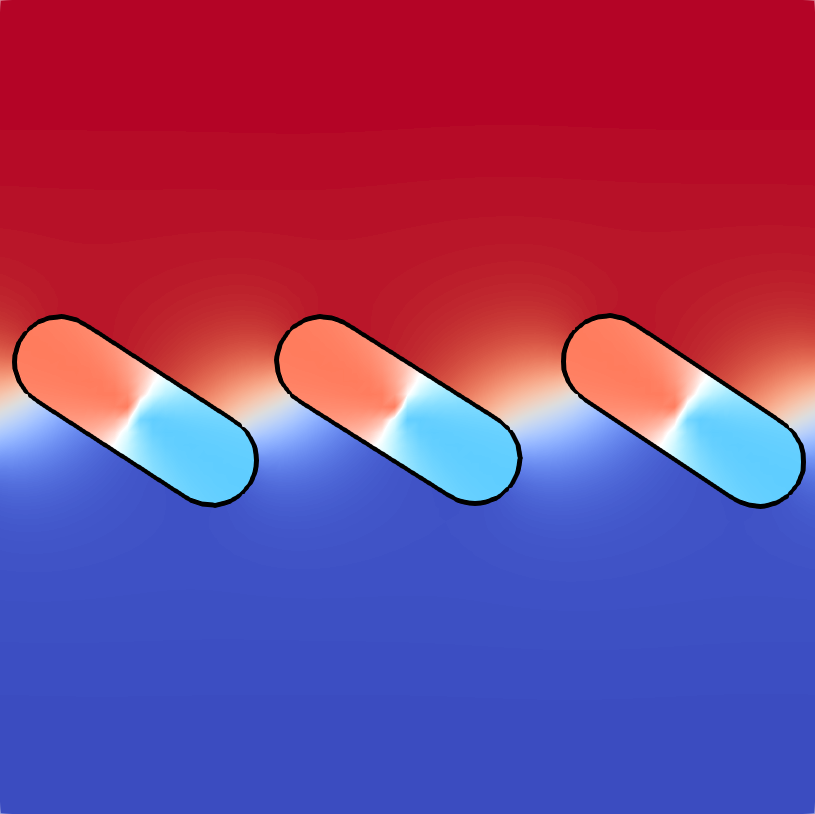} &
\includegraphics[scale=0.05]{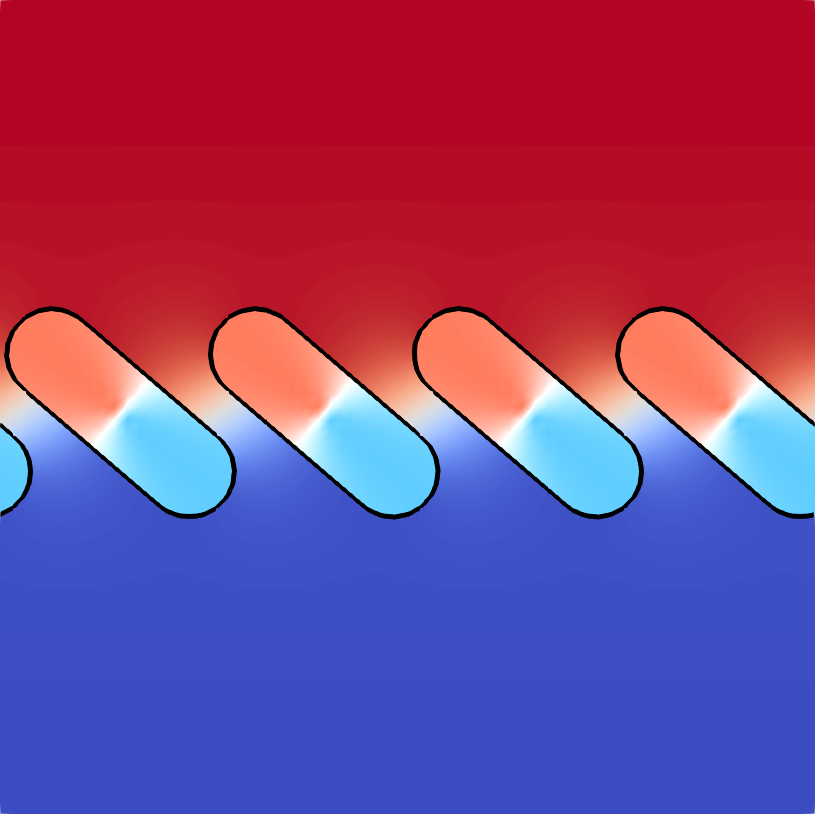} &
\includegraphics[scale=0.05]{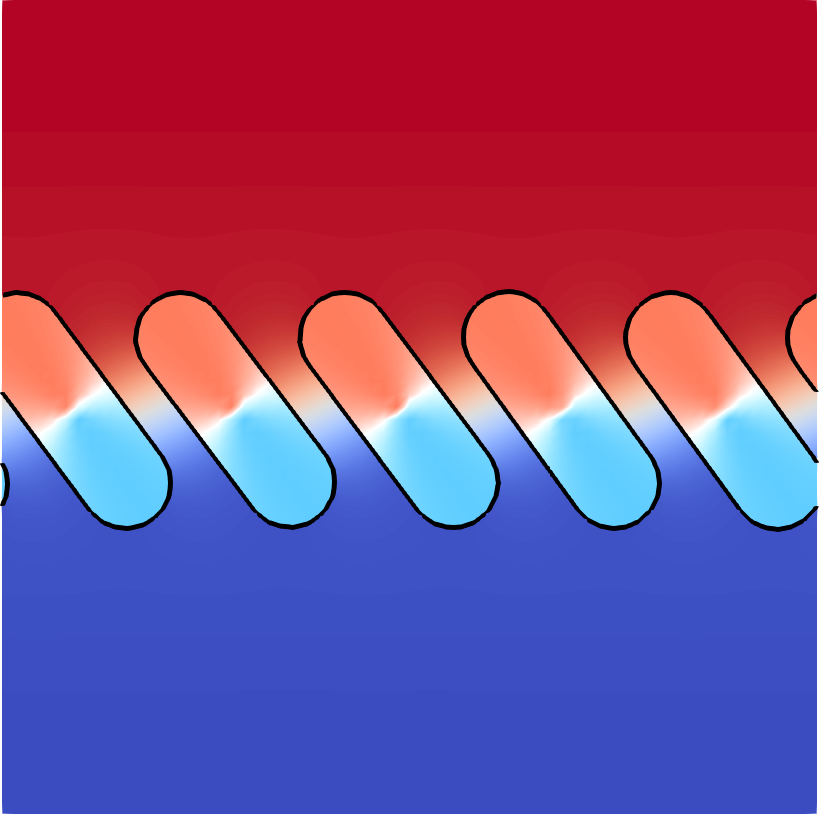} &
\includegraphics[scale=0.05]{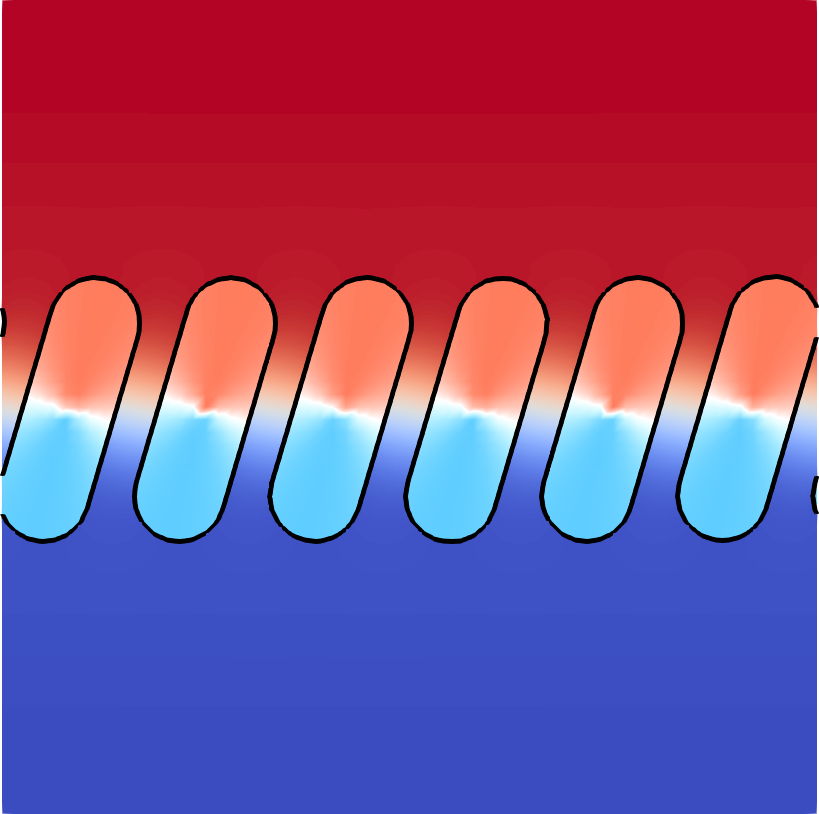} \\
$3$ &
\includegraphics[scale=0.05]{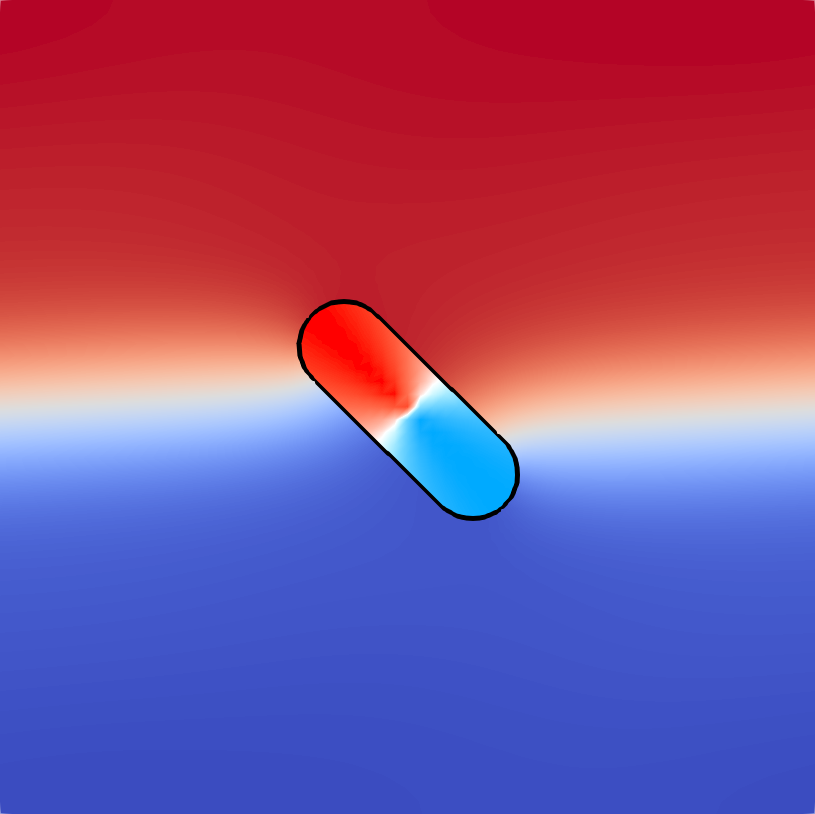} &
\includegraphics[scale=0.05]{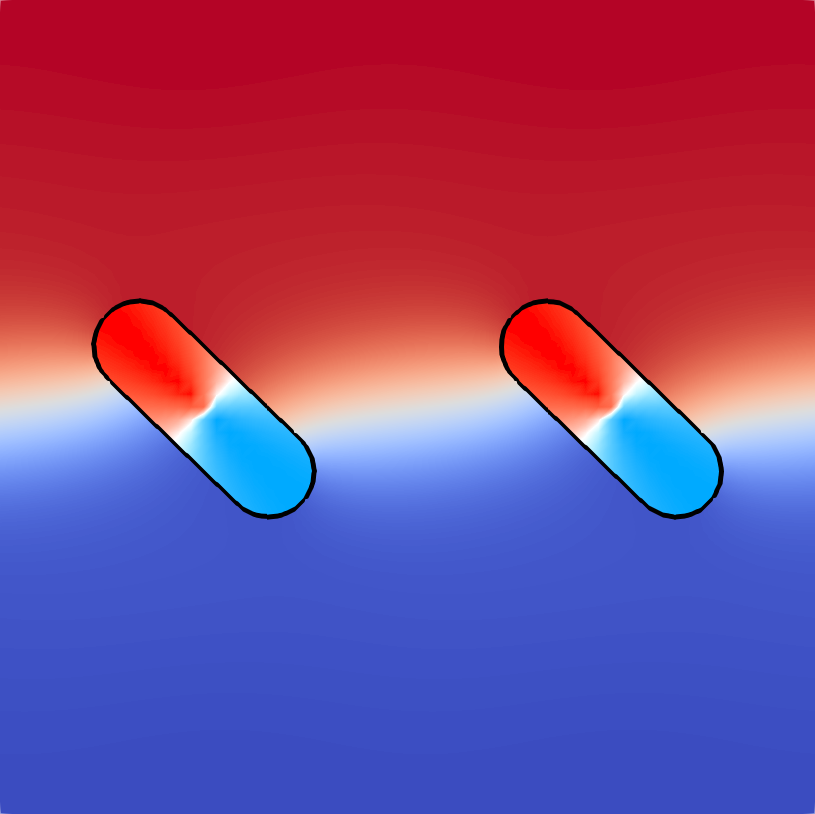} &
\includegraphics[scale=0.05]{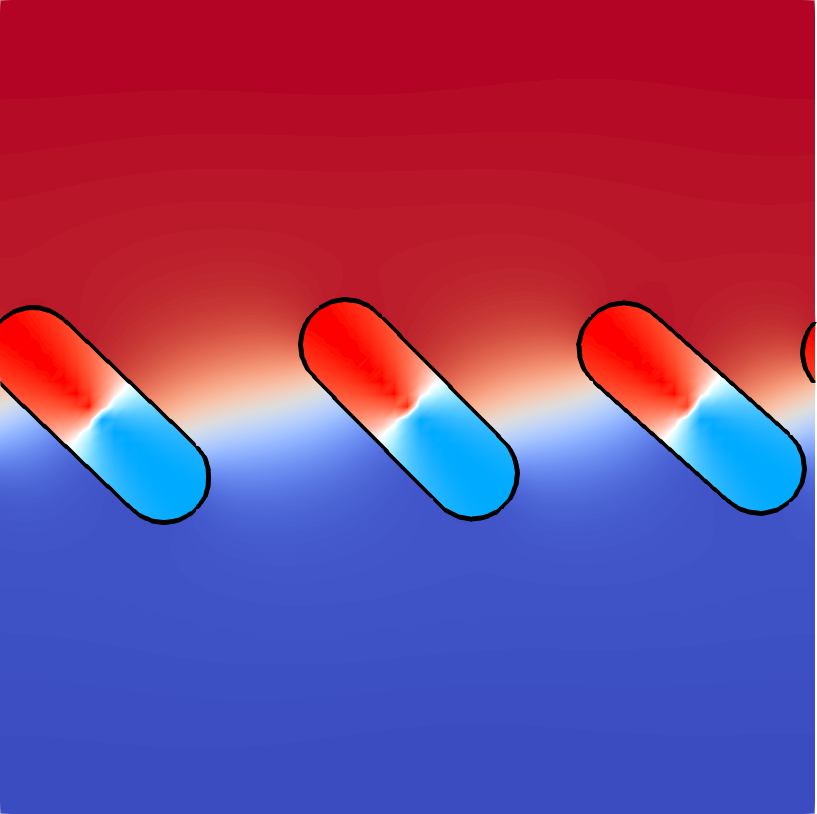} &
\includegraphics[scale=0.05]{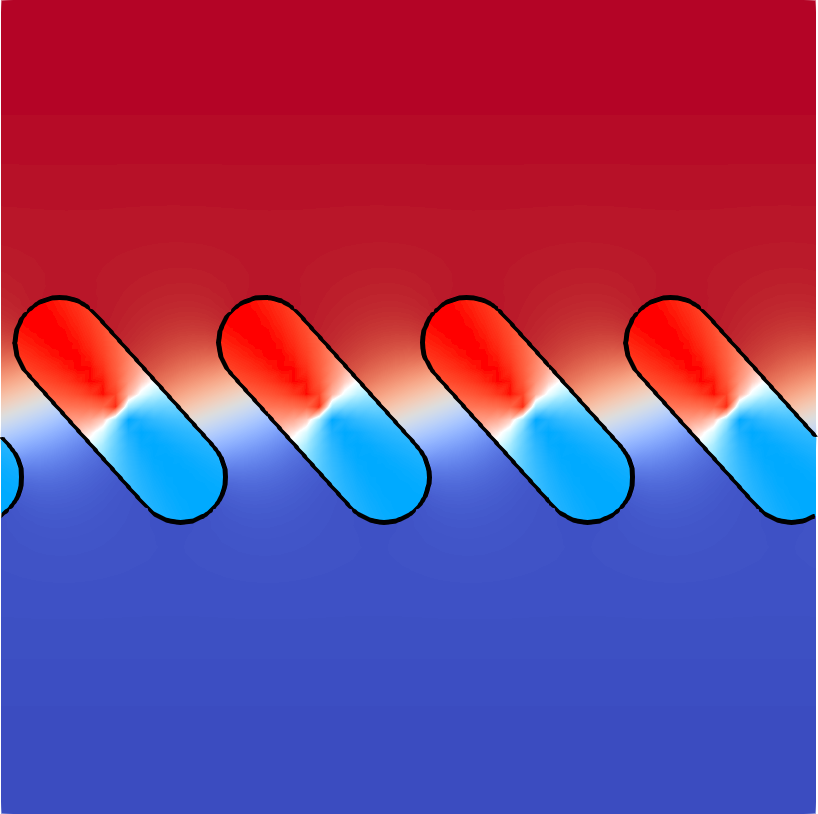} &
\includegraphics[scale=0.05]{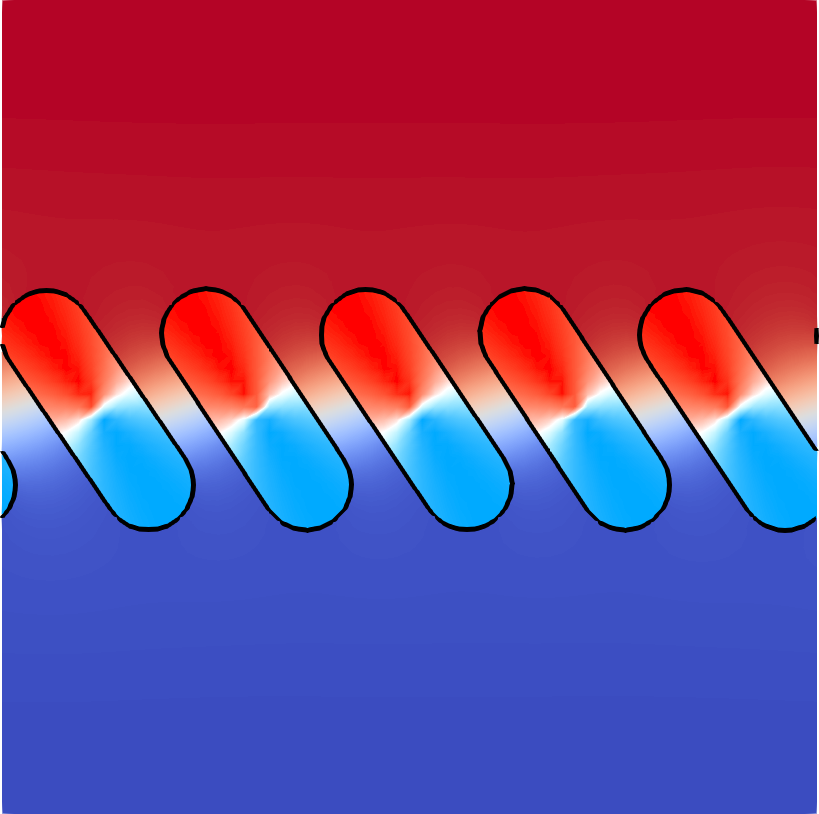} &
\includegraphics[scale=0.05]{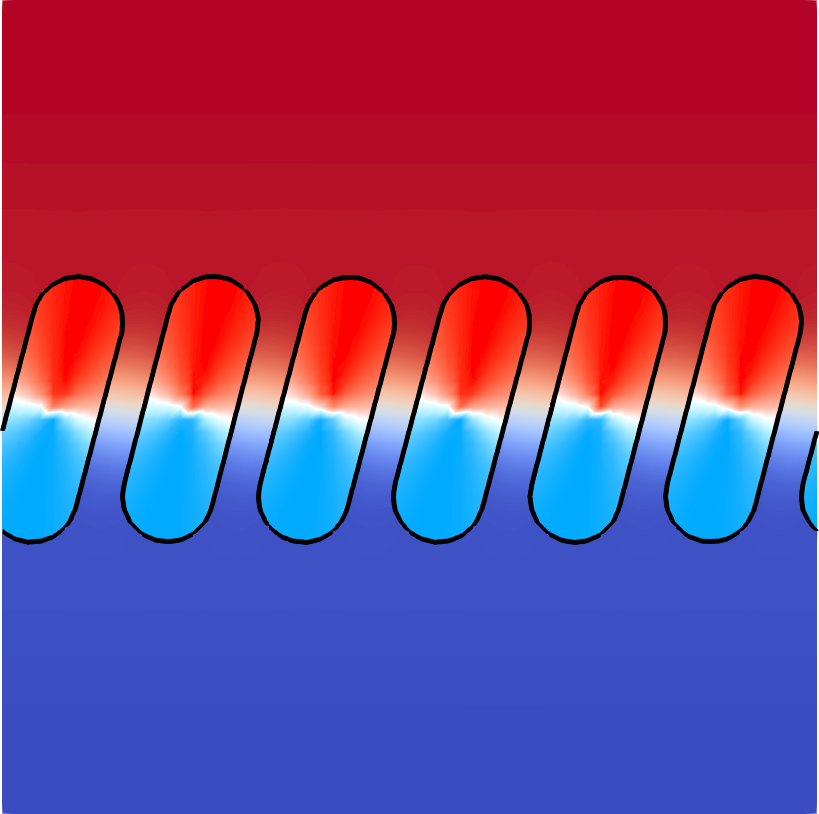} \\
$4$ &
\includegraphics[scale=0.05]{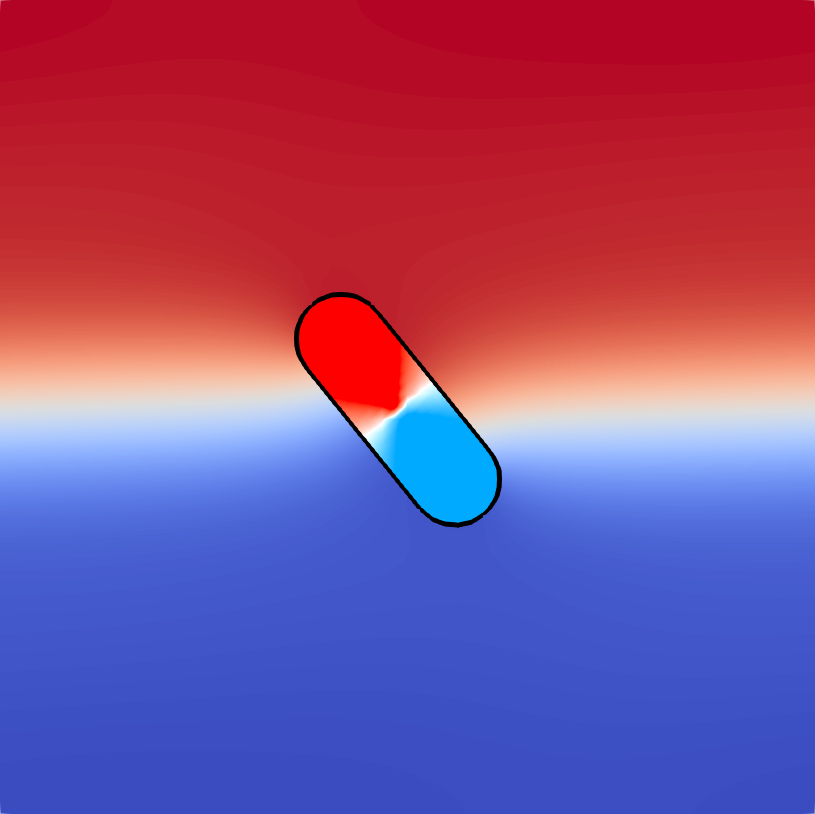} &
\includegraphics[scale=0.05]{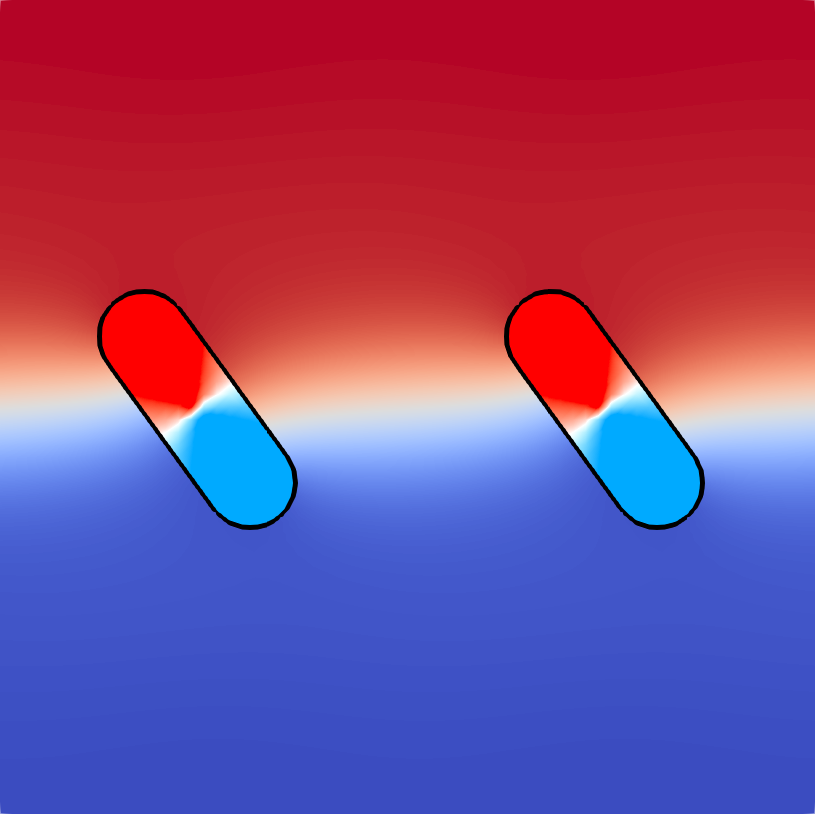} &
\includegraphics[scale=0.05]{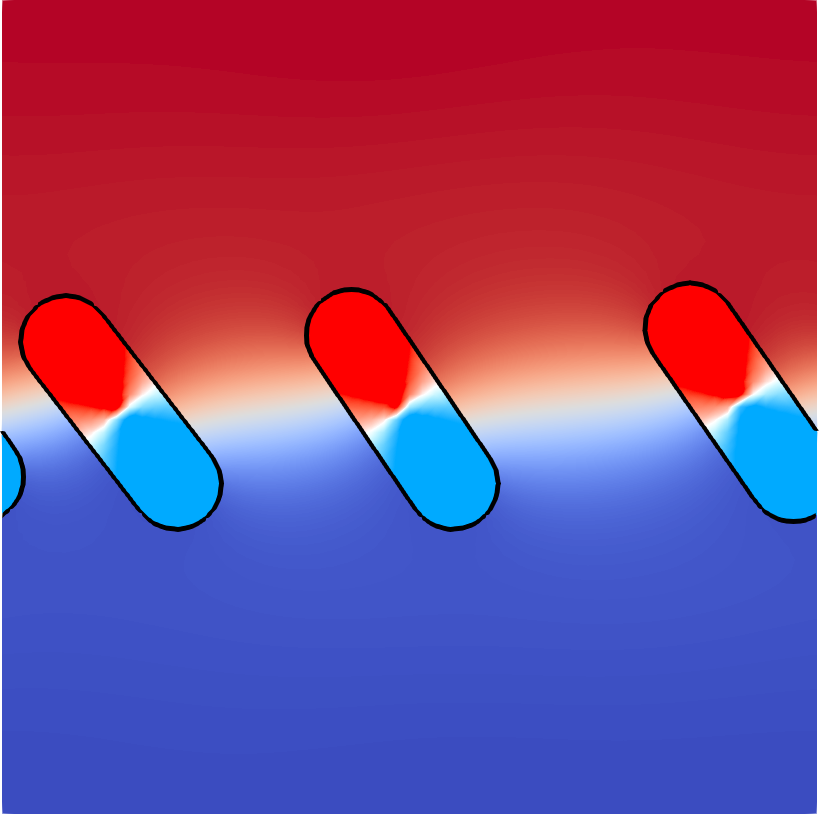} &
\includegraphics[scale=0.05]{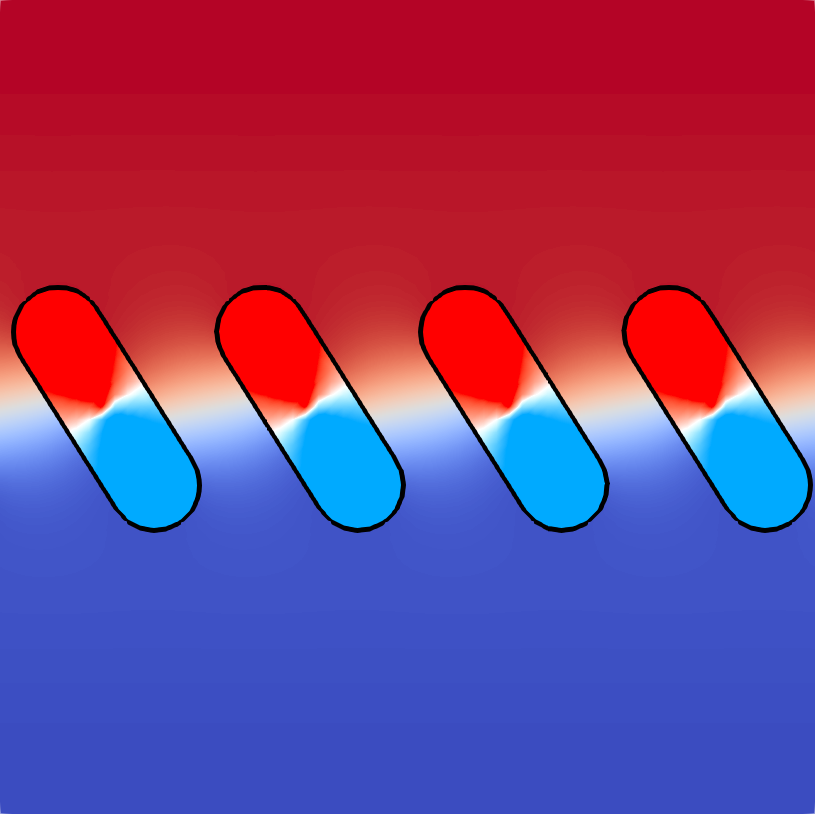} &
\includegraphics[scale=0.05]{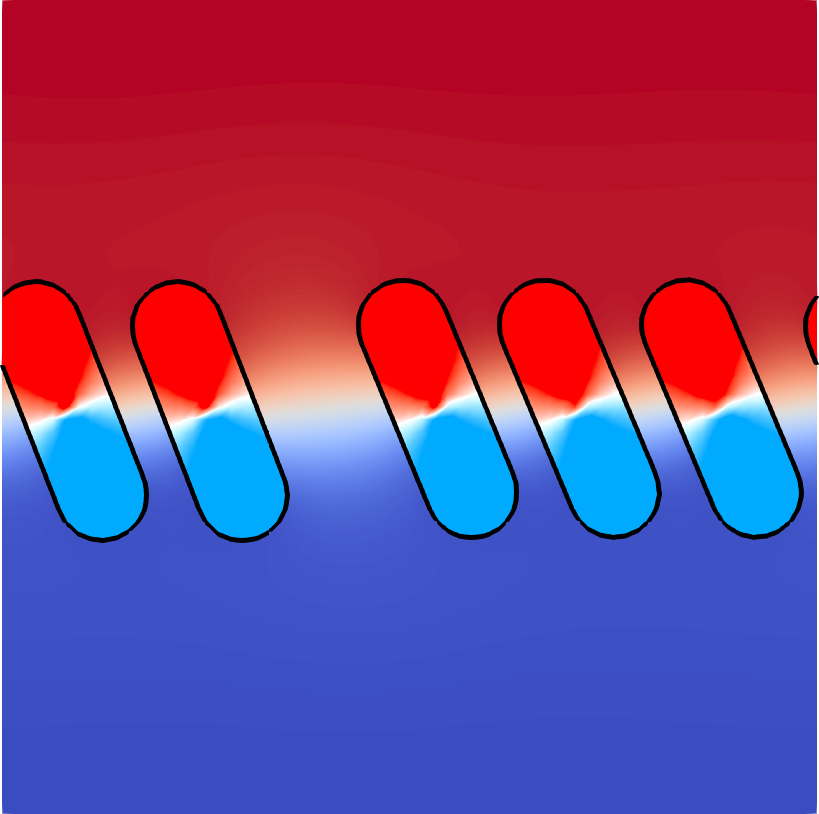} &
\includegraphics[scale=0.05]{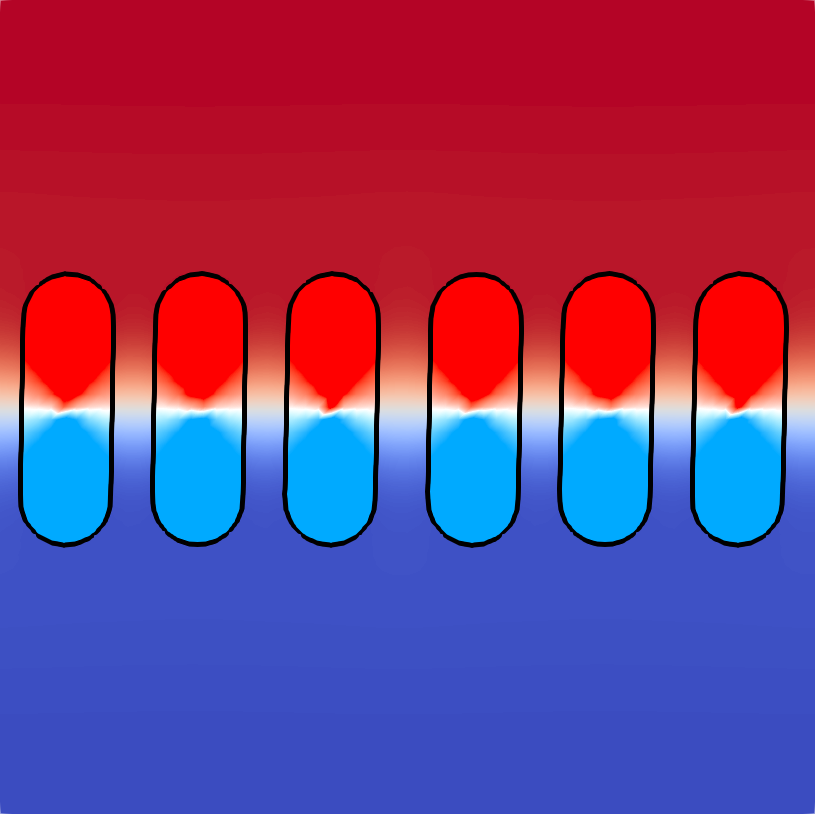} \\
$5$ &
\includegraphics[scale=0.05]{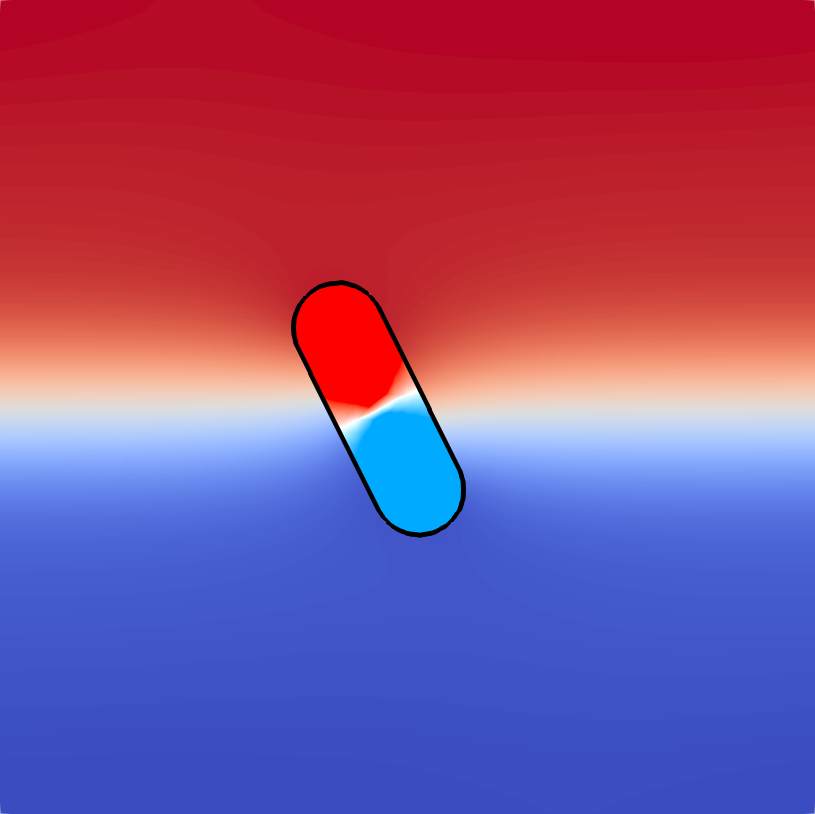} &
\includegraphics[scale=0.05]{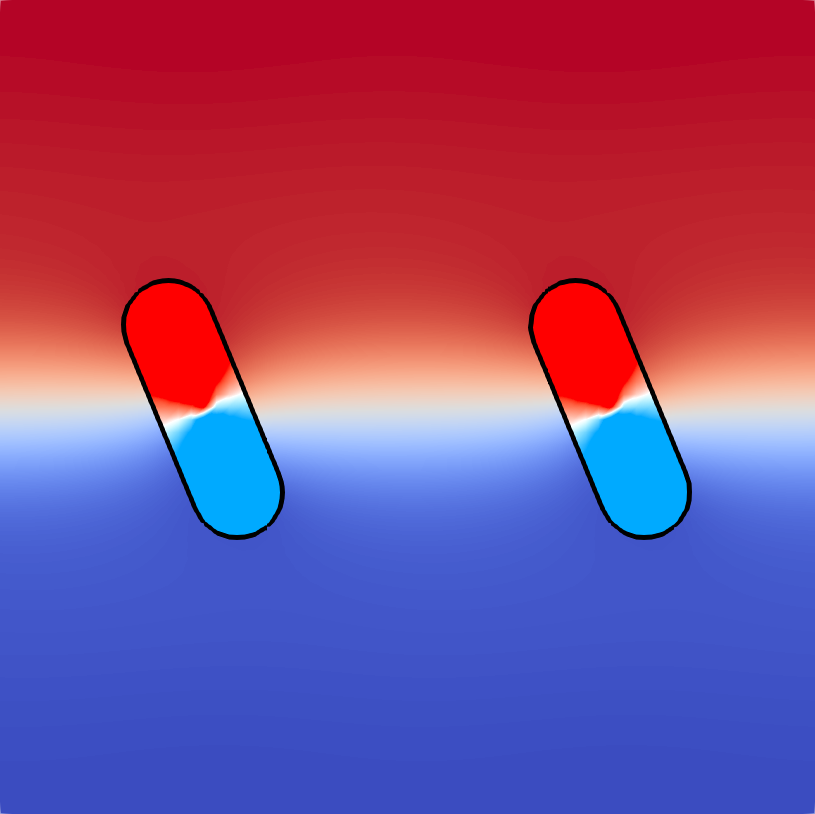} &
\includegraphics[scale=0.05]{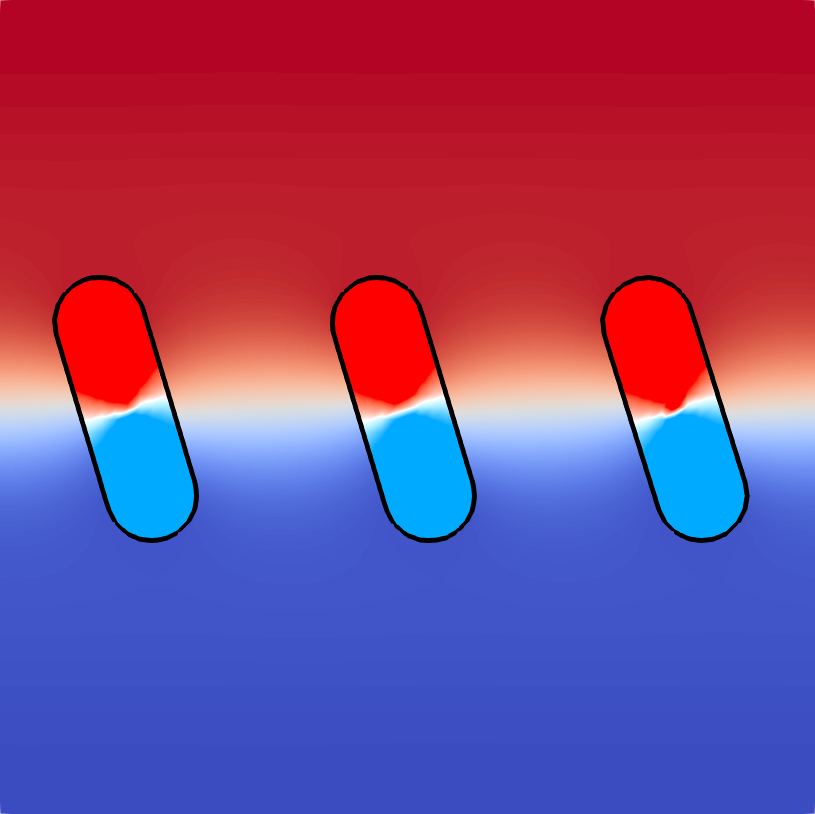} &
\includegraphics[scale=0.05]{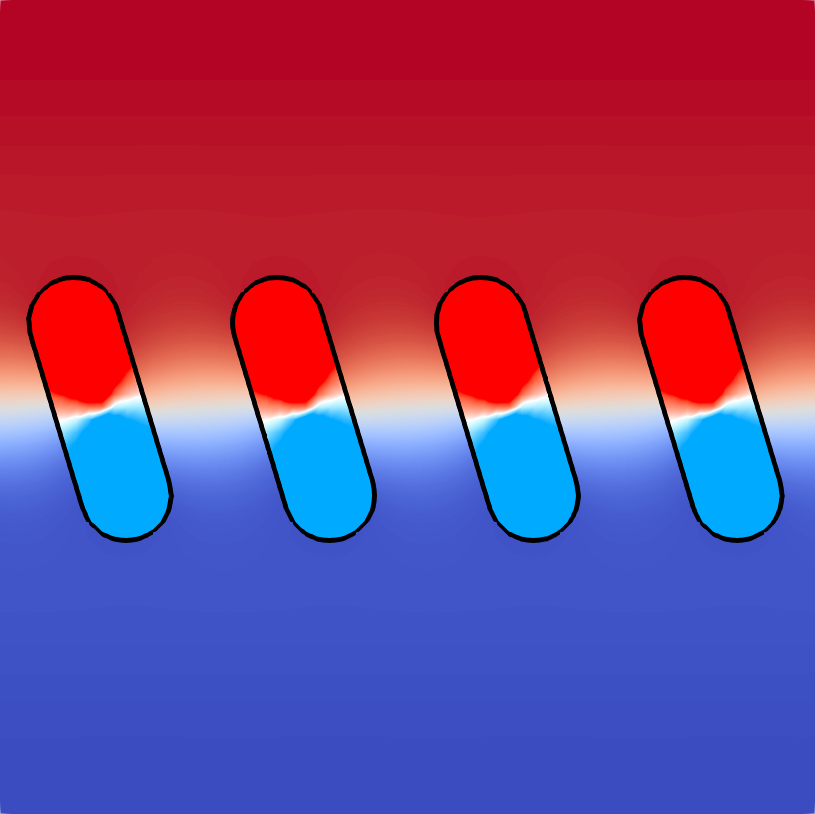} &
\includegraphics[scale=0.05]{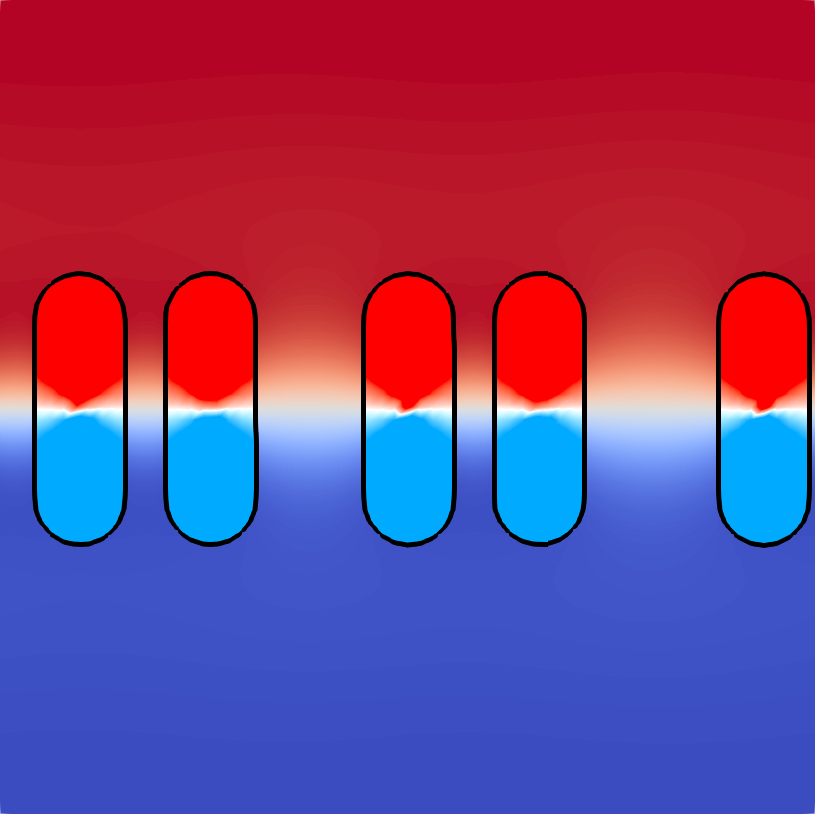} &
\includegraphics[scale=0.05]{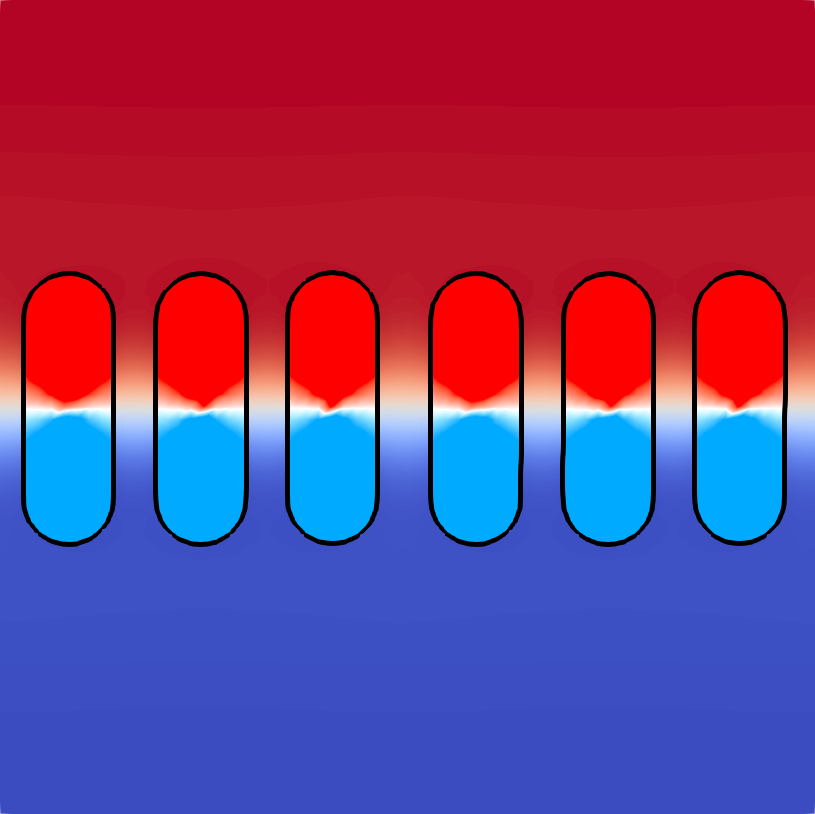} \\
$6$ &
\includegraphics[scale=0.05]{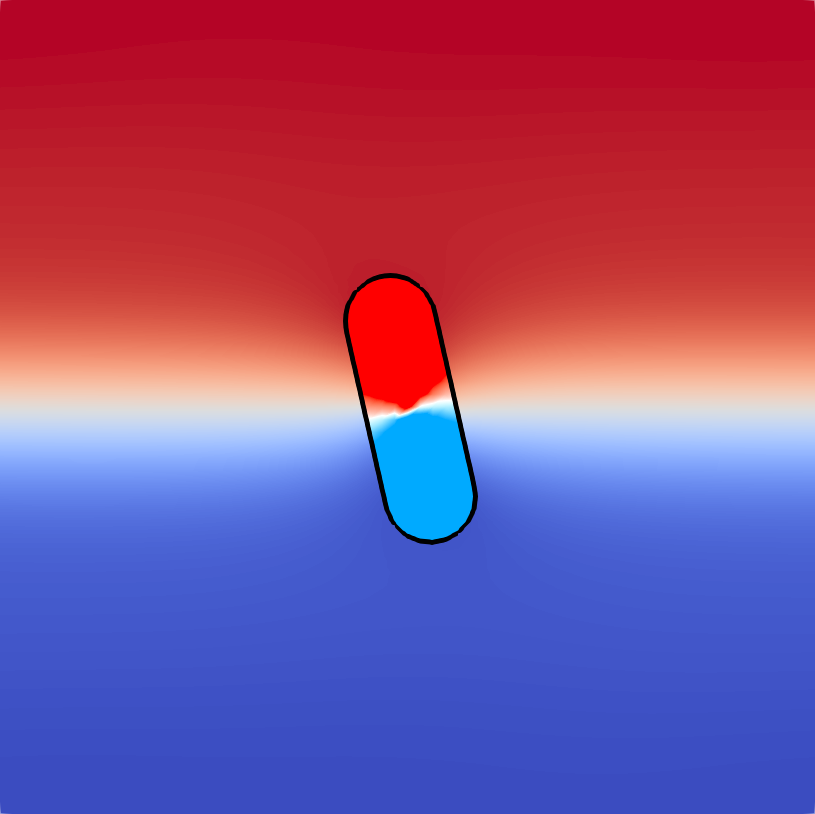} &
\includegraphics[scale=0.05]{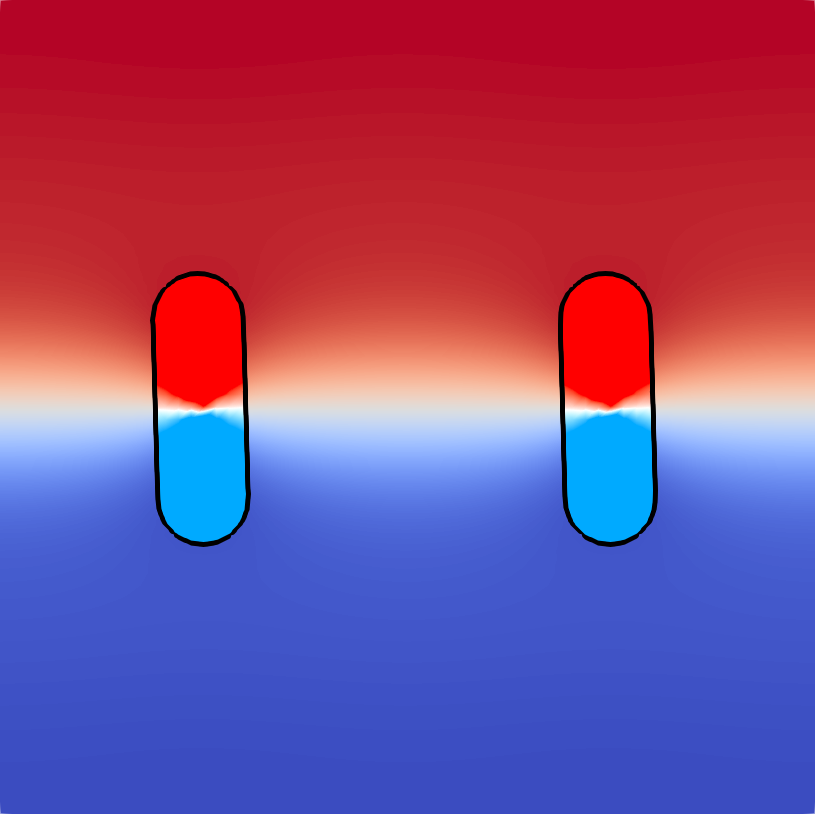} &
\includegraphics[scale=0.05]{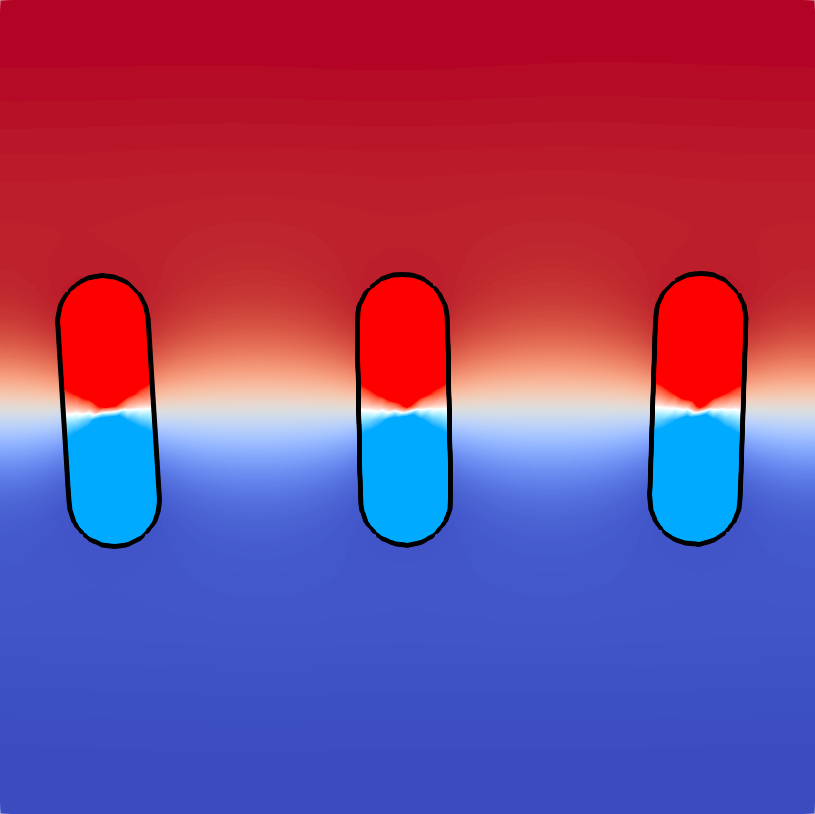} &
\includegraphics[scale=0.05]{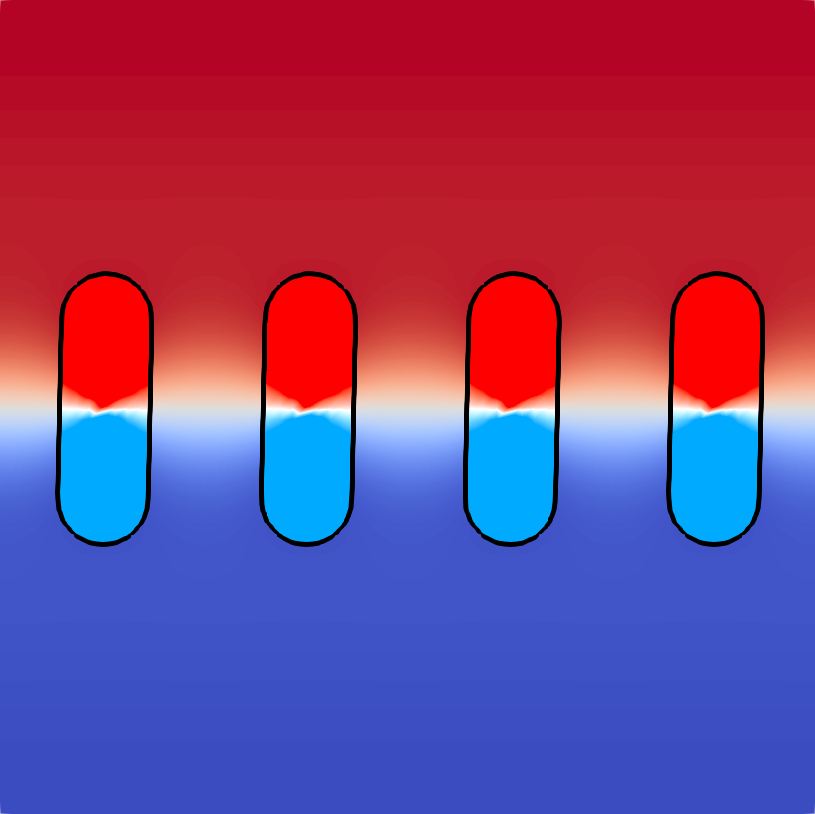} &
\includegraphics[scale=0.05]{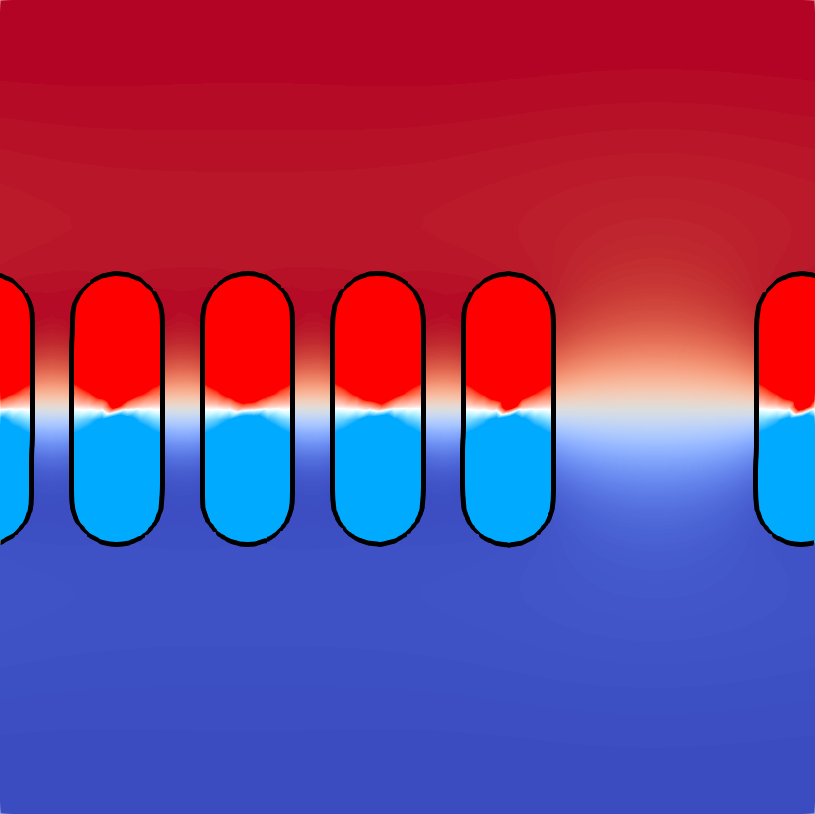} &
\includegraphics[scale=0.05]{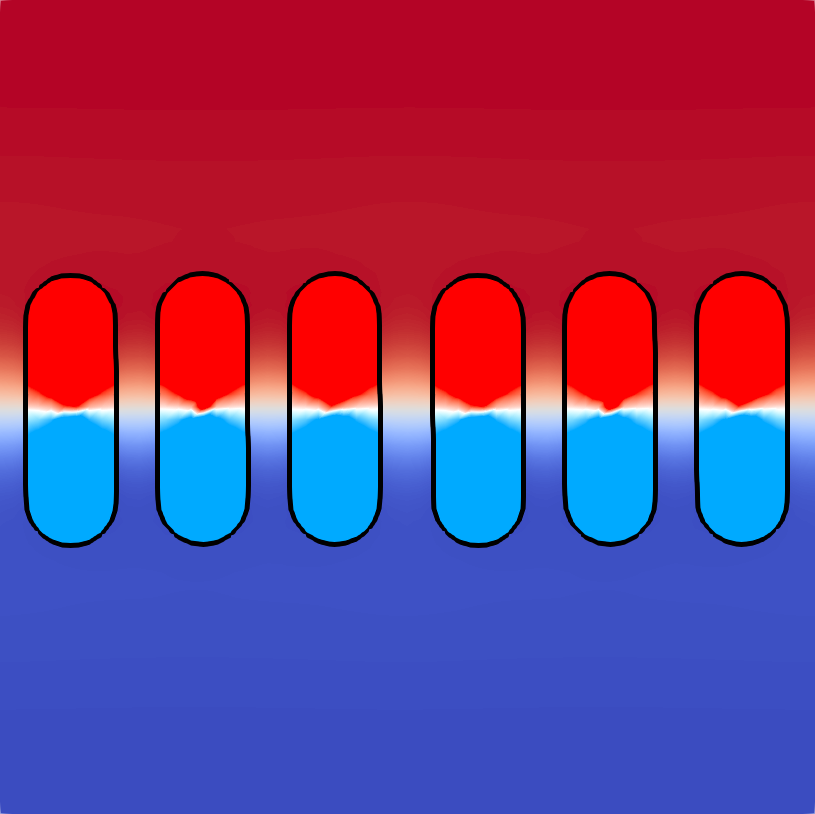} 
\end{tabular}
\caption{Equilibrium orientation of Janus nanorods at the interface between $A$ and $B$ blocks obtained for different values of the particles density and the polarization strength.}
\label{fig:results:nano:density}
\end{figure}

\begin{figure}[!h]
\centering  
\includegraphics[width=.6\textwidth]{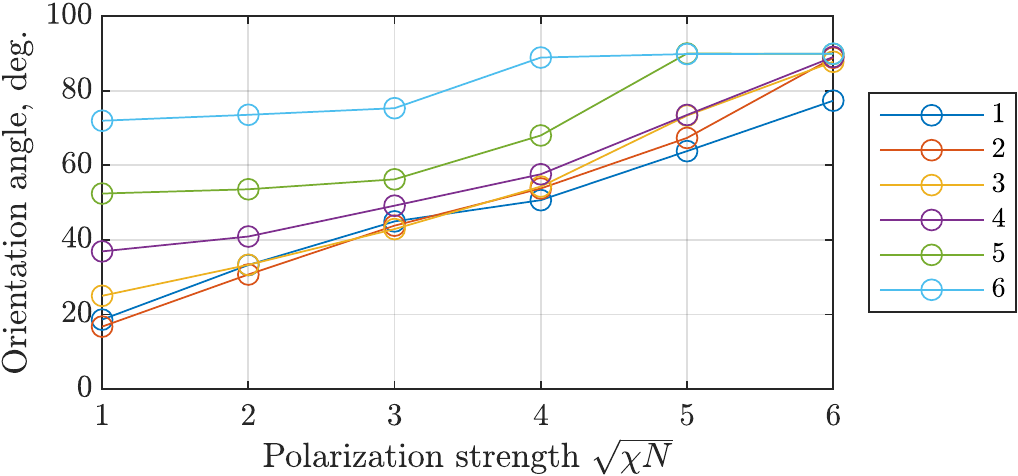}
\caption{Dependence of the Janus nanorod orientation on the polarization strength for several values of the particles density.}
\label{fig:results:nano:density:plot}
\end{figure}

Next, we perform similar simulations but this time with different numbers of nanoparticles inside the computational domain. Figures \ref{fig:results:nano:density} illustrate the obtained particles configurations and figure \ref{fig:results:nano:density:plot} presents the quantitative results describing these configurations. One can observe that, as long as the density of particles is low (1-3 particles) and no crowding occurs, the equilibrium orientations of nanorods are almost identical to those of a isolated nanorod. As the density of particles is increased, the equilibrium orientations of nanorods dramatically changes due to crowding effects. First, the equilibrium orientation angle increases for each polarization strength value as the particle density increases. Second, as the polarization strength decreases to low values the equilibrium orientation converges to a minimal nonzero value, which is dictated by the geometrical constraint due to crowding, and, thus, the accessible range of the particles' orientation is reduced.

To summarize the findings from numerical simulations of the co-assembly of Janus nanorods in a lamellar-forming diblock copolymer, we observed an aggregation of nanoparticles along interfaces between dissimilar polymer species oriented at a well defined angle with respect to these interfaces. The angle value of the equilibrium orientation continuously depend on the nanorod's length and on the polarization strength. High particle densities influence the equilibrium placement of nanorods due to the crowding effect, limiting the minimum achievable angle values. 

\section{Conclusion}\label{sec:conclusion}

In this work we have presented a generalized theory in the context of Self-Consistent Field Theory that describes equilibrium of polymer free surfaces and submerged nanoparticles. The theory applies to free and tethered polymer chains, substrates and nanoparticles of any shapes, and arbitrary non-uniform surface energies and surface grafting densities.
The surface tensions between the polymer material and its surroundings as well as tethered polymer chains are modeled using a consistent unified approach that results in pressure fields free of numerical singularities. The proposed theory has been validated on benchmark examples with imposed velocities and demonstrated a very close match between predicted and actual changes in the system energy. 

The proposed theory was applied to the investigation of the meniscus formation in the context of graphoepitaxy applications for lamellar and cylindrical diblock copolymers. It was shown that in case of the lamella-forming copolymer the polymer-air surface tension, as well as the contact angle value at the air-polymer-wall triple junction point affect the distribution of polymer material inside the guiding groove and also cause the bending of polymer lamellar domains. At the same time the cylinder-forming polymer morphology demonstrated less sensitivity to these parameters. In another example, we investigated the placement and orientation of ``polarized'' nanorod particle in lamellar diblock copolymer. It was demonstrated that such particles aggregate at the interface between dissimilar polymer blocks and that their resulting orientation is controlled by the balance of the particles' length, the ``polarization'' strength, and the particles density.

Future research will include the application of the presented methods to cases of grafted polymers, polymers of more complex architectures, and blends of copolymers. Another direction of future work will be the extension of the presented methods to a more accurate fluctuating field-theoretic description of block copolymers.

\begin{acknowledgement}
This research was supported by NSF DMS 1620471.
\end{acknowledgement}

\bibliographystyle{achemso}
\bibliography{./references}

\newpage
\appendix
\section{Weak formulations and equivalence of modified diffuion equations}\label{app:weak}

To demonstrate the equivalence of \eqref{eq:qf:delta} and \eqref{eq:qf:robin} we invoke their weak forms. Multiplying the diffusion equation in \eqref{eq:qf:delta} by an test function $\psi\of{s,\vecr}$ and integrating over domain $\Omega$ and contour variable range $s\in\range{0}{1}$ we obtain:
\begin{myalign}\label{eq:weak:start}
\myint_{\Omega}
\myint_0^1
\left(
\dds{\qf} + \left(\mu\of{s} + \eta \gamma\of{s}\delta_\Gamma \right) \qf - \lap \qf
\right) \psi 
\diff{\vecr}
\diff{s}
= 0.
\end{myalign}
Using integration by parts and the divergence theorem it is straightforward to show that:
\begin{mymultline}\label{eq:weak:bypartstime}
\myint_{\Omega}
\myint_0^1
\dds{\qf} \psi 
\diff{\vecr}
\diff{s}
= 
\myint_{\Omega}
\dds{ \left( \qf \psi \right)} 
\diff{\vecr}
\diff{s}
-
\myint_{\Omega}
\myint_0^1
 \qf \dds{\psi} 
\diff{\vecr}
\diff{s}
\\
= 
\myint_{\Omega}
\left( 
\qf\of{1,\vecr} \psi\of{1,\vecr} - 
\qf\of{0,\vecr} \psi\of{0,\vecr}
\right)
\diff{\vecr}
\diff{s}
-
\myint_{\Omega}
\myint_0^1
 \qf \dds{\psi} 
\diff{\vecr}
\diff{s}
\end{mymultline}
and
\begin{mymultline}\label{eq:weak:byparts}
\myint_{\Omega}
\myint_0^1
\lap q \psi 
\diff{\vecr}
\diff{s}
=
\myint_{\Omega}
\myint_0^1
\nabla \cdot \left( \nabla q \psi \right)
\diff{\vecr}
\diff{s}
-
\myint_{\Omega}
\myint_0^1
\nabla q \cdot \nabla \psi 
\diff{\vecr}
\diff{s}
\\
=
\myint_{\Gamma}
\myint_0^1
\ddn{q} \psi
\diff{\vecr}
\diff{s}
-
\myint_{\Omega}
\myint_0^1
\nabla q \cdot \nabla \psi 
\diff{\vecr}
\diff{s}.
\end{mymultline}
Using the above two expressions in \eqref{eq:weak:start} along with the initial and boundary conditions for $\qf$ produces the following weak form of \eqref{eq:qf:delta}:
\begin{mymultline*}
\myint_{\Omega}
\left( 
\qf\of{1,\vecr} \psi\of{1,\vecr} - 
\psi\of{0,\vecr}
\right)
\diff{\vecr}
\diff{s}
\\
+
\myint_{\Omega}
\myint_0^1
\left(
- \qf \dds{\psi}
+
\left(\mu\of{s} + \eta \gamma\of{s}\delta_\Gamma \right) \qf \psi 
+
\nabla q \cdot \nabla \psi 
\right)
\diff{\vecr}
\diff{s}
=0, \quad \forall \psi
\end{mymultline*}
Converting the term containing surface delta function $\delta_\Gamma$ into a boundary integral as
\begin{myalign*}
\myint_{\Omega}
\myint_0^1
\eta \gamma\of{s}\delta_\Gamma \qf \psi 
\diff{\vecr}
\diff{s}
=
\myint_{\Gamma}
\myint_0^1
\eta \gamma\of{s} \qf \psi 
\diff{\vecr}
\diff{s}
\end{myalign*}
the obtained weak form can be transformed into
\begin{mymultline*}
\myint_{\Omega}
\left( 
\qf\of{1,\vecr} \psi\of{1,\vecr} - 
\psi\of{0,\vecr}
\right)
\diff{\vecr}
\diff{s}
\\
+
\myint_{\Omega}
\myint_0^1
\left(
- \qf \dds{\psi}
+
\mu\of{s} \qf \psi 
+
\nabla q \cdot \nabla \psi 
\right)
\diff{\vecr}
\diff{s}
\\
+ 
\myint_{\Gamma}
\myint_0^1
\eta \gamma\of{s} \qf \psi 
\diff{\vecr}
\diff{s}
=0, \quad \forall \psi,
\end{mymultline*}
which represents a weak form an altered diffusion equation. Indeed, using formulas \eqref{eq:weak:bypartstime} and \eqref{eq:weak:byparts} in the reverse way, one can write the above expression as 
\begin{mymultline*}
\myint_{\Omega}
\left( 
\qf\of{0,\vecr} - 1
\right)
\psi\of{0,\vecr}
\diff{\vecr}
\diff{s}
\\
+
\myint_{\Omega}
\myint_0^1
\left(
\dds{\qf}
+
\mu\of{s} \qf 
-
\lap q
\right) \psi
\diff{\vecr}
\diff{s}
\\
+ 
\myint_{\Gamma}
\myint_0^1
\left(
\ddn{q} + 
\eta \gamma\of{s} \qf 
\right) \psi 
\diff{\vecr}
\diff{s}
=0, \quad \forall \psi
\end{mymultline*}
from which it immediately follows that $\qf$ satisfies \eqref{eq:qf:robin}. The equivalence of \eqref{eq:qb:delta}-\eqref{eq:qb:robin}, \eqref{eq:qf:robin}-\eqref{eq:qf:almost}, and \eqref{eq:qb:robin}-\eqref{eq:qb:almost} can be demonstrated in the analogous fashion. 

\section{Shape derivatives of integral quantities}\label{app:shape-deriv}
In order to calculate derivatives of integral quantities \eqref{eq:free:integrals} we employ the Level-Set methodology\cite{Sethian:98:Level-set-methods-an,osher2003level}.
\begin{figure}[!h]
\centering
\includegraphics[width=.9\textwidth]{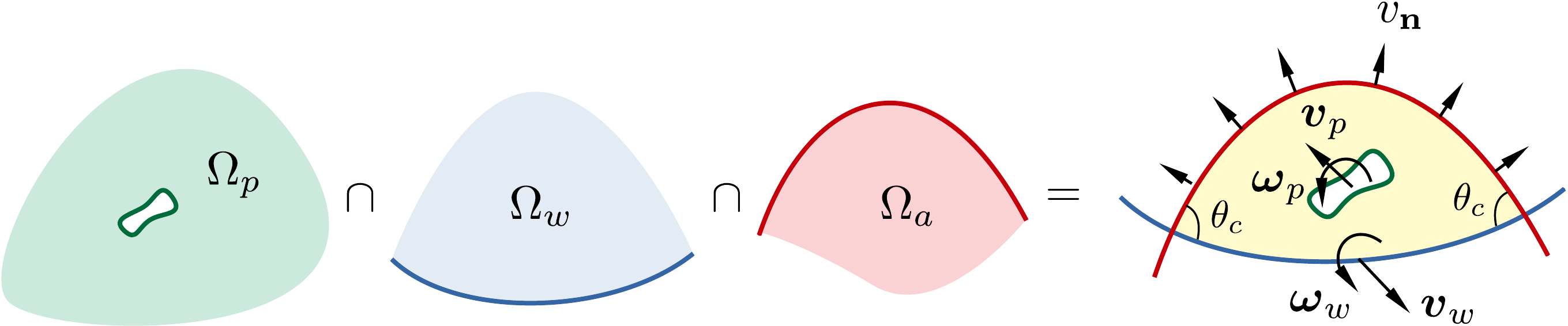}
\caption{Problem geometry and notation used in section \ref{app:shape-deriv}.}
\label{fig:shape-derive:problem}
\end{figure} 
Specifically, we represent domain $\Omega\of{\tau}$ as the intersection of three smooth domains $\Omega_a\of{\tau}$, $\Omega_p\of{\tau}$, and $\Omega_w$ as demonstrated in figure \ref{fig:shape-derive:problem}, and, second, that boundaries of domains $\Omega_a\of{\tau}$, $\Omega_p\of{\tau}$, and $\Omega_w$ are described as zero-isocontours of higher-dimensional functions $\phi_a\of{\tau}$, $\phi_p\of{\tau}$, and $\phi_w$, called \textit{level-set functions}, such that:
\begin{myalign*}
\phi_a\of{\tau}
\begin{cases}
< 0, &\vecr\in\Omega_a\of{\tau}, \\
> 0, &\vecr\notin \Omega_a\of{\tau},
\end{cases}
\quad
\phi_p\of{\tau}
\begin{cases}
< 0, &\vecr\in\Omega_p\of{\tau}, \\
> 0, &\vecr\notin \Omega_p\of{\tau},
\end{cases}
\quad
\tand
\quad
\phi_w
\begin{cases}
< 0, &\vecr\in\Omega_w, \\
> 0, &\vecr\notin \Omega_w.
\end{cases}
\end{myalign*}
In this setting the evolution of domain $\Omega_a\of{\tau}$ in the velocity field $\vn\of{\tau}$ and the motion of domain $\Omega_p\of{\tau}$ with translational and rotational velocities $\vect{v}_p$ and $\vect{\omega}_p$ are described by advection equations for function $\phi_a\of{\tau}$ and $\phi_p\of{\tau}$ as
\begin{myalign}\label{eq:shape-deriv:advection}
\pd{\phi_a}{\tau} + \vn \abs{\nabla \phi_a} &= 0,
\\
\pd{\phi_p}{\tau} + \left(\vect{v}_p - \vecr\times\vect{\omega}_p\right) \cdot \vect{n} \abs{\nabla \phi_p} &= 0,
\\
\pd{\phi_w}{\tau} + \left(\vect{v}_w - \vecr\times\vect{\omega}_w\right) \cdot \vect{n} \abs{\nabla \phi_w} &= 0.
\end{myalign}
The evolution of a field $b\of{\tau}$ that moves along with domain $\Omega_p\of{\tau}$ is described by a similar equation:
\begin{myalign}\label{eq:shape-deriv:advection2}
\pd{b}{\tau} + \left(\vect{v}_p - \vecr\times\vect{\omega}_p\right) \cdot \nabla b &= 0,
\\
\pd{c}{\tau} + \left(\vect{v}_w - \vecr\times\vect{\omega}_w\right) \cdot \nabla c &= 0.
\end{myalign}
The description of the problem geometry in this way allows to write the boundary and domain integrals using the Heaviside step functions $\heav\of{x}$, Dirac delta distribution $\dirac\of{x}$ and the co-area formula as
\begin{myalign*}
\myint_{\Omega} a \diff{\vecr} 
&= 
\myint_{\mathcal{R}^\dims} a \heav\of{-\phi_a} \heav\of{-\phi_p} \heav\of{-\phi_w} \diff{\vecr},
\\ 
\myint_{\Gamma_a} a \diff{\vecr} 
&= 
\myint_{\mathcal{R}^\dims} a \dirac\of{\phi_a} \abs{\nabla\phi_a} \heav\of{-\phi_p} \heav\of{-\phi_w} \diff{\vecr}, 
\\ 
\myint_{\Gamma_p} a b\of{\tau} \diff{\vecr} 
&= 
\myint_{\mathcal{R}^\dims} a b\of{\tau} \heav\of{-\phi_a} \dirac\of{\phi_p} \abs{\nabla\phi_p} \heav\of{-\phi_w} \diff{\vecr},
\\ 
\myint_{\Gamma_w} a c\of{\tau} \diff{\vecr} 
&=
\myint_{\mathcal{R}^\dims} a c\of{\tau} \heav\of{-\phi_a} \heav\of{-\phi_p} \dirac\of{\phi_w} \abs{\nabla\phi_w}  \diff{\vecr}, 
\end{myalign*}
where $\dims$ denotes the problem's dimensionality.
Taking the derivative of these expression with respect to $\tau$ leads to
\begin{multline}\label{eq:shape-deriv:raw}
\pd{}{\tau}\myint_{\Omega} a \diff{\vecr} 
=
-
\myint_{\mathcal{R}^\dims} a \dirac\of{-\phi_a} \pd{\phi_a}{\tau} \heav\of{-\phi_p} \heav\of{-\phi_w} \diff{\vecr}
-
\myint_{\mathcal{R}^\dims} a  \heav\of{-\phi_a} \dirac\of{-\phi_p} \pd{\phi_p}{\tau} \heav\of{-\phi_w} \diff{\vecr}
\\
-
\myint_{\mathcal{R}^\dims} a  \heav\of{-\phi_a} \heav\of{-\phi_p} \dirac\of{-\phi_w} \pd{\phi_w}{\tau}  \diff{\vecr},
\\
\pd{}{\tau}\myint_{\Gamma_a} a \diff{\vecr} 
= 
\myint_{\mathcal{R}^\dims} a \dirac^\prime\of{\phi_a} \pd{\phi_a}{\tau} \abs{\nabla\phi_a} \heav\of{-\phi_p} \heav\of{-\phi_w} \diff{\vecr} 
+ \myint_{\mathcal{R}^\dims} a \dirac\of{\phi_a} \frac{\nabla\phi_a}{\abs{\nabla\phi_a}} \cdot \nabla\pd{\phi_a}{\tau} \heav\of{-\phi_p} \heav\of{-\phi_w} \diff{\vecr}
\\ 
- 
\myint_{\mathcal{R}^\dims} a \dirac\of{\phi_a} \abs{\nabla\phi_a} \dirac\of{-\phi_p} \pd{\phi_p}{\tau} \heav\of{-\phi_w} \diff{\vecr}
- 
\myint_{\mathcal{R}^\dims} a \dirac\of{\phi_a} \abs{\nabla\phi_a} \heav\of{-\phi_p} \dirac\of{-\phi_w} \pd{\phi_w}{\tau} \diff{\vecr},
\\
\pd{}{\tau}\myint_{\Gamma_p} a b\of{\tau} \diff{\vecr} 
= 
\myint_{\mathcal{R}^\dims} a \pd{b}{\tau} \heav\of{-\phi_a} \dirac\of{\phi_p} \abs{\nabla\phi_p} \heav\of{-\phi_w} \diff{\vecr}
-
\myint_{\mathcal{R}^\dims} a b \delta\of{-\phi_a} \pd{\phi_a}{\tau} \dirac\of{\phi_p} \abs{\nabla\phi_p} \heav\of{-\phi_w} \diff{\vecr}
\\
+
\myint_{\mathcal{R}^\dims} a b \heav\of{-\phi_a} \dirac^\prime\of{\phi_p} \pd{\phi_p}{\tau} \abs{\nabla\phi_p} \heav\of{-\phi_w} \diff{\vecr}
+
\myint_{\mathcal{R}^\dims} a b \heav\of{-\phi_a} \dirac\of{\phi_p} \frac{\nabla\phi_p}{\abs{\nabla\phi_p}} \cdot \nabla\pd{\phi_p}{\tau} \heav\of{-\phi_w} \diff{\vecr},
\\
-
\myint_{\mathcal{R}^\dims} a b \heav\of{-\phi_a} \dirac\of{\phi_p} \abs{\nabla\phi_p} \delta\of{-\phi_w} \pd{\phi_w}{\tau} \diff{\vecr}
\\ 
\pd{}{\tau}\myint_{\Gamma_w} a c\of{\tau} \diff{\vecr} 
= 
\myint_{\mathcal{R}^\dims} a \pd{c}{\tau} \heav\of{-\phi_a} \dirac\of{\phi_w} \abs{\nabla\phi_w} \heav\of{-\phi_p} \diff{\vecr}
-
\myint_{\mathcal{R}^\dims} a c \delta\of{-\phi_a} \pd{\phi_a}{\tau} \heav\of{-\phi_p} \dirac\of{\phi_w} \abs{\nabla\phi_w} \diff{\vecr}
\\
+
\myint_{\mathcal{R}^\dims} a c \heav\of{-\phi_a} \heav\of{-\phi_p} \dirac^\prime\of{\phi_w} \pd{\phi_w}{\tau} \abs{\nabla\phi_w} \diff{\vecr}
+
\myint_{\mathcal{R}^\dims} a c \heav\of{-\phi_a} \heav\of{-\phi_p} \dirac\of{\phi_w} \frac{\nabla\phi_w}{\abs{\nabla\phi_w}} \cdot \nabla\pd{\phi_w}{\tau} \diff{\vecr},
\\
-
\myint_{\mathcal{R}^\dims} a c \heav\of{-\phi_a} \delta\of{-\phi_p} \pd{\phi_p}{\tau} \dirac\of{\phi_w} \abs{\nabla\phi_w} \diff{\vecr}, 
\end{multline}
Using the integration by parts\cite{zhao1996variational} the terms containing gradients of temporal derivatives can be transformed as
\begin{mymultline*}
\myint_{\mathcal{R}^\dims} a \dirac\of{\phi_a} \frac{\nabla\phi_a}{\abs{\nabla\phi_a}}  \cdot \nabla\pd{\phi_a}{\tau} \heav\of{-\phi_p} \heav\of{-\phi_w} \diff{\vecr}
=
-
\myint_{\mathcal{R}^\dims} \pd{\phi_a}{\tau} \nabla \cdot \left( a \dirac\of{\phi_a} \frac{\nabla\phi_a}{\abs{\nabla\phi_a}} \heav\of{-\phi_p} \heav\of{-\phi_w} \right) \diff{\vecr}
\\
=
-
\myint_{\mathcal{R}^\dims} \pd{\phi_a}{\tau} \Bigg( \dirac\of{\phi_a} \heav\of{-\phi_p} \heav\of{-\phi_w} 
\left(\nabla \cdot \frac{\nabla\phi_a}{\abs{\nabla\phi_a}} + \frac{\nabla\phi_a}{\abs{\nabla\phi_a}} \cdot \nabla \right) a
\\
+
a \dirac^\prime\of{\phi_a} \abs{\nabla\phi_a} \heav\of{-\phi_p} \heav\of{-\phi_w}
\\
-
a \dirac\of{\phi_a} \dirac\of{-\phi_p} \nabla \phi_p \cdot \frac{\nabla\phi_a}{\abs{\nabla\phi_a}} \heav\of{-\phi_w} 
-
a \dirac\of{\phi_a} \heav\of{-\phi_p} \dirac\of{-\phi_w} \nabla \phi_w \cdot \frac{\nabla\phi_a}{\abs{\nabla\phi_a}} \Bigg) \diff{\vecr}
\\
=
-
\myint_{\mathcal{R}^\dims} \pd{\phi_a}{\tau} \Bigg( \dirac\of{\phi_a} \heav\of{-\phi_p} \heav\of{-\phi_w} 
\left(\nabla \cdot \frac{\nabla\phi_a}{\abs{\nabla\phi_a}} + \frac{\nabla\phi_a}{\abs{\nabla\phi_a}} \cdot \nabla \right) a
\\
+
a \dirac^\prime\of{\phi_a} \abs{\nabla\phi_a} \heav\of{-\phi_p} \heav\of{-\phi_w}
-
a \dirac\of{\phi_a} \heav\of{-\phi_p} \dirac\of{-\phi_w} \nabla \phi_w \cdot \frac{\nabla\phi_a}{\abs{\nabla\phi_a}} \Bigg) \diff{\vecr}.
\end{mymultline*}
\begin{mymultline*}
\myint_{\mathcal{R}^\dims} a b \heav\of{-\phi_a}  \dirac\of{\phi_p} \frac{\nabla\phi_p}{\abs{\nabla\phi_p}} \cdot \nabla\pd{\phi_a}{\tau} \heav\of{-\phi_w} \diff{\vecr}
=
-
\myint_{\mathcal{R}^\dims} \pd{\phi_p}{\tau} \nabla \cdot \left( a b \heav\of{-\phi_a} \dirac\of{\phi_p} \frac{\nabla\phi_p}{\abs{\nabla\phi_p}} \heav\of{-\phi_w} \right) \diff{\vecr}
\\
=
-
\myint_{\mathcal{R}^\dims} \pd{\phi_p}{\tau} \Bigg( 
\heav\of{-\phi_a} \dirac\of{\phi_p} \heav\of{-\phi_w}
\left(\nabla \cdot \frac{\nabla\phi_p}{\abs{\nabla\phi_p}} + \frac{\nabla\phi_p}{\abs{\nabla\phi_p}} \cdot \nabla \right) \left( a b \right)
\\
+
a b \heav\of{-\phi_a} \dirac^\prime\of{\phi_p} \abs{\nabla\phi_p} \heav\of{-\phi_w}
\\
-
a b \dirac\of{-\phi_a} \nabla \phi_a \cdot \frac{\nabla\phi_p}{\abs{\nabla\phi_p}} \dirac\of{\phi_p} \heav\of{-\phi_w} 
-
a b \heav\of{-\phi_a} \dirac\of{\phi_p} \dirac\of{-\phi_w} \nabla \phi_w \cdot \frac{\nabla\phi_p}{\abs{\nabla\phi_p}} \Bigg) \diff{\vecr}
\\
=
-
\myint_{\mathcal{R}^\dims} \pd{\phi_p}{\tau} \Bigg( 
\heav\of{-\phi_a} \dirac\of{\phi_p} \heav\of{-\phi_w}
\left(\nabla \cdot \frac{\nabla\phi_p}{\abs{\nabla\phi_p}} + \frac{\nabla\phi_p}{\abs{\nabla\phi_p}} \cdot \nabla \right) \left( a b \right)
\\
+
a b \heav\of{-\phi_a} \dirac^\prime\of{\phi_p} \abs{\nabla\phi_p} \heav\of{-\phi_w}
\Bigg) \diff{\vecr}.
\end{mymultline*}

\begin{figure}[!h]
\centering
\includegraphics[width=.25\textwidth]{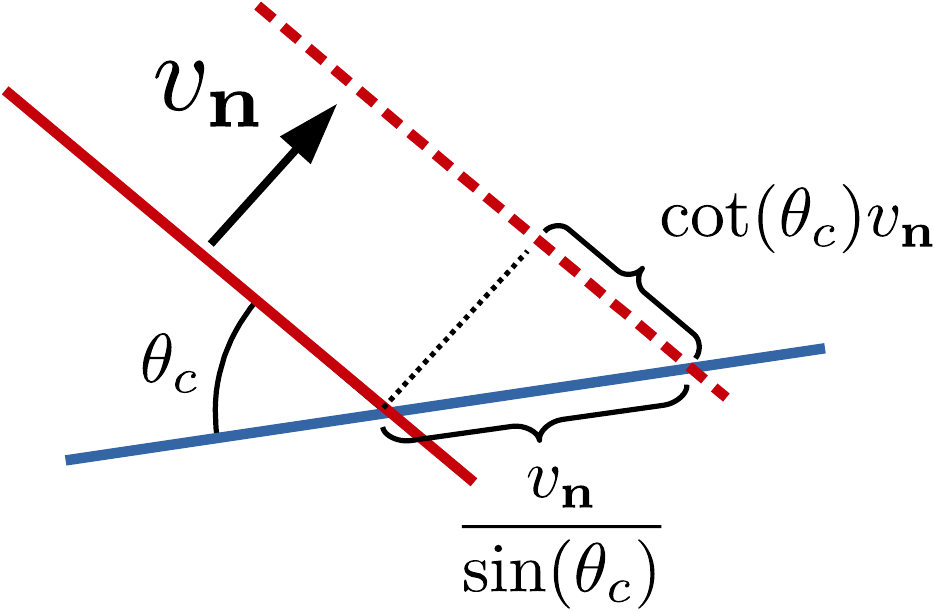}
\caption{Interpretation of resulting expression for shape derivatives in case of sharp features.}
\label{fig:shape-derive:meaning}
\end{figure}

\begin{mymultline*}
\myint_{\mathcal{R}^\dims} a c \dirac\of{\phi_w} \frac{\nabla\phi_w}{\abs{\nabla\phi_w}}  \cdot \nabla\pd{\phi_w}{\tau} \heav\of{-\phi_p} \heav\of{-\phi_a} \diff{\vecr}
=
-
\myint_{\mathcal{R}^\dims} \pd{\phi_w}{\tau} \nabla \cdot \left( a c \dirac\of{\phi_w} \frac{\nabla\phi_w}{\abs{\nabla\phi_w}} \heav\of{-\phi_p} \heav\of{-\phi_a} \right) \diff{\vecr}
\\
=
-
\myint_{\mathcal{R}^\dims} \pd{\phi_w}{\tau} \Bigg( \dirac\of{\phi_w} \heav\of{-\phi_p} \heav\of{-\phi_a} 
\left(\nabla \cdot \frac{\nabla\phi_w}{\abs{\nabla\phi_w}} + \frac{\nabla\phi_w}{\abs{\nabla\phi_w}} \cdot \nabla \right) a c
\\
+
a c \dirac^\prime\of{\phi_w} \abs{\nabla\phi_w} \heav\of{-\phi_p} \heav\of{-\phi_a}
\\
-
a c \dirac\of{\phi_w} \dirac\of{-\phi_p} \nabla \phi_p \cdot \frac{\nabla\phi_w}{\abs{\nabla\phi_w}} \heav\of{-\phi_a} 
-
a c \dirac\of{\phi_w} \heav\of{-\phi_p} \dirac\of{-\phi_a} \nabla \phi_a \cdot \frac{\nabla\phi_w}{\abs{\nabla\phi_w}} \Bigg) \diff{\vecr}
\\
=
-
\myint_{\mathcal{R}^\dims} \pd{\phi_w}{\tau} \Bigg( \dirac\of{\phi_w} \heav\of{-\phi_p} \heav\of{-\phi_a} 
\left(\nabla \cdot \frac{\nabla\phi_w}{\abs{\nabla\phi_w}} + \frac{\nabla\phi_w}{\abs{\nabla\phi_w}} \cdot \nabla \right) a c
\\
+
a c \dirac^\prime\of{\phi_w} \abs{\nabla\phi_w} \heav\of{-\phi_p} \heav\of{-\phi_a}
-
a c \dirac\of{\phi_w} \heav\of{-\phi_p} \dirac\of{-\phi_a} \nabla \phi_a \cdot \frac{\nabla\phi_w}{\abs{\nabla\phi_w}} 
\Bigg) \diff{\vecr}.
\end{mymultline*}
Substituting the above expression into \eqref{eq:shape-deriv:raw}, using advection equation \eqref{eq:shape-deriv:advection} and the fact that the normal vectors and curavtures of boundaries of $\Omega_a$, $\Omega_p$, and $\Omega_w$ are equal to 
\begin{alignat*}{2}
\vect{n}_a &= \frac{\nabla \phi_a}{\abs{\nabla \phi_a}}, \quad
&\kappa_a &= \nabla \cdot \left( \frac{\nabla \phi_a}{\abs{\nabla \phi_a}} \right), \\
\vect{n}_p &= \frac{\nabla \phi_p}{\abs{\nabla \phi_p}}, \quad
&\kappa_p &= \nabla \cdot \left( \frac{\nabla \phi_p}{\abs{\nabla \phi_p}} \right), \\
\vect{n}_w &= \frac{\nabla \phi_w}{\abs{\nabla \phi_w}}, \quad
&\kappa_w &= \nabla \cdot \left( \frac{\nabla \phi_w}{\abs{\nabla \phi_w}} \right),
\end{alignat*}
results in the following formulas
\begin{myalign*}
\begin{aligned}
\pd{}{\tau}\myint_{\Omega} a \diff{\vecr} 
&=
\myint_{\mathcal{R}^\dims} a \vn \dirac\of{\phi_a} \abs{\nabla{\phi_a}} \heav\of{-\phi_p} \heav\of{-\phi_w} \diff{\vecr}
\\
&+
\myint_{\mathcal{R}^\dims} a \left(\vect{v}_p - \vect{\omega}_p \times \vecr \right) \cdot \vect{n}_p \heav\of{-\phi_a} \dirac\of{\phi_p} \abs{\nabla{\phi_p}} \heav\of{-\phi_w} \diff{\vecr}
\\
&+
\myint_{\mathcal{R}^\dims} a \left(\vect{v}_w - \vect{\omega}_w \times \vecr \right) \cdot \vect{n}_w \heav\of{-\phi_a} \heav\of{-\phi_p} \dirac\of{\phi_w} \abs{\nabla{\phi_w}} \diff{\vecr},  
\\ 
\pd{}{\tau}\myint_{\Gamma_a} a \diff{\vecr} 
&= 
\myint_{\mathcal{R}^\dims} \vn \dirac\of{\phi_a} \abs{\nabla\phi_a} \heav\of{-\phi_p} \heav\of{-\phi_w} 
\left(\kappa + \ddn{} \right) a
\diff{\vecr}
\\
&+
\myint_{\mathcal{R}^\dims} 
\left[
\left(\vect{v}_w - \vect{\omega_w} \times \vecr \right) \cdot \vect{n}_w
- \vn \vect{n}_w \cdot \vect{n}_a
\right]
a \dirac\of{\phi_a} \abs{\nabla\phi_a} \heav\of{-\phi_p} \dirac\of{-\phi_w} \abs{\nabla \phi_w} 
\diff{\vecr},
\\ 
\pd{}{\tau}\myint_{\Gamma_p} a b\of{\tau} \diff{\vecr} 
&= 
\myint_{\mathcal{R}^\dims} 
\left(\vect{v}_p - \vect{\omega}_p \times \vecr \right) \cdot 
\Big[
a  \nabla b + 
\vect{n}_p
\left(\kappa + \ddn{} \right) 
\left( a b \right)
\Big]
\heav\of{-\phi_a} \dirac\of{\phi_p} \abs{\nabla\phi_p} \heav\of{-\phi_w} 
\diff{\vecr},
\\ 
\pd{}{\tau}\myint_{\Gamma_w} a c\of{\tau} \diff{\vecr} 
&=
\myint_{\mathcal{R}^\dims} a c 
\left( \vn - \left(\vect{v}_w - \vect{\omega}_w\times\vecr \right) \cdot \vect{n}_w \vect{n}_w\cdot\vect{n}_a \right)
\dirac\of{\phi_a} \abs{\nabla{\phi_a}} \heav\of{-\phi_p} \dirac\of{\phi_w} \abs{\nabla\phi_w} \diff{\vecr}
\\
&+
\myint_{\mathcal{R}^\dims} 
\left(\vect{v}_w - \vect{\omega}_w \times \vecr \right) \cdot 
\Big[
a  \nabla c + 
\vect{n}_w
\left(\kappa + \ddn{} \right) 
\left( a c \right)
\Big]
\heav\of{-\phi_a} \heav\of{-\phi_p} \dirac\of{\phi_w} \abs{\nabla\phi_w} 
\diff{\vecr},
\end{aligned}
\end{myalign*}
which after applying the coarea formulas and converting integrals into more traditional notation become \eqref{eq:free:integrals:derivs}. Note that these expressions contain integrals over the intersection of boundaries of $\Omega_a$ and $\Omega_w$. These terms can be interpreted as additional surface generation as the boundary of $\Omega_a$ travels along the boundary of $\Omega_w$ (see figure \ref{fig:shape-derive:meaning}).

\end{document}